\documentstyle[sprocl]{article}
\input psfig.sty

\newcounter{temp}
\def\baeq{\setcounter{temp}{\theequation}
\def\CS{{\cal{S}}}
\def\CF{{\cal{F}}}
\addtocounter{temp}{1}
\setcounter{equation}{0}
\renewcommand{\theequation}{\arabic{temp}\alph{equation}}}
\def\eaeq{\setcounter{equation}{\thetemp}
\renewcommand{\theequation}{\arabic{equation}}}
\def\dmu{\partial_{\mu}}
\def\psib{\bar{\psi}}
\def\tG{\tilde{G}}

\def\g5{\gamma_5}

\def\ab{\bar{\alpha}}
\def\be{\begin{equation}}
\def\ee{\end{equation}}
\def\bea{\begin{eqnarray}}
\def\eea{\end{eqnarray}}

\def\pslh{p\llap/}
\def\pslt{p\llap/_T}
\def\eslt{E\llap/_T}
\def\dsl{\partial \llap/}
\def\to{\rightarrow}
\def\Re{{\cal R \mskip-4mu \lower.1ex \hbox{\it e}}\,}
\def\Im{{\cal I \mskip-5mu \lower.1ex \hbox{\it m}}\,}

\def\te{\tilde e}
\def\tl{\tilde l}
\def\tf{\tilde f}
\def\tc{\tilde c}
\def\tu{\tilde u}
\def\td{\tilde d}
\def\ttau{\tilde \tau}
\def\tb{\tilde b}
\def\tg{\tilde g}

\def\tnu{\tilde\nu}
\def\tmu{\tilde\mu}
\def\tell{\tilde\ell}
\def\tq{\tilde q}
\def\tt{\tilde t}
\def\tw{\widetilde W}
\def\tz{\widetilde Z}

\def\sgn{\mathop{\rm sgn}}
\def\alt{\stackrel{<}{\sim}}
\def\agt{\stackrel{>}{\sim}}
\def\mhf{m_{\frac{1}{2}}}

\begin{document}
\hfill{UH-511-872-97}
\title{WHAT IS SUPERSYMMETRY AND HOW DO WE FIND IT?
\footnote{Lectures presented at the IX Jorge A.~Swieca
Summer School, Campos do Jord\~ao, Brazil, February 1997.}}

\author{XERXES TATA }

\address{Department of Physics and Astronomy, University of Hawaii, \\
Honolulu, HI 96822, USA}


\maketitle
\abstract{In these Lectures, we present a pedagogical introduction to weak
scale supersymmetry phenomenology. A
basic understanding of the Standard Model and of the ideas behind Grand
Unification, but no prior knowledge of supersymmetry, is assumed. 
Topics covered include:

\begin{itemize}
\item What is supersymmetry and why do we bother with it?
\item Working with a supersymmetric theory: A toy example
\item Construction of supersymmetric Lagrangians
\item The Minimal Supersymmetric Model 
\item The mSUGRA Model: A paradigm for SUSY phenomenology
\item Decays of supersymmetric particles
\item Production of supersymmetric particles at colliders
\item Observational constraints on supersymmetry
\item Supersymmetry searches at future colliders
\item Constraining supersymmetry models at future colliders
\item R-parity violation 
\item Gauge-mediated supersymmetry breaking
\end{itemize}
} 
\section{Why Is the TeV Scale Special?}

The 1970's witnessed the emergence of what has now become the Standard 
Model~\cite{STAND} (SM) of particle physics. This is a non-Abelian gauge theory based
on the gauge group $SU(3)_C \times SU(2)_L \times U(1)_Y$. The left- and 
right-handed components of the matter fermions are assigned to different
representations of the gauge group, thereby allowing a chiral structure
for the weak interactions. It is further assumed that the gauge symmetry is
spontaneously broken to the observed $SU(3)_C \times U(1)_{em}$ symmetry  
by a single $SU(2)_L$ doublet of spin zero fields that acquires a vacuum 
expectation value (VEV). The $SU(2)_L \times U(1)_Y$ structure of electroweak
interactions was strikingly confirmed with the discovery of the $W^{\pm}$
and $Z^0$ bosons at CERN. During the last few years, the beautiful
measurements~\cite{LEPC}
of the properties of the $Z^0$ boson have allowed us to test~\cite{HAG} electroweak 
theory at the $10^{-3}$ level. 
The QCD part of the SM has not been tested at the same level. Currently,
QCD tests are mostly confined to the domain where the theory can be
treated perturbatively. Unfortunately, this precludes the use of most of the
experimental data on strong interactions; {\it viz.} the observed
properties of hadrons, for QCD tests. In the future, lattice
computations may change this state of affairs. It would indeed be extremely
interesting if the lattice community could come up with an incisive test
that could (in principle) unambiguously falsify QCD. Despite this,
we should acknowledge that the SM has been spectacularly successful in
accounting for a variety of experimental data spanning a vast range of energy.
  
Why then do we entertain the possibility of any physics beyond the
SM?~\footnote{Sometimes physicists have entertained the possibility that
observed deviations between experiment and theoretical predictions point
toward new physics. Frequently, the effects discussed are at the
$2-3\sigma$ level. It should, however, be remembered that the chance of
a large number of independent measurements deviating by $\geq
2\sigma$ is significant; {\it e.g.} the chance that ten independent
measurements all yield agreement within $2\sigma$ is just
60\%. Moreover, the assessment of theoretical and experimental errors is
not straightforward. Here we conservatively assume that there is no
significant deviation between experimental observations and SM
predictions.}  First, we do not really know that electroweak symmetry is
broken by the VEV of a spin zero elementary field, let alone, that this
electroweak symmetry breaking (EWSB) sector consists of just one
$SU(2)_L$ doublet as is assumed in the SM. Understanding the mechanism
of EWSB is one of the most pressing questions of particle physics
today. There are also aesthetic reasons to believe that the SM is not
the complete story. It does not provide any explanation of particle
masses or mixing patterns. The SM, therefore, contains a large number of
arbitrary parameters. Moreover, one needs to make an {\it ad hoc} choice
of gauge group and particle multiplets. Also, the SM offers no
explanation for the replication of generations. Finally, we should
always keep in mind that the SM does not incorporate gravity.

Perhaps more to the point is a technical problem~\cite{SUSS} that arises
in quantum field theories with elementary spin zero fields. Ignoring
gravitational interactions (so that the vacuum energy is not relevant),
the largest quantum corrections are to scalar masses: the radiative
correction ($\delta m_H$) to the scalar boson mass diverges
quadratically as the internal momentum in the loop becomes very
large. This divergence is unphysical since our SM computation breaks
down for loop momenta $p^2 \sim \Lambda^2$, where $\Lambda$ denotes the
energy scale at which the SM ceases to be an adequate description of
nature. This breakdown could occur because of  
form factor effects which become important
at the scale $\Lambda$, or because there are new degrees of freedom at
this scale that are not part of the SM. For instance, $\Lambda \sim
M_{GUT}$ if the SM is embedded in a Grand Unified Theory (GUT) since the
effects of GUT boson exchanges become important for $p^2 \sim \Lambda^2$.
The scale $\Lambda$ thus serves as a cut-off on loop integrals in the
sense that effects not included in the SM become important above this scale,
and serve to dynamically regulate the integral. In lowest order in
perturbation theory, we would then write the physical scalar boson mass
as,
\begin{equation}
m_H^2 = m_0^2 +\delta m_H^2 \sim m_0^2 - g^2\Lambda^2,
\end{equation}
where $m_0$ is the bare Higgs boson mass parameter and $g$ a
dimensionless coupling constant. We will assume that all dimensionless
constants and ratios are of ${\cal{O}}(1)$. From perturbative unitarity arguments~\cite{DICUS} 
we believe that $m_H$ is not larger than a few hundred GeV, so that if
$\Lambda$ is indeed as large as $M_{GUT}$, the two terms on the right
hand side of the equation, each of which 
is $\sim 10^{30}$~GeV$^2$, 
have to combine to yield an answer $\leq 10^6$~GeV$^2$. While this
possibility cannot be logically excluded, the incredible sensitivity of
the theory to the input parameters is generally regarded as a
shortcoming of field theories with elementary scalars. 

If we turn this reasoning around, and require as a matter of principle 
that the theory should not require this incredible fine tuning of
parameters, we would be led to conclude that 
\begin{equation}
\Lambda \alt \ 1000 \  GeV.
\end{equation}
If this perturbative estimate is valid, we must conclude that new
physics effects not included in the SM must manifest themselves in
collisions of elementary particles at about the TeV energy scale. What
form this New Physics will take is unknown.
We do not even know
whether it will be in the form of direct production of new particles
or indication of structure (via form factors) for particles that we 
currently regard to be elementary. 
Possibilities that have been considered in the
literature include technicolour,~\cite{TECH} compositeness of leptons
and quarks~\cite{COMP} and
supersymmetry.~\cite{SUSY,WZ} It is the last of these alternatives that forms the
subject of these lectures.~\cite{LECTURES}

It is important to note that even though we do not know what the New
Physics might be, its scale has been fixed to be $\alt 1$~TeV. Along
with the search for the Higgs boson, the only missing ingredient of the
SM, the search for novel phenomena which are expected to occur at TeV
energy is the primary reason for the construction of supercolliders such
as the Large Hadron Collider~\cite{CMS,ATLAS} (LHC) or a 0.5-2~TeV
electron-positron collider.~\cite{ZDR,JLC} It is worth remarking that
strong interactions in the EWSB sector could invalidate the perturbative
argument that led to the bound (2). The search for effects of these new
strong interactions at colliders poses~\cite{STRONG} a formidable
experimental challenge.

Before closing this Section, we remark that the instability of the mass
to radiative corrections is endemic to spin zero fields. Chiral symmetry
and gauge symmetry, respectively protect fermions and 
gauge bosons from large radiative corrections to their masses.
For instance, in quantum electrodynamics, the corrections to the
electron mass is only logarithmically divergent, and hence, by
dimensional analysis must have the form,
\begin{displaymath}
\delta m \propto m \ln \Lambda,
\end{displaymath}
since $m$ is the only mass scale in the problem. Massless fermions,
therefore, are protected from acquiring masses, a property that can be
traced to the chiral symmetry of QED. Likewise, the photon remains
massless due to the gauge symmetry. In a generic quantum field theory,
however, there is no known symmetry that keeps scalars from acquiring
large masses by radiative corrections without resorting to fine-tuning.
There is, however, a special class of theories in which this is not
necessary. The price paid is that for every known particle, one has to
introduce a new partner with spin differing by $\frac{1}{2}$. The
properties of the known particles and their new partners are related by
a symmetry. This symmetry is unlike any known symmetry in that it
inter-relates properties of bosons and fermions. Such a
symmetry~\cite{REV} is known as a supersymmetry (SUSY). These
supersymmetric partners constitute the new physics that we alluded to
above. It should now be clear that if SUSY is to
ameliorate~\cite{KAUL,DIMO} the fine tuning problem, supersymmetric
partners should be lighter than $\sim 1$~TeV, so that they can be
searched for at supercolliders.

\section{An Introduction to Supersymmetry}
In order to describe what supersymmetric particles would look
like in experiments at high energy colliders, we have to understand how
they might be produced, and, if they are unstable, into what these
particles decay. In other words, we have to understand their
interactions. We should mention at the outset that as yet no compelling
model has emerged (primarily because of our ignorance of physics at high
energy scales). Nonetheless, there is a useful
(albeit cumbersome)  parametrization of the effective theory that can be
used for phenomenological analyses. Before delving into the
complications of constructing realistic SUSY field theories, we will
illustrate the essential ideas of supersymmetry using a simple
example first written down by Wess and Zumino \cite{WZ}.

\subsection{Working with a SUSY field Theory: A Toy Model}

Consider a field theory with the Lagrangian given by,
\setcounter{temp}{\theequation}
\addtocounter{temp}{1}
\setcounter{equation}{0}
\renewcommand{\theequation}{\arabic{temp}\alph{equation}}
\begin{equation}
\cal{L}=\cal{L}_{\it kin}+\cal{L}_{\it mass},
\end{equation}
where,
\begin{equation}
{\cal{L}}_{\it kin} = \frac{1}{2}(\partial_{\mu}A)^2 +
\frac{1}{2}(\partial_{\mu}B)^2 + \frac{i}{2}\bar{\psi}\dsl\psi +
\frac{1}{2}(F^2+G^2), 
\end{equation}
and 
\begin{equation}
{\cal{L}}_{mass} = -m \left [\frac{1}{2}\bar{\psi}\psi -GA -FB \right ].
\end{equation}
\setcounter{equation}{\thetemp}
\renewcommand{\theequation}{\arabic{equation}}
Here, $A$, $B$, $F$ and $G$ are real scalar fields, and $\psi$ is a
self-conjugate or Majorana spinor field satisfying,
\renewcommand{\theequation}{\arabic{equation}}
\setcounter{equation}{3}
\begin{equation}
\psi = C\bar{\psi}^T
\end{equation}
where the charge conjugation matrix $C$ satifies
\baeq
\begin{equation}
C\gamma_{\mu}^T C^{-1} = -\gamma_{\mu}, \nonumber,
\end{equation}
\begin{equation}
C^T=C^{-1}=-C, 
\end{equation}
and
\begin{equation}
[C,\gamma_5 ]=0.
\end{equation}
\eaeq
Notice that (4) is a constraint equation that tells us
that only two of the four components of $\psi$ are independent. This can
be easily seen by projecting out the right-handed component in (4) to
get
\begin{equation}
\psi_R=C\gamma_0\psi_L^*
\end{equation}

Bilinears of Majorana spinors also have very special properties. For
instance,
\baeq
\begin{equation}
\bar{\psi}\chi =\psi^T C\chi = \psi_{\alpha}C_{\alpha\beta}\chi_{\beta}
= -\chi_{\beta}(-C_{\beta\alpha})\psi_{\alpha} = \chi^T C \psi =
\bar{\chi}\psi, 
\end{equation}
where the first minus sign in the fourth step is due to the
anticommutativity of the spinor fields and the second one due to the
antisymmetry (5b) of the matrix $C$. Similarly, one can show that
\begin{eqnarray}
\bar{\psi}\gamma_5 \chi = \bar{\chi}\gamma_5 \psi, \\
\bar{\psi}\gamma_{\mu} \chi = -\bar{\chi}\gamma_{\mu} \psi, \\
\bar{\psi}\gamma_{\mu}\gamma_5 \chi = \bar{\chi}\gamma_{\mu}\gamma_5
\psi, \\
\bar{\psi}\sigma_{\mu\nu} \chi = -\bar{\chi}\sigma_{\mu\nu} \psi. 
\end{eqnarray}
\eaeq

Wess and Zumino \cite{WZ} observed that under the transformations,
\baeq
\begin{eqnarray}
\delta A &=&i\bar{\alpha}\gamma_5\psi ,\\
\delta B &=&-\bar{\alpha}\psi ,\\
\delta\psi &=&-F\alpha+iG\gamma_5\alpha+\dsl\gamma_5 A\alpha +i\dsl B\alpha ,\\
\delta F &=&i\bar{\alpha}\dsl \psi ,\\
\delta G &=&\bar{\alpha}\gamma_5 \dsl \psi
\end{eqnarray}
\eaeq
the Lagrangian density (3) changes by a total derivative. The action then
changes by just a surface term, and the equations of motion remain unchanged.
Before verifying this, we note that the transformations (8) mix boson
and fermion fields; {\it i.e.} the invariance of the equations of motion is
the result of a supersymmetry. The parameter of the transformation
$\alpha$ is thus spinorial. Furthermore, to preserve the reality of the
bosonic fields $A$, $B$, $F$ and $G$, as well as the Majorana nature of
$\psi$, $\alpha$ itself must satisfy the Majorana property (4). To
verify that (3) indeed changes by a total derivative under the
transformations (8), we note that
\baeq
\begin{eqnarray}
{1\over 2}\delta [(\dmu A)^2] &=&(\partial^{\mu} A)\dmu\delta A
=i\partial^{\mu} 
A\bar{\alpha}\gamma_5\dmu\psi ,\\
{1\over 2}\delta [(\dmu B)^2] &=&-\partial^{\mu} B\bar{\alpha}\dmu\psi ,\\
{i\over 2}\delta [\psib\dsl\psi ] &=&
{i\over 2}[\delta\psib\dsl\psi +
\psib\dsl\delta\psi ] \nonumber \\
& &={i\over 2}\dmu [\delta\psib\gamma_{\mu}\psi ]-{i\over
2}(\dmu\delta\psib )
\gamma_{\mu}\psi
+{i\over 2}\psib\dsl\delta\psi \nonumber \\
& &={i\over 2}\dmu [\delta \psib\gamma_{\mu}\psi ]+i\psib\dsl\delta
\psi, 
\end{eqnarray}
where in the last step we have used (7c) for the Majorana spinors $\psi$
and $\delta\psi$.
Continuing, we have
\begin{eqnarray}
{1\over 2}\delta (F^2) &=& iF\bar{\alpha}\dsl\psi, \\
{1\over 2}\delta (G^2) &=& G\bar{\alpha}\g5\dsl\psi .
\end{eqnarray}
\eaeq
We thus find that apart from a total derivative,
\begin{eqnarray*}
\delta {\cal{L}}_{kin} & = & -i\Box A \bar{\alpha}\g5 \psi + 
\Box B \bar{\alpha}\psi \\
&  & +i\bar{\psi}[-\dsl F\alpha +i\dsl G \g5 \alpha + \Box A \g5\alpha 
+i\Box B\alpha] \\
&  & +iF\bar{\alpha}\dsl \psi + G \bar{\alpha}\g5\dsl \psi. \\ 
\end{eqnarray*}
Using (7a) and (7b), we see that the terms involving $\Box A$ and $\Box
B$ exactly cancel, leaving the remainder which,
using (7c) and (7d), can be written as a total derivative. It will be
left as an exercise for the reader to verify that
$\delta{\cal{L}}_{mass}$ is also a total derivative.

In order to further understand the supersymmetric transformations, we
consider the effect of two successive SUSY transformations with
parameters, $\alpha_1$ and $\alpha_2$. Starting from (8a) followed by
(8b) and judicuously using (7), it is simple to show that,
\begin{equation}
(\delta_2\delta_1-\delta_1\delta_2)A =
2i\ab_2\gamma^{\mu}\alpha_1 \dmu A,
\end{equation}
In order to find the algebra satisfied by the  Majorana spinor 
supersymmetry generators $Q$, we write $\delta \equiv i\bar{\alpha}Q$,
and find from (10) that
\begin{displaymath}
-(\ab_2 Q \ab_1 Q \  -  \ \ab_1 Q \ab_2 Q)A =
-\ab_{2b}\alpha_{1a}(Q_b\bar{Q}_a \ + \bar{Q}_aQ_b)A = 
2i\ab_{2b}\alpha_{1a}(\dsl A)_{ba},
\end{displaymath}
where $a$ and $b$ are spinor indices. In the first step we have used
(7a) together with the fact that the parameters $\alpha_{ia}$
anticommute amongst themselves and also with the components $Q_{a}$ of
the SUSY generators. We will leave it to the reader to verify that the
same relation~\footnote{For 
the bosonic fields, the required steps 
are identical to the ones above; the verification with $\psi$ involves
judicious use of Fierz rearrangement and the relations (7).}
holds for successive action of SUSY transformations on the
fields $B$, $\psi$, $F$ and $G$. We can thus
write,
\begin{equation}
\{Q_a,\bar{Q}_b\} \ = \ -2(\gamma_{\mu}P^{\mu})_{ab}
\end{equation}
where $P^{\mu}$ is the translation generator of the Poincar\'e
group, and the curly brackets denote the anti-commutator. The presence
of the translation generator in (11) shows
that supersymmetry is a spacetime symmetry. Conservation of
supersymmetry implies
\baeq
\begin{equation}
[Q_a,P^0] = 0,
\end{equation}
or, from Lorentz covariance,
\begin{equation}
[Q_a,P^{\mu}] = 0.
\end{equation}
\eaeq
The commutators of $Q$ with the Lorentz group generators $J_{\mu\nu}$ are fixed
because we have already declared $Q$ to be a spin $\frac{1}{2}$ Majorana
spinor. 

The Supersymmetry algebra described above is not a Lie Algebra since it
includes anti-commutators. Such algebras are referred to as Graded Lie
Algebras. Haag, Lopuzanski and Sohnius \cite{HLS} have shown that (except
for the possibility of neutral elements and of more than one spinorial
charge $Q$ \cite{OLIVE}) the algebra that we have obtained above is the
most general graded Lie Algebra consistent with rather reasonable
physical assumptions. Models with more than one SUSY charge in the low
energy theory do not lead to chiral fermions and so are excluded for
phenomenological reasons. We will henceforth assume that there is just
a single super-charge.

We immediately note that (12a) implies all states but the zero energy
ground state (the vacuum) come in degenerate pairs, with one member of
the pair being a boson and the other a fermion. Thus, in any
supersymmetric theory, every particle has a partner with the same mass
but with a spin differing by $\frac{1}{2}$ (since $Q$ carries
$\frac{1}{2}$ unit of spin). A study of how the partners of the known
SM particles would manifest themselves in experiments at colliders forms
the main subject of these Lectures. But continuing our study of
the basics, we note that SUSY acts independently of any internal
symmetry. In other words, the generators of supersymmetry commute with
all internal symmetry generators. We immediately conclude that {\it any
particle and its superpartner have identical internal quantum numbers}
such as electric charge, isospin, colour, {\it etc.}

In order to see how supersymmetry is realized in the model defined by
(3), we note that the fields $F$ and $G$ are not dynamically independent
as the Lagrangian has no kinetic terms for these fields (which,
therefore, do not propagate). The Euler-Lagrange equations of these
fields are,
\begin{equation}
F = - mB, \hspace{5mm} G = -mA.
\end{equation}
If we substitute these back into the Lagrangian (3), we obtain,
\begin{equation}
{\cal{L}} = \frac{1}{2}(\dmu A)^2  + \frac{1}{2}(\dmu B)^2 +   
\frac{i}{2}\psib \dsl \psi  -  \frac{1}{2}m^2(A^2+B^2) -   
\frac{1}{2} m\psib\psi .
\end{equation}
This Lagrangian describes a non-interacting theory and, as such, is not
terribly interesting. Notice, however, that there are two real scalar
fields $A$ and $B$ and one spin $\frac{1}{2}$ Majorana fermion field,
all with mass $m$. We thus see that the number of bosonic degrees of
freedom (two) matches the fermionic degrees of freedom (recall that (6)
shows that just two of the four components of $\psi$ are dynamically
independent) at each space-time point.

In order to make the model more interesting, we include an interaction
term given by
\begin{equation}
{\cal{L}}_{int} = -\frac{g}{\sqrt{2}}A\psib\psi +
\frac{ig}{\sqrt{2}}B\psib\g5\psi + \frac{g}{\sqrt{2}}(A^2-B^2)G
+ g\sqrt{2}ABF,
\end{equation}
to the Lagrangian (3). The courageous reader can verify that
${\cal{L}}_{int}$ is invariant up to a total derivative under the
transformations (8). Once again we can eliminate the auxiliary fields
$F$ and $G$ via their Euler-Lagrange equations which get modified to,
\begin{eqnarray*}
F & = & - mB - g\sqrt{2}AB \\
G & = & - mA - \frac{g}{\sqrt{2}}(A^2-B^2), 
\end{eqnarray*} 
and obtain the total Lagrangian in terms of the dynamical fields as, 
\begin{eqnarray}
{\cal{L}} & = & \frac{1}{2}(\partial_{\mu}A)^2 +
\frac{1}{2}(\partial_{\mu}B)^2 + \frac{i}{2}\bar{\psi}\dsl\psi - 
\frac{1}{2}m^2(A^2+B^2) - \frac{1}{2} m\psib\psi \nonumber \\
 &  & -\frac{g}{\sqrt{2}}A\psib\psi + \frac{ig}{\sqrt{2}}B\psib\g5\psi -
gm\sqrt{2}AB^2 - \frac{gm}{\sqrt{2}}A(A^2-B^2) \nonumber \\
&  & -g^2A^2B^2 - \frac{1}{4}g^2(A^2-B^2)^2.
\end{eqnarray}

Several features of the Lagrangian in (16) are worth stressing.
\begin{enumerate}
\item It describes the interaction of two real spin zero fields and a
Majorana field with spin half. As before, the number of bosonic and
fermionic degrees of freedom match.

\item There is a single mass parameter $m$ common to all the fields.

\item Although the interaction structure of the model is very rich and
includes scalar and pseudoscalar interactions of the fermion as well as
a variety of trilinear and quartic scalar interactions, there is just
one single coupling constant $g$. We thus see that supersymmetry is like
other familiar symmetries in that it relates the various interactions as
well as masses. The mass and coupling constant relationships inherent in
(16) are completely analogous to the familiar (approximate) 
equality of neutron and
proton masses or the relationships between their interactions with the
various pions implied by (approximate) isospin invariance.
\end{enumerate}

\subsection{How Supersymmetry Removes Quadratic Divergences}
\begin{figure}
\centerline{\psfig{file=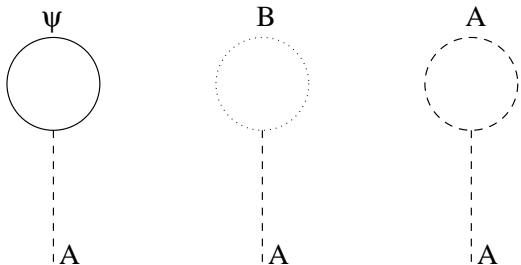,height=3.5cm,angle=270}}
\caption[]{ Lowest order diagrams contributing to quadratic divergences in the
one-point function of $A$.}
\label{fig1}
\end{figure}
We have already mentioned that the existence of supersymmetric partners
serves to remove the quadratic divergences that destabilize the scalar
sector of a generic field theory. We will illustrate this cancellation
of quadratic divergences in the toy model that we have been
studying. Consider the corrections to the ``one point function'' of the
field $A$ to first order in the coupling constant $g$ in (16). These
corrections, which are represented by tadpole diagrams shown in
Fig.~\ref{fig1}, come from 
trilinear couplings in the second line of the Lagrangian (16). 
A simple computation gives,
\begin{eqnarray}
\langle 0 | {\cal{L}}_{int} | A \rangle \sim & 
{\frac{g}{\sqrt{2}}}\left \{Tr\int \frac{d^4p}{\pslh-m_{\psi}} 
-m \int \frac{d^4p}{p^2-m_B^2} -3m \int \frac{d^4p}{p^2-m_A^2}\right\} 
\nonumber \\
=& {\frac{g}{\sqrt{2}}} \left \{\int \frac{d^4p}{p^2-m_{\psi}}4m_{\psi}
-m \int \frac{d^4p}{p^2-m_B^2} -3m \int \frac{d^4p}{p^2-m_A^2}\right\}. 
\end{eqnarray}
The factor 3 in the last term arises since any one of the three fields
in the $A^3$ interactions could annihilate the external particle.
Here, we have deliberately denoted the masses that enter via the
propagators by $m_A$, $m_B$ and $m_{\psi}$ although these are exactly
the same as the mass parameter $m$ that enters via the trilinear scalar
couplings in Eq.~(16). We first see that because all these masses are
exactly equal in a supersymmetric theory, the three contributions in
(17) add to zero. Thus although each diagram is separately quadratically
divergent, the divergence from the fermion loop exactly cancels the sum
of divergences from the boson loops. Two remarks are in order.
\begin{enumerate}
\item In order for this cancellation to occur, it is crucial that the 
$A^3$, $AB^2$ and $A\psib\psi$ couplings be exactly as given in (16).

\item The {\it quadratic} divergence in the expression (17) is
independent of the scalar masses, $m_A$ and $m_B$. It is, however, crucial that
the fermion mass $m_{\psi}$ is exactly equal to the mass $m$ that enters
via the trilinear scalar interactions in order for the cancellation of
the quadratic divergence to be maintained. If the boson masses
differ from the fermion mass $m_{\psi}$, the expression in (17) is at
most logarithmically divergent. As we have discussed, logarithmic
divergences do not severely destabilize scalar masses.
\end{enumerate}
\begin{figure}
\centerline{\psfig{file=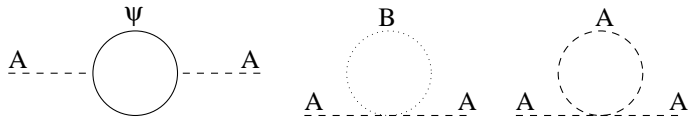,height=1.5cm,angle=270}}
\caption[]{ Lowest order diagrams contributing to quadratic divergences in the
two-point function of $A$.}
\label{fig2}
\end{figure}
It is also instructive to inspect the lowest order quadratic divergences
in the two-point function of $A$. 
The one loop contributions to the
quadratic divergences are shown in Fig.~\ref{fig2}.~\footnote{There are
additional quadratic divergences in the two point function from the
tadpoles of Fig.~\ref{fig1} which, as we have just seen, separately
cancel.}  
It is left as an exercise for the reader to check that while each one of
the diagrams in Fig.~\ref{fig2} is individually quadratically divergent,
this divergence cancels when the three diagrams are summed. Once again,
the contributions from the fermion loop cancel those from the boson
loops. Moreover,
this cancellation occurs for {\it all} values of particle
masses. This is because trilinear scalar interactions do not contribute
to the quadratic divergence that we have just computed. 
It is, however, crucial that the fermion Yukawa coupling
($\frac{g}{\sqrt{2}}$) is related to the quartic scalar couplings on the
last line of (16).

\subsection{Soft Supersymmetry Breaking}

The fact that the quadratic divergences continue to cancel even if the
scalar boson masses are not exactly equal to fermion masses (as implied
by SUSY) is absolutely critical for the construction of
phenomenologically viable models. We know from observation that SUSY
cannot be an exact symmetry of nature. Otherwise, there would have to
exist a spin zero or spin one particle {\it with exactly the mass and charge
of an electron.} Such a particle could not have evaded experimental
detection. The only way out of this conundrum (if we are to continue
with these Lectures) is to admit that supersymmetric partners cannot be
degenerate with the usual particles. Thus, supersymmetry must be a
broken symmetry.

Does this mess up our solution to the fine-tuning problem that got us
interested in SUSY in the first instance? Fortunately, it does not. We
have just seen (by the two examples above) that if SUSY is explicitly
broken because scalar masses differ from their fermion counterparts, no
new quadratic divergences occur. We will state without proof that this
is true for all processes, and to all orders in perturbation theory. It
is, therefore, possible to introduce new terms such as independent
additional masses for the scalars which break SUSY without the
reappearance of quadratic divergences. Such terms are said to break SUSY
{\it softly}. Not all SUSY breaking terms are soft. We have already seen
that if $m_{\psi} \not= m$, the expression in (17) is quadratically
divergent. Thus additional contributions to the fermion mass in the
Wess-Zumino model results in a {\it hard} breaking of
supersymmetry. Similarly, any additional contribution to just the
quartic scalar interactions will result in the reappearance of a quadratic
divergence in the correction to $m_A^2$

Are there other soft SUSY breaking terms in the toy theory that we have
been considering? Recall that the combinatorial factor 3 in the last term in
(17). This tells us that the contribution of the $A$ loop from the
trilinear $A^3$ interaction is exactly three times bigger than the
contribution from the $B$ loop from the $AB^2$ interaction (the coupling
constants for these interactions are exactly equal). Thus, there will be
no net quadratic divergence in the expression (17) even if we add a term
of the form,
\begin{displaymath}
{\cal{L}}_{soft} = k(A^3-3AB^2)
\end{displaymath}
to our model, where $k$ is a dimensional coupling constant. Obviously,
this interaction does not give a quadratically divergent correction 
to the one loop contribution to $m_A^2$. It is an example of a soft
supersymmetry breaking interaction term. 
We remark that this term can be written 
in terms of $\CS =\frac{A+iB}{\sqrt{2}}$ as
\baeq
\begin{equation}
{\cal{L}}_{soft} = \sqrt{2}k(\CS^3 \ + \ h.c.)
\end{equation}
while an arbitrary splitting in the masses of $A$ and $B$ can be
incorporated by including a term,
\begin{equation}
{\cal{L}}_{soft}= m'^2(\CS^2 \ + \ h.c.)
\end{equation}
\eaeq
into the Lagrangian.
It will turn out that super-renormalizable terms that are analytic in
$\CS$ are soft 
while terms that involve products of $\CS$ and $\CS^*$ (except
supersymmetric terms such as $\CS^*\CS$ already present in (16)) 
result in a hard breaking of SUSY. The reader can, for
instance, easily check that an interaction proportional to
($\CS^2\CS^*+h.c.)=2(A^2+B^2)A$ leads to a quadratically divergent
contribution to the expression in (17).

Although we have illustrated the cancellation of quadratic divergences
with just a couple of examples, it is important to stress that this is a
general feature of supersymmetric theories. The reader is also urged to
verify that the quadratic divergence cancels in the one loop tadpole and mass
corrections to the $B$ field. Furthermore, we have already noted that this
cancellation of quadratic divergences is true to all orders in
perturbation theory. The SUSY resolution of the fine-tuning issue rests
upon this important propery of supersymmetric models.

\section{Construction of Supersymmetric Lagrangians}
\subsection{Non-Gauge Theory}
The fields in the model we have been considering can be re-written in
terms of
\begin{eqnarray}
\CS & = & \frac{1}{\sqrt{2}}(A+iB) \nonumber\\
\psi  &   & \\
{\cal{F}} & = & \frac{1}{\sqrt{2}}(F+iG) \nonumber
\end{eqnarray}
where $\CS$, $\psi$ and \cal{F} transform into one another under the SUSY
transformations (8) which can be re-written as,~\footnote{Since $\psi_R$
is not independent of $\psi_L$, we only have to specify how $\psi_L$
transforms.}
\baeq
\begin{eqnarray}
\delta\CS &=& -\sqrt{2}i\ab\psi_L, \\
\delta\psi_L& =& -\sqrt{2}{\cal{F}}\alpha_L +\sqrt{2}\dsl\CS\alpha_R, \\
\delta{\cal{F}}& =& \sqrt{2}\ab\dsl\psi_L 
\end{eqnarray}
\eaeq Thus ($\CS$, $\psi_L$ and $\cal{F}$) together constitute an
irreducible supermultiplet in exactly the same way that the proton and
neutron form a doublet of isospin. Further, analogous to the isospin
formalism that treats the nucleon doublet as a single entity, there is a
formalism known as the superfield formalism~\cite{SUPERF} that combines
all three components of the supermultiplet into a superfield
$\hat{S}$. Since only one chiral component of the Majorana spinor $\psi$
enters the transformations, such superfields are referred to as (left)
chiral superfields. Further, because the lowest spin component of the
multiplet has spin zero, this superfield is known as a left chiral
scalar superfield. It is easy to check that the Hermitean conjugate of a
left chiral superfield is a right chiral superfield. There are, of
course, other irreducible multiplets of supersymmetry just as there are
other representations of isospin symmetry.

The superfield formalism~\cite{SUPERF} is the most convenient way of
discussing how to write supersymmetric Lagrangians. As we do not have
time to discuss it during these Lectures, we will content ourselves by
stating clearly (but without proof) those features that will be useful
to us.
\begin{enumerate}
\item There is a multiplication rule $\hat{S}=\hat{S_1}\hat{S_2}$ which
allows us to compute the components of the ``product superfield'' in
terms of the components of $\hat{S_1}$ and $\hat{S_2}$, and

\item The product of two (and hence, several) left (right) chiral
superfields is itself a left (right) chiral superfield, but the product
of a left chiral superfield and a right chiral superfield is neither a
left nor a right chiral superfield. 

\end{enumerate}

The strategy for the construction of supersymmetric Lagrangians is
straightforward once we observe from (8d,e), or equivalently from (20c)
that the $\cal{F}$ component of a left-chiral superfield changes by a
total derivative under a SUSY transformation. By (2) above, since
any product of left-chiral superfields is itself a left-chiral
superfield, any {\it analytic function}
$f(\hat{S_1},\hat{S_2},....\hat{S_N}$) of left chiral superfields is a
composite left chiral superfield. The $\CF$ component of this composite
superfield is thus a function of the component fields in
$\hat{S_1},\hat{S_2},....{\hat{S}}_N$ which changes by a total derivative
under supersymmetry transformations. This function, therefore, has
exactly the properties that we want from a supersymmetric Lagrangian
density. It is a function of various fields $\CS_i,\psi_i$ and
${\cal{F}}_i $ 
that remains invariant up to a total derivative under SUSY, and is thus a
candidate for our Lagrangian. The function $f$ is referred to as the
{\it superpotential.} The reason that it has to be analytic is that if
it involves both $\hat{S}_i$ and $\hat{S}_i^*$, it will no longer be a
chiral superfield and its $\CF$-component will no longer be a SUSY
invariant. 

The Lagrangian density, {\it i.e.} the $\CF$-component of the
superpotential, can be readily computed using the rules for superfield
multiplication. The computation is somewhat tedious. It is a function
of the component fields  $\CS_i,\psi_i$ and ${\cal{F}}_i$ of the
superfields $\hat{S}_i$ that appear in the superpotential. The auxiliary
fields ${\cal{F}}_i$ can be eliminated using the algebraic constraints
(analogous to the equation below (15)) from their Euler-Lagrange
equations. The resulting Lagrangian takes the form,
\begin{eqnarray}
{\cal{L}} &=& -\sum_i \left|\frac{\partial f}{\partial
\hat{S}_i}\right|^2_{\hat{S_i}=\CS_i}\nonumber\\ & &
-\frac{1}{2}\sum_{i,j}\left\{\psib_i\left[\frac{1-\gamma_5}{2}
\right]\left(\frac{\partial^2 f}{\partial \hat{S_i} \partial
\hat{S_j}}\right)_{\hat{S_i}=\CS_i}\psi_j +h.c.\right\}
\end{eqnarray}
The terms involving derivatives of the superpotential are functions of
just the scalar fields $\CS_i$ since in the expression we set
$\hat{S_i}=\CS_i$ after differentiation. The first term in (21) is the
scalar potential while the second term describes the interaction of the 
scalars with the fermions. Notice that the bilinear terms in the
superpotential become mass terms for both the scalars and the fermions.

It is apparent from the first term in (21) that a term of degree $n$ in
the superpotential leads to a Lagrangian density with mass dimension
$d=2(n-1)$. For the theory is to be power-counting renormalizable, we
must have $2(n-1) \leq 4$, and the superpotential at most
cubic in the superfields.

Before proceeding further, let us illustrate the use of (21) by a simple
example where the superpotential is a function of just one superfield. Choose
\baeq
\begin{equation}
f=\frac{1}{2}m\hat{S}^2 + \frac{1}{3}g\hat{S}^3.
\end{equation}
Then, using (21) it is easy to see that
\begin{eqnarray}
{\cal{L}} & = & -\left |m\CS+g\CS^2 \right |^2 \nonumber \\
& & -\frac{1}{2}\left \{\psib\left[\frac{1-\g5}{2}\right ](m+2g\CS)\psi +
h.c\right \},
\end{eqnarray}
\eaeq
which, using (19) reduces to the Lagrangian (16) except that the kinetic
energy terms are missing.

These have their origin in a different source. For our purposes it is
sufficient to recall that we saw that they were separately
supersymmetric, so that the Lagrangian in (21) only needs to be
supplemented~\footnote{We are oversimplifying at this point. The
Lagrangian for the kinetic terms as well as the one in (21) involves the
auxiliary fields $\CF$ (see Eq.~(3), as an example of this). It is only
after the auxiliary fields are eliminated that we end up with a sum of
(21) and (23).} by,
\begin{equation}
{\cal{L}}_{kin} = 
\sum_{i} (\partial_{\mu}\CS_i)^{\dagger}(\partial^{\mu}\CS_i) + 
\frac{i}{2}\sum_{i}\bar{\psi_i}\dsl \psi_i,
\end{equation}
which are the canonically normalized kinetic energies for complex scalar
and Majorana fermion fields. The Lagrangian given by the sum of (21) and
(23) is the most general globally supersymmetric Lagrangian for
non-gauge theories. We now turn to the corresponding formula for the
Lagrangian in gauge theory.

\subsection{The Lagrangian for Supersymmetric Gauge Theories}
In order to write down a locally gauge invariant supersymmetric
Lagrangian, we have to introduce a gauge covariant derivative. As in the
usual Yang-Mills construction of gauge theories, this is done by
introducing a set of massless vector fields which, under gauge
transformations, transform as the adjoint representation of the gauge
group. In a supersymmetric theory, this cannot be done without also
including some additional fermions to match the
gauge bosons that we had to introduce. 
We have to introduce a complete supermultiplet of gauge potentials.
However, 
this supermultiplet differs from the multiplet (19) of the Wess Zumino model, 
in that the gauge 
field (unlike the scalar field $\CS$ in the
chiral supermultiplet) is real. As a result, the gauge supermultiplet is
neither a left nor a right chiral superfield.  Although we 
have not shown this in these Lectures, it can be demonstrated
that all but three components of the gauge supermultiplet
can be chosen to be zero.~\footnote{In general, a real superfield has more
than three non-vanishing components. In a SUSY gauge theory, however,
the gauge parameter itself can be chosen as the scalar component of a
chiral superfield. By a judicious choice of this gauge-parameter
superfield, all but three of the components of the original real
superfield can be gauged away. This choice is not supersymmetric, and
vanishing components are resurrected by a SUSY transformation. A
combination of a SUSY and a gauge transformation leaves the form of the
gauge field unaltered.} This choice of the
supermultiplet is known as the Wess--Zumino gauge in the
literature.  In this gauge, the gauge supermultiplet
consists of $(V_\mu, \lambda, {\cal D})$, where $V_\mu$ is the
usual Yang--Mills gauge potential, $\lambda$ is a Majorana spinor
field, and ${\cal D}$, like the field ${\cal F}$ in (19) is an
auxiliary non--propagating field that can be algebraically
eliminated via its Euler--Lagrange equations.

Notice that once again the number of dynamical bosonic 
degrees of freedom (two for the gauge field) matches the 
number for the dynamical fermionic degrees of freedom (two 
for the Majorana fermion $\lambda)$, in agreement with our 
general considerations.  This new fermion, called the 
gaugino, is the supersymmetric partner of the gauge boson 
and so, under gauge transformations,  
transforms as a member of the adjoint representation of the gauge 
group.

As before, we 
will content ourselves by presenting a general 
formula for the couplings of ``matter'' particles and their superpartners
to gauge bosons and gauginos in a globally supersymmetric gauge theory.
Matter particles belong to chiral supermultiplets such as (19), while
the gauge bosons and their gaugino partners reside in the gauge
multiplet that we have just introduced.
After elimination of
the auxiliary fields ${\cal F}_i$ and ${\cal D}_A$ the globally
supersymmetric Yang-Mills Lagrangian takes the form,
\begin{eqnarray}
{\cal{L}} & =& 
\sum_{i} (D_{\mu}\CS_i)^{\dagger}(D^{\mu}\CS_i) + 
\frac{i}{2}\sum_{i}\bar{\psi_i}\gamma^{\mu}D_{\mu} \psi_i \nonumber\\
  &  & -\frac{1}{4}\sum_A F_{\mu \nu A}F^{\mu \nu}_A + 
\frac{i}{2}\sum_A \bar{\lambda}_A\gamma^{\mu}D_{\mu} \lambda_A\nonumber\\
 &  & -\sqrt{2}\sum_{i,A}\left[\CS_i^\dagger(g_{\alpha} t_{\alpha A})\bar{\psi_i} 
\frac{1-\gamma_5}{2} \lambda_A + h. c.\right]\nonumber\\
  &  & -\frac{1}{2}\sum_A\left[\sum_i \CS_i^\dagger g_{\alpha} t_{\alpha A} \CS_i + 
\xi_A\right]^2\nonumber\\
  &  & -\sum_i \left|\frac{\partial f}{\partial
\hat{S}_i}\right|^2_{\hat{S_i}=\CS_i} \nonumber\\
  &  & -\frac{1}{2}\sum_{i,j}\left\{\bar{\psi_i}
\left[\frac{1-\gamma_5}{2} \right] 
\left(\frac{\partial^2 f}{\partial \hat{S}_i 
 \partial \hat{S}_j}\right)_{\hat{S}_i=\CS_i}\psi_j +h.c.\right\}
\label{eq:lagrangian}
\end{eqnarray}
Here, $\CS_i$ ($\psi_i$) denotes the scalar 
(Majorana fermion) component 
of the {\it i}th chiral superfield,
$F_{\mu \nu A}$ is the Yang-Mills gauge field, $\lambda_A$ is 
the Majorana gaugino superpartner of the corresponding 
gauge boson and $\xi_A$ are constants~\cite{FI} which can be non-zero only for 
$U(1)$ factors of the gauge group.
In anticipation of simple grand unification, we will set these to zero. 
The last two lines of (24) come from the superpotential interactions and
are identical to the Lagrangian in Eq.~(21).

We note the following:
\begin{enumerate}
\item The first two lines are the gauge invariant kinetic energies for
the components of the chiral and gauge superfields. The derivatives that
appear are gauge covariant derivatives appropriate to the particular
representation in which the field belongs.  For example, if we are
talking about SUSY QCD, for quark fields in the first line of
Eq.~(\ref{eq:lagrangian}) the covariant derivative contains triplet $SU(3)_C$
matrices; {\it i.e.}
$D_{\mu}=\partial_{\mu}+ig_s\frac{\lambda_A}{2}V_A^{\mu}$,
whereas the covariant derivative acting on the gauginos in the
following line will contain octet matrices. These terms completely
determine how all particles interact with gauge bosons.

\item The third line describes the interactions of gauginos with matter
and Higgs multiplets (we will soon see that quarks, leptons as well as Higgs
bosons (and their superpartners) belong to chiral supermultiplets). 
Notice that these interactions are also
determined by the gauge couplings. Here $t_{\alpha A}$ is the appropriate
dimensional matrix represention of the group generators for the
$\alpha$th factor of the gauge group, while
$g_{\alpha}$ 
are the corresponding gauge
coupling constants (one for each factor of the gauge group). 
Matrix multiplication is implied. To see that these
terms are gauge invariant, recall that $\psi_{iR}$ which is fixed by the
Majorana condition, transforms according to the conjugate representation
to $\psi_{iL}$.

\item Line four describes quartic couplings of scalar matter. Notice
that these are determined by the gauge interactions. The interactions on
this line are referred to as $D$-terms.

\item Finally, the last two lines in Eq.~(\ref{eq:lagrangian}) describe
the non-gauge, superpotential interactions of matter and Higgs fields,
such as the Yukawa interactions responsible for matter fermion masses in
the SM. Since these interactions do not involve any spacetime
derivatives, choosing the superpotential to be a globally gauge
invariant function of superfields is sufficient to guarantee the gauge
invariance of the Lagrangian. For a renormalizable theory, the
superpotential must be a polynomial of degree $\leq 3$.

\end{enumerate}

Since the procedure that we have described is crucial for the
construction of SUSY models, 
we summarize by presenting a recipe for constructing an arbitrary
supersymmetric gauge theory.

\begin{itemize}
\item[($a$)] Choose a gauge group and the representations for
the various supermultiplets, taking care to ensure that the theory is
free of chiral anomalies. Matter fermions and Higgs bosons form parts
of chiral scalar supermultiplets, while gauge bosons reside in the real
gauge supermultiplet. 
\item[($b$)] Choose a superpotential function which is a
globally gauge
invariant polynomial (of degree $\leq$ 3 for renormalizable
interactions) 
of the various
left chiral superfields.
\item[($c$)] The interactions of all particles with gauge
bosons are given by the usual ``minimal coupling'' prescription.
\item[($d$)] Couple the gauginos to matter via the gauge
interactions given in (\ref{eq:lagrangian}).
\item[($e$)] Write down the additional self interactions of
the scalar matter fields as given by (\ref{eq:lagrangian}).
\item[($f$)] Write down the non--gauge interactions of matter 
fields coming from the superpotential.  The form of these is
as given by (21), or equally well, by the last two lines of (24).
\end{itemize}
The final step is to write down soft supersymmetry breaking terms which are
crucial for the construction of realistic models. 

Before closing our discussion of exact supersymmetry we briefly comment
(without proof) on how supersymmetry protects scalar masses from large
radiative corrections.  Using perturbation theory, it can be
shown~\cite{NONREN} that radiative corrections can alter masses and
couplings in the superpotential only through wave function
renormalization, provided supersymmetry is unbroken: in other words, if
any parameter in the superpotential is zero to begin with, it will not
be generated at any order in perturbation theory unless quantum
corrections in the propagator induce mixing in the kinetic energy terms
($D$-terms) of the superfields.~\footnote{What is really shown is that
radiative corrections can only induce $D$-terms, and not $F$-terms. If
off-diagonal bilinear $D$-terms are induced by radiative corrections,
``new'' superpotential interactions may be generated. We should also
caution that some $D$-terms can be rewritten as $F$-terms. Such terms
may also be radiatively generated. Generally, these involve
non-renormalizable interactions. However, there are interesting
cases~\cite{FTERMS} where renormalizable superpotential interactions may
be generated in the effective low energy theory obtained by integrating
out super-heavy fields.} This statement is most easily proven using
supergraphs,~\cite{SUPERF,NONREN} which is a diagramatic technique that
keeps the underlying supersymmetry manifest in the course of the
calculation. We have seen, however, that the mass parameter for the
scalar component of a chiral superfield (the Higgs field is just such a
scalar) arises from the superpotential and hence, gets radiatively
corrected only due to the (at most logarithmically divergent) wave function
renormalization.  In terms of a calculation involving usual Feynman
graphs with components of the superfield, this is equivalent to a
cancellation between graphs involving internal boson loops and those
involving loops of the fermionic partners of the bosons. Individually,
these contributions are all very large, but supersymmetry leads to a
precise 
 cancellation of the bosonic and fermionic contributions,
order by order in perturbation theory.  If supersymmetry is broken at a
scale $\Lambda_{SUSY}$, the cancellations are not complete and we are
left with $\delta m^2 \sim ${\cal O}($\Lambda^2_{SUSY}$), which fixes
$\Lambda_{SUSY}$ to be smaller than {\cal O}(1)~TeV, as we have already
seen.

\subsection{Supersymmetry Breaking}

{\it a. Spontaneous Supersymmetry Breaking}

Supersymmetry cannot be an exact symmetry of nature. Aesthetically, it
would be most pleasing if SUSY is spontaneously broken. As with gauge
symmetry breaking, this would then preserve the coupling constant
relationships needed for the cancellations of quadratic divergences but
break the unwanted degeneracy between the masses of particles and their
superpartners. 

The action of any symmetry transformation on a field operator $\phi$
can be schematically written as,
\baeq
\begin{equation}
\delta \phi \ = [ \bar{\alpha} Q, \phi ].
\end{equation}
In order for the symmetry {\it not}
to be spontaneously broken, the symmetry generator $Q$ should 
annihilate
the vaccum, so that 
\begin{equation}
\langle 0 | \delta \phi | 0 \rangle = \ 0.
\end{equation}
\eaeq
If this is not the case, the symmetry will be spontaneously broken.

If $Q$ above is a generator of supersymmetry it is 
clear that $\delta \phi$ must be a spin zero field 
(otherwise rotational invariance automatically ensures 
$\langle 0 | \delta \phi | 0 \rangle = \ 0$, and we obtain no new
information), or from Eq.~(20), $\phi \ = \psi$. 
We thus see that in order to break supersymmetry spontaneously without breaking
Lorentz invariance we must have 
\begin{displaymath}
\langle 0 |\CF | 0 \rangle \neq 0,
\end{displaymath}
where $\CF$ is the auxiliary field of a chiral supermultiplet.

We now recall that the auxiliary fields $\CF_i$ are algebraically
eliminated via their Euler-Lagrange equations. This then suggests a way
of breaking SUSY spontaneously: choose the superpotential so that the
system of equations,
\begin{displaymath}
\langle 0 |\CF_i |0 \rangle  = 0,
\end{displaymath}
is inconsistent. This is equivalent to the statement that the set of equations,
\begin{equation}
\langle 0 | \left [ \frac{\partial f}{\partial \hat{S}_i} 
\right ]_{\hat{S}_i=\CS_i}
| 0 \rangle \qquad = \ 0,
\end{equation}
has no consistent solution.
This mechanism, due to O'Raifeartaigh,~\cite{ORAF} is
also known in the literature as ${\cal F}$--type breaking. 
As an example, we leave it to the reader to check that the 
superpotential,
\begin{eqnarray}
f (\hat{X}, \hat{Y}, \hat{Z}) \ = \lambda (\hat{X}^2 - \mu^2) \hat{Y} \ 
+ m \hat{X}\hat{Z}, \nonumber
\end{eqnarray}
leads to the spontaneous breakdown of SUSY.

Although we have not discussed the transformation of real 
superfields in detail, we should mention that supersymmetry is also
spontaneously 
broken if the corresponding auxiliary field
(denoted by ${\cal D}$) develops a vacuum expectation value.  This 
gives us the other known way of breaking SUSY \cite{FI,DBREAK}
spontaneously: $D$-term or Fayet-Illiopoulos breaking.

{\it b. Practical Supersymmetry Breaking}

Much as we would like to have it, 
a compelling model where SUSY is broken spontaneously
has not yet been constructed. 
From many phenomenological analyses, it is fortunate that one does not
need to know the details of the physics of SUSY breaking since they
are not currently understood.
The best that
we can do at present is to provide a useful parametrization of SUSY-breaking
effects. Our guiding principle is that the
SUSY breaking terms should not destabilize scalar masses by
reintroducing the quadratic divergences that SUSY was introduced to
eliminate in the first place. In other words, SUSY breaking effects can
be incorporated by including all soft SUSY breaking masses and
interactions consistent with the known symmetries (Poincar\'e invariance,
SM gauge invariance and any other global symmetries that we might
impose).   

Girardello and Grisaru~\cite{GG} have
classified all renormalizable soft SUSY breaking operators. For our
purposes, it is sufficient to know that these consist of,
\begin{itemize}

\item explicit masses for the scalar members of chiral multiplets; {\it
i.e.} squarks, sleptons and Higgs bosons;

\item an independent gaugino mass
for each factor of a direct product gauge group: for instance, 
we would have  masses $M_1$, $M_2$ and $M_3$ for the $U(1)_Y$, $SU(2)_L$
and $SU(3)_C$ gauginos, respectively;

\item new super-renormalizable scalar interactions: for each trilinear
(bilinear) term in the superpotential of the form
$C_{ijk}\hat{S_i}\hat{S_j}\hat{S_k}$ ($C_{ij}\hat{S_i}\hat{S_j}$), we
can introduce a soft supersymmetry breaking scalar interaction
$A_{ijk}C_{ijk}{\CS_i}{\CS_j}{\CS_k}$ ($B_{ij}C_{ij}{\CS_i}{\CS_j}$) where the
$A$'s and $B$'s are constants. These terms are often referred to as $A$-
and $B$-terms, respectively.

\end{itemize}
We have already seen examples of the cubic and quadratic soft breaking
interactions at the end of Sec.~2 (see Eq.~(18)).
The scalar and gaugino masses obviously serve to break the undesired
degeneracy between the masses of sparticles and particles. We will see
later that the explicit trilinear scalar interactions mainly affect the
phenomenology of the third family.

\section{The Minimal Supersymmetric Model}\label{sec:framework}

We now have the necessary background to begin discussing particle
physics in a supersymmetric world, where SUSY is somehow (softly) broken
at the weak scale. We start with a discussion of what we will term as
the Minimal Supersymmetric Model~\cite{DIMO,HK} (MSSM). As the
name~\footnote{We warn the reader that the term MSSM does not have a
standard usage in the literature. Therefore, one should always pay
attention to explicit and implicit assumptions that are made by each
group of authors analysing the MSSM.}
suggests, it is the simplest phenomenologically viable supersymmetric
theory in that it contains the fewest number of new particles and new
interactions.

\subsection{Field Content}

We have already seen that in order to construct
supersymmetric theories, we have to introduce a partner for
every particle of the SM, with a spin differing by 
$\frac{1}{2}$ but with the same internal quantum numbers. Furthermore,
we have seen that matter fermions and Higgs bosons are members of chiral
scalar supermultiplets.
Thus, the SUSY partners of matter fermions must have spin
zero
bosons as their partners. The
dynamical matter fields of our model are thus given by,
\baeq
\begin{equation}
\left (
\begin{array}{c}
\nu \\ e
\end{array} \right )_L  \  \ , \  e_R 
\ \ , 
\left ( 
\begin{array}{c}
u \\ d 
\end{array} \right )_L  \  \ , \ u_R 
\  , \ d_R \ ,
\end{equation}
\begin{equation}
\left ( 
\begin{array}{c}
\widetilde{\nu}_L \\ \widetilde{e}_L 
\end{array} \right )  \ \ , \ \widetilde{e}_R \ , 
\left ( 
\begin{array}{c}
\widetilde{u}_L \\ \widetilde{d}_L 
\end{array} \right )  \ \  , \ \widetilde{u}_R 
\   , \ \widetilde{d}_R \ , 
\end{equation}
\eaeq
for the first family. The other families are copies of this 
exactly as in the SM: SUSY sheds no light on the reason for
the replication of generations. Note that for the lepton 
(quark) doublet there is a scalar lepton or slepton (scalar 
quark or squark) doublet, and likewise for the singlets.  
Thus, corresponding to each massive Dirac fermion $f$, there are {\it two }
complex SUSY fields $\widetilde{f}_L$ and $\widetilde{f}_R$, 
the partners of the left and right chiral 
projections of the fermion. This is in keeping with our 
counting of the number of degrees of freedom --- a massive Dirac 
fermion has four degrees of freedom corresponding to two 
spin states of the particle and antiparticle. Note also 
that $\widetilde{f}_L$ and $\widetilde{f}_R$ have exactly the same
gauge quantum numbers as their SM fermion partners.

In the gauge field sector, we have a gauge supermultiplet for each
factor of the gauge group; {\it i.e.}, the dynamical fields are,
\baeq
\begin{equation}
A_0 \; , \ \stackrel{\rightarrow}{A}_\mu \; , \
         \stackrel{\rightarrow}{g}_\mu  
\end{equation}
\hspace{5 cm} and
\begin{equation}
{\widetilde{\lambda}}_{0}
\; , \ \stackrel{^{\widetilde{\rightarrow}}}{\lambda} \; , \
        \stackrel{^{\widetilde{\rightarrow}}}{g}. 
\end{equation}
\eaeq
The vector fields above are the $U(1)_Y$, $SU(2)_L$ and $SU(3)_C$ gauge
potentials whereas the fermion fields are the spin $\frac{1}{2}$
Majorana supersymmetric partners of these fields.  Like the gauge
fields, these fermion fields transform as the adjoint representation of
the appropriate group factor.  Once $SU(2)_L \times U(1)_Y$ is broken,
fields with the same spin and charge can mix to form the
observed particles ({\it e.g}. $\gamma$ and $Z^0)$ as discussed in the
next subsection.

Finally, we come to the electroweak symmetry breaking sector. 
In the SM, the $SU(2)_L \times U(1)_Y$ symmetry is broken by 
a single doublet of Higgs fields which acquires a 
non--vanishing value in the ground state. Moreover, the same 
field, by virtue of its Yukawa interactions with fermions 
gives rise to a mass term for all the fermions of the SM. 
Technically, this is possible because the doublet (the 
complex conjugate of the doublet) can couple to the 
$T_3 \ = + \frac{1}{2} \ (T_3 \ = -\frac{1}{2}$) fermions in 
a gauge invariant way. In a supersymmetric theory, however, 
Yukawa interactions come from a superpotential which, 
as we have seen, cannot depend on a field as well as its 
complex conjugate.  As a result, any doublet can give mass 
either to a $T_3 \ = +\frac{1}{2}$ or a $T_3 \ = 
-\frac{1}{2}$ fermion, but not both.  Thus, in order to give 
masses to all the fermions, we are forced to introduce two 
Higgs doublet chiral superfields $\hat{h}_u$ and $\hat{h}_d$ which 
interact with $T_3 \ = +\frac{1}{2}$ and $T_3 = -\frac{1}{2}$
fermions, respectively.  
This sector then consists of the dynamical boson fields with 
hypercharge $Y_{h_u} \ = 1$, $Y_{h_d} = -1$,
\baeq
\begin{equation}
\left(
\begin{array}{c}
\ h_u^+ \\
h^0_u 
\end{array} \right ) , \qquad
\left ( 
\begin{array}{c}
h_d^- \\
h_d^0 
\end{array} \right ) , \qquad 
\end{equation}
and their fermionic partners (the Higgsino doublets)
\begin{equation}
\left(
\begin{array}{c}
\widetilde{h}_u^+ \\
\widetilde{h}_u^0 
\end{array} \right ) , \qquad
\left ( 
\begin{array}{c}
\widetilde{h}_d^{-} \\
\widetilde{h}_d^0 
\end{array} \right ) . \qquad 
\end{equation}
\eaeq
The fermion spinor fields that appear are Majorana. The charge shown
corresponds to that of its left chiral component; the right-handed part
has the opposite charge.
Notice also that the upper component of the doublet $\hat{h}_d$ has
been written with an electric charge 
$Q = -1$.  In other words, we have taken the $\hat{h}_u$ and 
$\hat{h}_d$ doublets to transform as the \underline{2} and \underline{2*} 
representations, respectively.  Since these are equivalent, it should 
be clear that this is done only for convenience.~\footnote
{Specifically, this makes it easy to 
embed these fields into the \underline{5} and \underline{5*} representations of
$SU(5)$ if the theory is embedded in a GUT.}

\subsection{Interactions}
The supersymmetric interactions for these fields can now be readily
worked out using Eq.~(\ref{eq:lagrangian}). The interactions of the
matter and Higgs fields (and their superpartners) with gauge bosons and
gauginos are as given by the first four lines, and are thus
model-independent, except for the constant $\xi$ which  we set to zero. In
particular, because of supersymmetry, the gauge couplings also determine
all the interactions of gauginos.  Given the field content, model
dependence arises via the choice of the superpotential $f$ which, for
the MSSM, is taken to be,

\begin{eqnarray}
f_{MSSM} &=& \mu(\hat{h}_u^0\hat{h}_d^0 + \hat{h}_u^+\hat{h}_d^-) + 
f_u(\hat{u}\hat{h}_u^0 -\hat{d}\hat{h}_u^+)\hat{U}^c\nonumber \\
    & & +f_d(\hat{u}\hat{h}_d^- + \hat{d}\hat{h}_d^0)\hat{D}^c 
+ f_e(\hat{\nu}\hat{h}_d^- + \hat{e}\hat{h}_d^0)\hat{E^c} + \ldots.
\label{eq:supR}
\end{eqnarray}    
Since
the superpotential is a function of only the 
left--chiral superfields,
we work with the (left--handed) conjugates of 
the $SU(2)_L$ singlet fermions and their partners, which together
constitute a left chiral supermultiplet with the quantum 
numbers of the representation conjugate to that of the usual
(right-handed) singlet fermions.
In Eq.~(\ref{eq:supR}), $\hat{u}$ and $\hat{d}$ denote the $SU(2)_L$ components of
the doublet quark
superfield. A similar notation is used for leptons. The minus sign in
the second term is because it is the anti-symmetric combination of two
doublets that forms an $SU(2)_L$ singlet. Since $\hat{h}_d$ is defined to
transform according to the \underline{2*} representation, the symmetric
combination appears in other terms.  Finally, $f_u$, $f_d$ and $f_e$ are
the coupling constants for the Yukawa interactions that give rise to
first generation quark and lepton masses. The ellipses denote similar
terms for other generations. 

The observant reader will have noticed that we have {\it not} written
the most general $SU(3)_C \times SU(2)_L \times U(1)_Y$ invariant
renormalizable superpotential in Eq.~(\ref{eq:supR}). In particular, we
could have included the terms given by, \baeq
\begin{equation}
g_1 = \sum_i\mu'_i\hat{L_i}\hat{h}_u+
\sum_{i,j,k}\left[\lambda_{ijk}\hat{L_i}\hat{L_j}\hat{E_k}^c +
\lambda^{'}_{ijk}\hat{L_i}\hat{Q_j}\hat{D_k}^c\right],
\label{eq:supL}
\end{equation}

and, 

\begin{equation}
g_2 = \sum_{i,j,k} \lambda^{''}_{ijk}\hat{U_i}^c\hat{D_j}^c\hat{D_k}^c.
\label{eq:supB}
\end{equation}
\eaeq 
in the superpotential $f$.
In the Eq.~(\ref{eq:supL}) and
(\ref{eq:supB}), $i,j$ and $k$ denote generation indices, while the
$\lambda$'s are coupling constants. 
We have, for brevity, not expanded out the gauge invariant
product of doublets in Eq.~(\ref{eq:supL}).  

The Lagrangian interactions
can now be obtained by substituting the appropriate superpotential 
into Eq.~(\ref{eq:lagrangian}).  It is easy to check that the terms
obtained from $g_1$ and $g_2$ lead to the violation of lepton and baryon
number conservation, respectively. This can also be seen directly from
the superpotential. For instance, with the usual assignment of lepton
number of one unit to $\hat{L}$ and $\hat{E}$ (so that $f_{MSSM}$ remains
invariant), $g_1$ clearly is not globally invariant under the
corresponding $U(1)$ transformations. In other words, the MSSM framework
in which the couplings in $g_1$ and $g_2$ are all set to zero, {\it
assumes} that there are no renormalizable baryon or lepton number
violating operators in the superpotential.

In addition to the supersymmetric interactions discussed above, we also
need to include soft supersymmetry breaking interactions. These include
an independent mass (or, allowing for flavour mixing, mass matrix) for
each squark, slepton and Higgs boson multiplet: {\it i.e.} 9 squark + 6
slepton + 2 Higgs boson masses neglecting inter-generation mixing, a gaugino
mass for each of the $SU(3)_C$, $SU(2)_L$ and $U(1)_Y$ gauginos, and
finally the trilinear and bilinear $A$- and $B$-terms for scalars. Again
ignoring inter-generation mixing, there are nine $A$-terms for the nine
Yukawa interactions in the superpotential (\ref{eq:supR}) but just one
$B$-term. We further assume that baryon and lepton number are conserved
by {\it all}
renormalizable interactions of the MSSM, and set the trilinear and
bilinear soft SUSY breaking terms corresponding to operators in $g_1$
and $g_2$ to zero.  The MSSM thus contains thirty soft SUSY breaking
parameters~\footnote{We will leave it as an exercise for the reader to
figure out how many parameters there would be if we allowed
inter-generation mixing.} together with the supersymmetric parameter
$\mu$ in addition to the arbitrary parameters of the SM. One of these
parameters can be eliminated in favour of $M_W$, so that we have thirty
independent SUSY parameters left over.

With these assumptions, it is easy to check that 
$R$-parity, defined~\cite{RPAR} to be $+1$ for leptons,
quarks, gauge bosons and Higgs bosons, and $-1$ for their supersymmetric
partners, is automatically conserved in the interactions of gauge bosons
and gauginos as given by the first four lines of (\ref{eq:lagrangian}).
Whether or not it is a good symmetry depends on the choice of the
superpotential and the soft SUSY breaking interactions. 
The reader can easily verify
using (\ref{eq:lagrangian}) that the interactions from the $f_{MSSM}$ term in
the superpotential also conserve $R$-parity,~\footnote{For the MSSM
fields, it is easy to check that $R = (-1)^{2S+L+3B}$, where $L$ and $B$
denote the lepton- and baryon-number, respectively and $S$ is the
spin. Since the MSSM conserves $B$, $L$ and angular momentum, $R$ is
conserved. Spontaneous $R$ violation via a VEV of a doublet sneutrino is
excluded by the measurement of the $Z$ width at LEP.} while those from
$g_1$ or $g_2$ do not.
Thus $R$-parity is conserved by the
renormalizable interactions (including soft SUSY breaking terms) of the
MSSM. It is {\it assumed} that $R$-parity invariance is an {\it exact} 
symmetry of the model. This has important implications as we will see later.

\subsection{Mass Eigenstates in the MSSM}

{\it SUSY Scalars:} The scalar partners $\tf_L$ and $\tf_R$ have the
same electric charge and colour, and so can mix if $SU(2)_L \times
U(1)_Y$ is broken. It is simple to check that the gauge interactions
conserve chiral flavour in that they couple only left (right) multiplets
with one another, {\it i.e.}  $\tf_L$ couples only to $\tf_L$ ($f_L$)
via gauge boson (gaugino) interactions. Unless this ``extended chiral
symmetry'' is broken, there can be no $\tf_L-\tf_R$ mixing. This
symmmetry is, however, explicitly broken by the Yukawa interactions in
the superpotential.~\footnote{Without the assumption of $R$-parity
conservation there would also be mixing between the $\hat{h}$ and
$\hat{L}$ supermultiplets. Such a mixing, which is absent in the MSSM,
can have significant phenomenological impact.} We thus conclude that
$\tf_L-\tf_R$ mixing is proportional to the corresponding Yukawa
coupling and hence to the corresponding {\it fermion} mass. For the
purposes of collider signals that will be our main concern, this mixing
is generally negligible except for third generation sfermions. We will,
therefore, neglect this intra-generational mixing for the first two
generations, and for simplicity, also any inter-generational mixing.

{\it SUSY Fermions:} The gauginos and Higgsinos are the only
spin-$\frac{1}{2}$ fermions.  Of these, the gluinos being the only
colour octet fermions, remain unmixed and have a mass $m_{\tg} =
|M_3|$.

Electroweak gauginos and Higgsinos of the same charge can mix, once
electroweak gauge invariance is broken.  The MSSM mass matrices can be
readily worked out using Eq.~(\ref{eq:lagrangian}) and
Eq.~(\ref{eq:supR}).
We first focus on the mass terms in 
the charged gaugino--Higgsino sector.  Since the field operator for 
the charged eigenstate must be a Dirac spinor, we first combine 
the Majorana gauginos (Higgsinos) into a Dirac gaugino 
(Higgsino) field with definite charge $Q = -1$.  These 
combinations are,
\baeq
\begin{equation}
\widetilde{\lambda} =  \frac{1}{\sqrt{2}} \ 
(\widetilde{\lambda}_1 + i \widetilde{\lambda}_2)  
\end{equation}
and
\begin{equation}
\widetilde{\chi} =  P_L \ \widetilde{h}_d^- - P_R \ \widetilde{h}_u^+
\end{equation}
\eaeq
where $P_L$ and $P_R$ are the left and right chiral 
projectors, respectively. We stress again that ${\widetilde{h}_d}^-$ and 
$\widetilde{h}_u^+$ are Majorana spinors 
{\sl{whose left chiral components have the charge denoted by 
the spinor.}}  The right chiral component is fixed by (6), 
and because of the complex conjugation, has the opposite 
charge as the left--chiral component. Hence, $P_R \widetilde{h}_u^+$ is
negatively charged, so that
$\widetilde{\chi}$ is indeed a Dirac field with charge 
$Q = -1$.  The interactions can then be written in terms of 
the spinors $\widetilde{\lambda}$ and $\widetilde{\chi}$ and 
the corresponding mass matrix read off from the bilinear 
terms involving these fields.  In the Lagrangian, these take 
the form,
\baeq
\begin{equation}
-( \overline{\widetilde{\lambda}} \ , 
\overline{\widetilde{\chi}} )
\left [ M_{(charge)} \ P_L \ + M^T_{(charge)} \ P_R \right ] \
\left ( 
\begin{array}{c}
\widetilde{\lambda} \nonumber \\
\widetilde{\chi}
\end{array} \right )
\end{equation}
with,
\begin{equation}
M_{(charge)} \ =
\left (
\begin{array}{cc}
M_2 \ & -gv_d \nonumber \\
-gv_u \ & -\mu
\end{array} \right )
\label{eq:chargino}
\end{equation}
\eaeq
In Eq.~(\ref{eq:chargino}) the $-\mu$ entry comes from the superpotential
whereas the off--diagonal entries come the interactions 
involving the Higgs supermultiplet and the gauginos
(the third term in Eq.(\ref{eq:lagrangian})); 
this trilinear term becomes an 
off--diagonal mass term if the scalars acquire VEVs.
In (\ref{eq:chargino}), $g$ and $g'$ are the $SU(2)_L$ and 
$U(1)_Y$ gauge coupling constants, respectively. Finally, $M_2$ is the 
soft SUSY breaking $SU(2)_L$ gaugino mass.

The mass matrices for the neutral gaugino--Higgsino sector can
be similarly worked out.  In this case, we find that the 
Lagrangian contains the terms
\baeq
\begin{equation}
- \frac{1}{2} ( \overline{\widetilde{h}_u^0}, \
\overline{\widetilde{h}_d^0}, \ 
\overline{\widetilde{\lambda}}_3, \
\overline{\widetilde{\lambda}}_0 )
\left [ M_{(neutral)} \ P_L \ + M^T_{(neutral)} 
\ P_R \right ] \ 
\left (
\begin{array}{c}
\widetilde{h}_u^0 \nonumber \\
\widetilde{h}_d^{0} \nonumber \\
\widetilde{\lambda}_3 \nonumber \\
\widetilde{\lambda}_0 
\end{array} \right )
\end{equation}
with
\begin{equation}
M_{(neutral)} \ = \ \left (
\begin{array}{cccc}
0    \ & \mu \ & -\frac{g v_u}{\sqrt{2}} \ & \frac{g'v_u}{\sqrt{2}} 
\nonumber \\
\mu \ & 0    \ &  \frac{g v_d}{\sqrt{2}} \ & -\frac{g'v_d}{\sqrt{2}} 
\nonumber \\
-\frac{g v_u}{\sqrt{2}}  \ &  \frac{gv_d}{\sqrt{2}} \ & M_2 \ 
& 0 \nonumber \\
\frac{g'v_u}{\sqrt{2}}  \ & -\frac{g'v_d}{\sqrt{2}}  \ & 0 \ & M_1
\end{array} \right )
\label{eq:neutralino}
\end{equation}
\eaeq
The sources of the various terms are more or less as in 
(\ref{eq:chargino}).  Note that in addition to the soft supersymmetry
breaking mass term $M_2$ for the 
$SU(2)_L$ gaugino which also appears in (\ref{eq:chargino}), there is
now an independent mass term $M_1$ for the $U(1)_Y$ gaugino.
Note also that $R$-parity conservation precludes any mixing of the charged
leptons (neutrinos) with the charged (neutral) gaugino-Higgsino sector.
 
The mass eigenstates can now be obtained by diagonalizing these
matrices.~\footnote{If an eigenvalue of the mass matrix for any state
$\psi$ turns out to be negative, one can always define a new spinor
$\psi^{'}=\gamma_5\psi$ which will have a positive mass. For
neutralinos, $\psi^{'}$ should be defined with an additional factor $i$
to preserve its Majorana nature after the $\gamma_5$ transformation.}
In the MSSM, the charged Dirac Higgsino (composed of the charged
components of the doublets $\tilde{h}_u$ and $\tilde{h}_d$) and the
charged gaugino (the partner of the charged $W$ boson) mix to form two
Dirac charginos, $\tw_1$ and $\tw_2$, while the two neutral Higgsinos
and the neutral $SU(2)_L$ and $U(1)_Y$ gauginos mix to form four Majorana
neutralinos $\tz_1 \ldots \tz_4$, in order of increasing mass. In
general, the mixing patterns are complex and depend on several
parameters: $\mu$, $M_{1,2}$ and $\tan\beta \equiv \frac{v_u}{v_d}$, the
ratio of the vacuum expectation values of the two Higgs fields
introduced above. If either $|\mu|$ or $|M_1|$ and $|M_2|$ are very
large compared to $M_W$, the mixing becomes small. For $|\mu| >> M_W,
|M_{1,2}|$, the lighter chargino is essentially a gaugino while the
heavier one is a Higgsino with mass $|\mu|$; also, the two lighter
neutralinos are gaugino-like while $\tz_{3,4}$ are dominantly Higgsinos
with mass $\sim |\mu|$. If instead, the gaugino masses are very large,
it is the heavier chargino and neutralinos that become gaugino-like.

Without further assumptions, the three gaugino masses are independent
parameters. It is, however, traditional to assume that there is an
underlying Grand Unification, and that these masses derive from a common
gaugino mass parameter defined at the unification scale.  The
differences between the various gaugino masses then come from the fact
that they have different interactions, and so undergo different
renormalization when these are evolved down from the GUT scale to the
weak scale. The gaugino masses are then related by,

\begin{equation}
\frac{3M_1}{5\alpha_1} = \frac{M_2}{\alpha_2} = \frac{M_3}{\alpha_3}. 
\label{eq:gaugino}
\end{equation}
Here the $\alpha_i$ are the fine structure constants for the different
factors of the gauge group.  With this GUT assumption, $\tw_1$ and
$\tz_{1,2}$ will always be substantially lighter than gluinos.  It is
for this reason that future $e^+e^-$ colliders operating at $\sqrt{s}
\simeq 500$-1000~GeV are expected to be competitive with hadron
supercolliders such as the LHC which has much higher energy. We also
mention that for not too small values of $|\mu|$, $M_{1,2}$ the
lightest neutralino tends to be dominantly the hypercharge gaugino.

{\it The Electroweak Symmetry Breaking Sector:} Although this is not in
the mainstream of what we will discuss, we should mention that because
there are two doublets in the MSSM, after the Higgs mechanism there are
five physical spin zero Higgs sector particles left over in the
spectrum. Assuming that there are no CP violating interactions in this
sector, these are two neutral CP even eigenstates ($h$ and $H$, with
$m_h \leq m_H$)
which behave as scalars as far as their couplings to fermions go, a neutral
``pseudoscalar'' CP odd particle $A$, and a pair of charged particles
$H^{\pm}$.

The Higgs boson sector~\cite{HHG} of the MSSM is greatly restricted by
SUSY. At tree-level, it is completely specified by the parameters
$m_{H_u}^2$, $m_{H_d}^2$, $\mu$ and $B$. The soft masses and the
$B$-parameter can be eliminated in favour of the VEVs (or equivalently,
$M_W^2 = \frac{1}{2}g^2(v_u^2+v_d^2)$ and $\tan\beta$) and one of the
physical Higgs boson masses, usually chosen to be $m_A$. The 
parameters $\tan\beta$ and $\mu$ also enter into the SUSY fermion mass
matrices, so that the SUSY Higgs sector is completely characterized by
just one additional parameter.
In particular, the Higgs
quartic self-couplings are all given by those on line four of
Eq.~(\ref{eq:lagrangian}) and so are fixed to be \cal{O}($g^2$). This
leads to the famous (tree-level) bound, $m_h < \min [M_Z,
m_A]|\cos2\beta|$. 

This bound receives important corrections from $t$ and
$\tt$ loops because of the rather large value of the top Yukawa coupling
and the bound is weakened~\cite{OK} to about 120-130~GeV depending on the
value of $m_t$. Thus, in contrast to early expectations, $h$ may
well escape detection at LEP2. It is worth mentioning that if we assume
that all couplings remain perturbative up to the GUT scale, then the
mass of the lightest Higgs boson is bounded by 145-150~GeV in {\it any}
weak scale SUSY model.~\cite{KOLD} The physics behind this is the same
as that behind the bound~\cite{CAB} $m_{H_{SM}} \alt 200$~GeV on the
mass of the SM Higgs boson, obtained under the assumption that the Higgs
self-coupling not blow up below the GUT scale; the numerical difference
between the bounds comes from the difference in the evolution of the
running couplings in SUSY and the SM. An $e^+e^-$ collider operating at
a centre of mass energy $\sim 300$~GeV would thus be certain~\cite{OKAD}
to find a Higgs boson if these arguments are valid.

\subsection{Implications of R-parity Conservation}

In any realistic SUSY theory, the
existence of scalar quarks and leptons 
admits the possibility  of  gauge-invariant, renormalizable baryon and lepton
number violating interactions and
so  forces us to impose additional
global symmetries. To see this, note that if all the
dimensionless couplings $\lambda$, $\lambda'$ and $\lambda''$ that occur
in Eq.~(\ref{eq:supL}) and Eq.~(\ref{eq:supB}) are of similar strength
as the gauge couplings, and if supersymmetric particles are indeed
lighter than $\sim 1$~TeV, we would be led to conclude that the proton
would decay at the weak rate at complete variance with our very
existence! 
Furthermore, the decays $\mu \to e\gamma$
and $\mu \to ee\bar{e}$, or processes such as $\mu N \to e N$
on which there are stringent bounds~\cite{PDG} from experiment,
would certainly have been observed. This situation is quite different
from the SM where gauge invariance guarantees the absence of any
renormalizable lepton- or baryon-number violating interactions. Within
the MSSM, we introduce a discrete symmetry 
($R$-parity invariance) to ensure that both $g_1$
and $g_2$ vanish. Other alternatives will be discussed toward the end of
these Lectures.

The conservation of $R$-parity has important implications for
phenomenology.
\begin{itemize}
\item SUSY ($R$-odd) particles have their own identity and do not mix
with $R$-even SM particles. We will refer to these as sparticles and denote
them with twiddles.
\item Sparticles can only be pair produced in collisions of ordinary
particles.
\item A sparticle must decay into an odd number of sparticles.
\item As a result, the lightest supersymmetric particle (LSP) must be
absolutely stable.
\end{itemize}

There are strong limits on the existence of stable or even 
very long--lived ($\tau \stackrel{>}{\sim}$ age of the 
universe) charged or coloured particles.  Such particles 
would have been produced in the Big Bang and would bind with 
ordinary particles resulting in exotic heavy atoms or 
nuclei.  For masses up to about $1\; TeV$, estimates~\cite{WOLFRAM} of 
their expected abundance are in the range 
{\cal O}($10^{-10} - 10^{-6}$) whereas the empirical abundance \cite{COSMO}
is 
$\alt$ {\cal O}($10^{-12} - 10^{-29}$) depending on
the element whose exotic isotope is searched for.
This null result is taken to imply that a stable LSP must be 
a weakly interacting, neutral sparticle.  Within the MSSM, 
the LSP can then only be either the lightest neutralino or one of the scalar
neutrinos. We will see later that a sneutrino is disfavoured
if we also assume that the LSP also forms the dark matter in
our galactic halo. On the other hand, 
it has been shown that a stable neutralino is a
promising candidate for cosmological cold dark matter.
In a supergravity theory, the SUSY partner 
of the graviton could also be the LSP.  Unless it is 
extremely light, however, it couples to other 
particles only with gravitational strength couplings so that 
it effectively decoupled for the purposes of collider
phenomenology.  In this case, the next lightest sparticle 
plays the role of the LSP and the constraints discussed 
above apply to it as long as the lifetime for its decay into 
the gravitino exceeds the age of the universe.~\footnote{Moroi
\cite{MOROI} has pointed out that late decay
of this effective LSP can potentially spoil the successful predictions
of Big Bang Nucleosynthesis} If this is 
not the case, or if $R$-parity is not conserved,
the ``effective LSP'' may even be charged or 
coloured. Throughout most of these Lectures we will assume that the LSP
is the lightest neutralino.

We note here that regardless of what the LSP is (as long as it is
neutral and lives to travel at least a few metres), the LSP's 
produced in the decays of sparticles in SUSY events behave 
like neutrinos in the experimental apparatus: {\it i.e.} they 
escape without depositing any energy.  Thus, in any model where
$R$-parity is assumed to be conserved, apparent 
missing energy $(\rlap/E)$ and an imbalance of transverse 
momentum $(\rlap/p_T)$ are generally regarded as characteristic 
signatures of SUSY.

\subsection{Is the MSSM a Practical Framework for SUSY Phenomenology?}

The MSSM is the simplest framework for SUSY
phenomenology. A big advantage of this framework is that except for the
assumption of $R$-parity conservation (and, of course, supersymmetry!),
we have assumed very little else: a minimal particle content, Poincar\'e
invariance and gauge invariance. The price that we have to pay for such
an agnostic view is the large number of free parameters to parametrize
SUSY breaking. We saw that even in the simplified case where we neglect
generational mixing between sfermions, there were 30 new parameters.
This is not necessarily a problem for SUSY phenomenology if we are
studying a SUSY process where just a few sparticles
are relevant: if this is so, only a handful of these
thirty parameters would be relevant for the analysis of the process in
question.~\footnote{An example of this would be $e^+e^- \to
\tmu_R\tmu_R$, if $\tmu_R \to \mu \tz_1$ all the time. In this case,
the SUSY reaction can be completely described by
$m_{\tmu_R}$ and $m_{\tz_1}$.} We will, however, see that this is generally not
the case. Typically, heavy sparticles decay into lighter sparticles which
further decay until this cascade terminates in the stable LSP. This, of
course, means that to describe completely even a single SUSY reaction
may require the knowledge of properties of several particles (the
parent, along with all the daughters that are part of the decay
cascade) which, in turn, depend on the large number of MSSM
parameters. This renders the MSSM rather unwieldy for many
phenomenogical analyses.

Assuming grand unification
ameliorates the situation to some extent: there are then only two scalar
masses per generation of sfermions in $SU(5)$ and only one gaugino mass
parameter, but the parameter space is still too large.
In the future, a deeper understanding of the mechanism of supersymmetry
breaking may relate the many SUSY breaking parameters of the MSSM,
resulting in a significant reduction of the parameter space. For the
present, however, we have to rely on assumptions about the nature of
physics at the high energy scale to reduce the number of parameters. It
is important to keep in mind that these assumptions may prove to be
incorrect. For this reason, one should always be careful to test the
sensitivity of phenomenological predictions to the various assumptions,
particularly when using the models to guide our thinking about the
design of future experiments. 

Historically, inspired by supergravity model studies, many early
phenomenological studies assumed that all squarks (often, sleptons were either
assumed to be degenerate with squarks, or to have masses related to
$m_{\tq}$) were degenerate except for $D$-term splitting. They also
incorporated the GUT assumption (\ref{eq:gaugino}) for gaugino
masses. The masses and couplings of all sparticles were 
determined by the SM parameters together with 
relatively few additional (SUSY) parameters which were
frequently taken to be,
\begin{equation}
m_{\tg}, m_{\tq}, (m_{\tell}), \mu, \tan\beta, A_t, m_A.
\label{eq:mssm}
\end{equation}
The parameter $A_t$ mainly affects top squark phenomenology, and so, was
frequently irrelevant.  Other $A$-terms, being proportional to the light
fermion masses, are irrelevant for collider phenomenology.

In view of the fact that additional assumptions are necessary, and
further, that assumptions based on supergravity models are incorporated
into phenomenological analyses, it seems reasonable to explore the
implications of these models more seriously. Toward this end, we
describe the underlying framework in the following section.

\section{The mSUGRA framework: A Model Paradigm}\label{sec:sugra}

When supersymmetry is promoted to a local symmetry, additional fields
have to be introduced.  The resulting theory~\cite{GRAV} which includes
gravitation is known as supergravity (SUGRA). It is not our purpose here
to study SUGRA models~\cite{NILLES,DMART} in any detail. Our purpose is solely 
to provide motivation for an economic and elegant framework that
has recently become very popular for phenomenological
analysis and to carefully spell out its underlying assumptions.

It was recognised rather early~\cite{WEIN} that it is very difficult to construct
globally supersymmetric models where SUSY is spontaneously broken at the
weak scale. This led to the development of geometric hierarchy models
where SUSY is broken in a ``hidden'' sector at a scale $\mu_s \gg
M_W$. This sector is assumed to interact with ordinary particles and
their superpartners (the ``observable'' sector) only via exchange of
superheavy particles $X$. This then suppresses the couplings of the
Goldstone fermion (which resides in the hidden sector) to the observable
sector: as a result, the effective mass gap in the observable sector is
$\mu \sim \frac{\mu_{s}^{2}}{M_X}$ which can easily be $\alt 1$~TeV even
if $\mu_s$ is much larger.

An especially attractive realization of this idea stems from the
assumption that the hidden and observable sectors interact only
gravitationally, so that the scale $M_X$ is $\sim M_{Planck}$.  This led
to the development of supergravity GUT models~\cite{SUGGUT} of particle
physics. Because supergravity is not a renormalizable theory, we should
look upon the resulting Lagrangian, with heavy degrees of freedom
integrated out, as an effective theory valid below some ultra-high scale
$M_X$ around $M_{GUT}$ or $M_{Planck}$, in the same way that chiral
dynamics describes interactions of pions below the scale of chiral
symmetry breaking.  Remarkably, this Lagrangian turns out to be just the
same~\cite{HLW} as that of a globally supersymmetric
$SU(3)_C \times SU(2)_L \times U(1)_Y$ model, together with soft SUSY breaking
masses and $A$- and $B$-parameters of ${\cal{O}}(M_{Weak})$.

The economy of the {\it minimal} supergravity (mSUGRA)
GUT framework~\footnote{Here the
term minimal refers to a technical assumption:
the canonical choice of kinetic energy terms for
matter and gauge fields. In this case, there is a global $U(N)$ symmetry
in a model with $N$ chiral super-multiplets.
However, since supergravity is a non-renormalizable
theory, in principle, kinetic terms may also arise from higher dimensional
operators, and need not take the canonical form.} 
stems from the fact that because of the assumed symmetries,
various soft SUSY breaking parameters become related independent of the
details of the hidden sector and the low energy effective Lagrangian can
be parametrized in terms of just a few parameters. For instance,
the chiral multiplets all acquire the same soft SUSY
breaking scalar mass $m_0$. Likewise, there is a universal
$A$-parameter, common to all trilinear interactions. 
The GUT assumption,
of course, implies that the soft SUSY breaking gaugino masses are
related as in Eq.~(\ref{eq:gaugino}). The only role played by the
``minimal'' in mSUGRA is to provide a rationale for the universality of
soft SUSY breaking parameters. It is worth stressing~\cite{DINE} that the
universality does not follow from an established principle such as
general covariance. For phenomenological purposes, one can
forget about the origins of the model and simply view it as a version of
the MSSM with universal scalar and gaugino masses and $A$-parameters at
some ultra-high scale.

The
universality of the scalar masses does not imply that the physical
scalar masses of all sfermions are the same. The point is that the
parameters in the Lagrangian obtained by integrating out heavy fields
should be regarded as renormalized at the high scale $M_X$ at which
these symmetries are unbroken.  If we use this Lagrangian to compute
processes at the 100~GeV energy scale relevant for phenomenology, large
logarithms \cal{O}($\ln\frac{M_X}{M_W}$) due to the disparity between
the two scales invalidate the perturbation expansion. These logarithms
can be straightforwardly summed by evolving the Lagrangian parameters
down to the weak scale.  This is most conveniently done~\cite{INOUE}
using renormalization group equations (RGE).

The renormalization group evolution leads to a definite pattern of
sparticle masses, evaluated at the weak scale.~\footnote{These running
masses evaluated at the sparticle mass, or more crudely, at a scale
$\sim M_Z$, are not identical to, but are frequently close to the
physical masses which are given by the pole of the renormalized
propagator.~\cite{POLE}}  For example, gauge boson-gaugino loops result
in increased sfermion masses as we evolve these down from $M_X$ to $M_W$
while superpotential Yukawa couplings (which are negligible for the two
lightest generations) have just the opposite effect. Since squarks have
strong interactions in addition to the electroweak interactions common
to all sfermions, the weak scale squark masses are larger than those of
sleptons. Neglecting Yukawa couplings in the RGE, 
we have to a good approximation,
\baeq
\begin{eqnarray}
m_{\tq}^2 & = & m_0^2 + m_q^2 + (5-6)\mhf^2 + D-terms,  \\
m_{\tell}^2 & = & m_0^2 + m_{\ell}^2 + (0.15-0.5)\mhf^2 + D-terms.
\label{eq:sfermions}
\end{eqnarray}
\eaeq
In Eq.~(\ref{eq:sfermions}), $\mhf$ is the common gaugino mass at the
scale $M_X$. Notice that squarks and sleptons within the same $SU(2)_L$
doublets are split only by the $D$-terms, whose scale is set by $\frac{1}{2}M_Z^2$.
In contrast, various flavours of left- (and separately, right-)
type squarks of the first two generations are essentially degenerate,
consistent~\cite{FCNC} with flavour changing neutral current (FCNC)
constraints from the observed properties of 
$K$, $D$ and $B$ mesons.~\footnote{This is a non-trivial
observation since alternative mechanisms to suppress FCNC based on
different symmetry considerations have been proposed.~\cite{NS}} 
The difference in the
coefficients of the $\mhf$ terms reflects the difference between the
strong and electroweak interactions alluded to above. Although we have
not shown this explicitly, $\tell_R$ which has only hypercharge
interactions tends to be somewhat lighter than $\tell_L$ as well as $\tnu_L$
unless D-term effects are significant. Since $m_{\tg} = (2.5-3)\mhf$,
it is easy to see that squark and slepton masses are related by,

\begin{equation}
m_{\tq}^2 = m_{\tell}^2 + (0.7-0.8)m_{\tg}^2.
\label{eq:squark}
\end{equation} 
Here, $m_{\tq}^2$ and $m_{\tell}^2$ are the squared masses averaged over
the squarks or sleptons of the first (or second) generation. In the
second term, the unification of gaugino masses has been assumed. 

The Yukawa couplings of the top family are certainly not negligible. 
These Yukawa
interactions tend to reduce the scalar masses at the weak scale. The RGE
effects from these can overcome the additional $m_t^2$ in
Eq.~(37a), so that $\tt_L$ and $\tt_R$ tend to be
significantly lighter than other squarks (of course, by $SU(2)_L$
invariance, the soft-breaking mass for $\tb_L$ is the same as that for
$\tt_L$). In fact, we can say more: because $\tt_R$ receives top quark
Yukawa corrections from both charged and neutral Higgs loops in contrast
to $\tt_L$ which gets corrections just from the neutral Higgs, its
squared mass is reduced (approximately) twice as much as that of
$\tt_L$.  Moreover, as we have already mentioned, these same Yukawa
interactions lead to $\tt_L-\tt_R$ mixing, which further depresses the
mass of the lighter of the two $t$-squarks (sometimes referred to as the
stop) which we will denote by $\tt_1$. In fact, care must be exercised
in the choice of input parameters: otherwise $m_{\tt_1}^2$ is driven
negative, leading to the spontaneous breakdown of electric charge and
colour. For
very large values of $\tan\beta \sim \frac{m_t}{m_b}$, bottom and tau Yukawa
couplings are also important; these affect $b$-squark and $\tau$ slepton
masses and mixings in an analogous way.

The real beauty and economy of this picture comes from the fact that
{\it these same Yukawa radiative corrections drive~\cite{RAD} electroweak
symmetry breaking}. Since the Higgs bosons are part of chiral
supermultiplets, they also have a common mass $m_0$ at the scale $M_X$
and undergo similar renormalization as doublet sleptons due to gauge
interactions: {\it i.e.} these positive contributions are not very
large. The squared mass $m_{h_u}^2$ of the Higgs boson doublet which
couples to the top family, however, receives large negative
contributions (thrice those of the $\tt_L$ squark since there are three
different colours running in the loop) from Yukawa interactions, and so
can become negative, leading to the correct pattern of gauge symmetry
breaking. Furthermore, because $f_t > f_b$, $\tan\beta > 1$.  While this
mechanism is indeed very pretty, it is not a complete
explanation of the observed scale of spontaneous symmetry breakdown
since it requires that $m_0$, the scalar mass at the very large scale
$M_X$ be chosen to be $\leq 1$~TeV: in other words, the small
dimensionless ratio $\frac{m_0}{M_X}$ remains unexplained. Also, for
some choices of model parameters, it is
possible to get ground states where $SU(3)_C$ is broken.

Let us compare the model parameters with our list (\ref{eq:mssm}) for
the MSSM. Here, we start with GUT scale parameters, $m_0$,
$\mhf$, $A_0$, $B_0$ and $\mu_0$. The weak scale parameter $\mu$
(actually, $\mu^2$) is adjusted to give the experimental value of $M_Z$.
It is convenient to eliminate $B_0$ in favour of $\tan\beta$ so that the
model is completely specified by just a four parameter set (with a sign
ambiguity for $\mu$),

\begin{equation}
m_0, \mhf, \tan\beta, A_0, sgn(\mu),
\label{eq:sugra}
\end{equation}
(together with SM parameters) 
without the need of additional {\it ad hoc} assumptions as in the
MSSM. 

The mSUGRA model leads to a rather characteristic pattern of sparticle
masses~\cite{SPECTRA} and mixings. We have already seen that the first
two generations of squarks are approximately degenerate, while the
lighter of the $t$-squarks, and also $\tb_L$ can be substantially
lighter. If $\tan\beta$ is large, the lighter of the two stau states will
be considerably lighter than other sleptons.
Also, from Eq.~(\ref{eq:squark}) it follows that sleptons may
be significantly lighter than the first two generations of squarks if
$m_{\tg} \simeq m_{\tq}$, and have comparable masses if squarks are
significantly heavier than gluinos. We also see that gluinos can never
be much heavier than squarks. Furthermore, because the top quark is very
massive, the value of $|\mu|$ obtained from the radiative symmetry
breaking constraint generally tends to be much larger than the
electroweak gaugino masses, so that the lighter (heavier) charginos and
neutralinos tend to be gaugino-like (Higgsino-like). As a result, except
when $\tan\beta$ is very large, the
additional Higgs bosons $H$, $A$ and $H^\pm$ also become very heavy,
and $h$ couples like the SM Higgs boson. 

We should stress that while the mSUGRA GUT framework
provides a very attractive and economic picture, it hinges upon untested
assumptions about the symmetries of physics at very high energies. It
could be that the GUT assumption is incorrect though this would then
require the unification implied by the observed values of
gauge couplings at LEP to be either purely fortituous, or due to some sort of
string unification.~\cite{IBAN} It could be that the
assumption of universal scalar masses (or $A$-parameters) is
wrong. Recall that our arguments for this hinged upon the existence of
an additional global $U(N)$ symmetry among the $N$ chiral
multiplets.~\footnote{We warn the reader that there is a folk 
theorem that says that effective
field theories for the classical ground states of superstring theories
cannot have continuous global symmetries except for Peccei-Quinn type
symmetries associated with axion-like fields; {\it i.e.} all other continuous
symmetries (except, possibly, an accidental
symmetry of lower dimensional operators)
are gauge symmetries. If this is the case, then the
introduction of a global symmetry to obtain universal SUSY breaking
parameters as discussed above would be questionable. This theorem has,
however, been proven~\cite{BANKS} using 2D superconformal theory on the
string world sheet. It is presently unclear whether such a proof would
survive the recent theoretical developments where non-perturbative effects
play a critical role. The reader may, of course, take the view that the mSUGRA
model is not derived from string theory, in which case these
considerations are not relevant.
It is, perhaps, also worth mentioning~\cite{FERNANDO} that while 
many supersting models lead to non-universal SUSY breaking parameters at 
the string scale, 
universal soft-breaking parameters are at least possible in the so-called 
dilaton dominated scenario.
I am grateful to Fernando Quevedo for discussions about global symmetries 
in string theory.} This is, 
perhaps, reasonable as
long as we are near the Planck scale where gravitation presumably
dominates gauge or Yukawa interactions. Non-universal masses could result if
this $U(N)$ is broken by the
explicit introduction of non-canonical kinetic terms for chiral
supermultiplets.~\cite{DINE} 
We should also remember that in the absence of a theory about physics at
the high scale, we do not have a really good principle for choosing the
scale $M_X$ at which the scalar masses are universal. In practice, most
phenomenological calculations set this to be the scale of GUT symmetry
breaking where the gauge couplings unify. If, instead, this scale were
closer to $M_{Planck}$ the evolution between these scales~\cite{PP}
could result in non-universal scalar masses at $M_{GUT}$: this could
have significant impact, particularly on the condition for electroweak
symmetry breaking. 

This mismatch between the GUT scale and the scale at which scalar masses
are assumed to unify also yields a novel source of lepton flavour
violation in SUSY GUTS. The point is that if lepton and slepton vertices
are not diagonalized by the same rotation, the lepton-slepton-gaugino
vertex will not conserve lepton flavour. This is not an issue if
sleptons have universal masses since any rotation leaves the identity
matrix invariant. Barbieri and Hall~\cite{BARHAL} have pointed out that
if the scale at which scalar masses unify is substantially larger than
the GUT scale, radiative corrections due to large top quark Yukawa
interactions split the third generation slepton (defined to be the
slepton in the same supermultiplet as the top) mass from that of other
sleptons. The resulting mismatch of the lepton and slepton mass
matrices, they note, leads to leptonic flavour violation that might be
detectable in the next round of experiments.

Despite these shortcomings, this framework at the very least should be
expected to provide a useful guide to our thinking about supersymmetry
phenomenology. In spite of the fact that it is theoretically rather
constrained, it is consistent with all experimental and even
cosmological constraints and even, as we will see, contains a candidate
for galactic and cosmological dark matter. 
Indeed mSUGRA provides a 
reasonably flexible yet tractable framework whose underlying
assumptions, as we will see, can be subject to direct tests at future
colliders.

\section{Decays of Supersymmetric Particles}\label{sec:decays}

Before we can discuss signatures via which sparticle production might be
detectable at colliders, we need to understand how sparticles decay. 
We will assume that $R$-parity is conserved and that $\tz_1$ is the LSP.

\subsection{Sfermion Decays}

We have seen in Sec. \ref{sec:framework} that gauginos and Higgsinos
couple sfermions to fermions.  Since we have also assumed that $\tz_1$
is the LSP, the decay $\tf_{L,R} \to f\tz_1$ ($f \not = t$) is always
allowed. Depending on sparticle masses, the decays
\begin{equation}
\tf_{L,R} \to f\tz_i, \  \tf_{L} \to f'\tw_i
\label{eq:sfermiondk}
\end{equation}
to other neutralinos or to charginos may also be allowed. The chargino
decay modes of $\tf_R$ only proceed via Yukawa interactions, and so are
negligible for all but $t$-squarks, except for large values of
$\tan\beta$ for which the effects of the bottom and tau Yukawa
interactions become important.~\footnote{The Yukawa coupling of the
upper (lower) member of the weak doublet is given by $f_f =
\frac{gm_f}{\sqrt{2}M_W\sin\beta} (\frac{gm_f}{\sqrt{2}M_W\cos\beta})$,
where $m_f$ is the mass of the fermion $f$ which may be either a quark or a
lepton.}

Unlike sleptons, squarks also have
strong interactions, and so can also decay into gluinos via,
\begin{equation}
\tq_{L,R} \to q\tg, 
\label{eq:squarkdk}
\end{equation}
if $m_{\tq}>m_{\tg}$. Unless suppressed by phase space, the gluino decay
mode of squarks dominates, so that squark signatures are then mainly determined
by the decay pattern of gluinos. If $m_{\tq} < m_{\tg}$, squarks, like
sleptons, decay\cite{CAS,BBKT} to charginos and neutralinos. The
important thing to remember is that sfermions dominantly decay via a
two-body mode.

The various partial decay widths can be easily computed using the
Lagrangian we have described above. Numerical results may be found in
the literature for both sleptons~\cite{BBKMT} and squarks~\cite{BBKT}
and will not be repeated here. The following features, however, are
worthy of note:

\begin{itemize}
 
\item The electroweak decay rates are $\sim \alpha m_{\tf}$
corresponding to lifetimes of about $10^{-22}(\frac{100 \
GeV}{m_{\tf}})$ seconds. Thus sfermions decay without leaving any tracks
in the detector. The reader can check that the same
is true for the decays of other sparticles discussed below.

\item Light sfermions directly decay to the LSP. For heavier sfermions,
other decays also become accessible. Decays which proceed via the larger
$SU(2)_L$ gauge coupling are more frequent than those which proceed via the
smaller $U(1)_Y$ coupling (we assume Higgs couplings are negligible). Thus, for
$\tf_L$, the decays to charginos dominate unless they are kinematically
suppressed, whereas $\tf_R$ ($f \not= t$)mainly decays into the
neutralino with the largest hypercharge gaugino component.

\item Very heavy sleptons (and squarks, if the gluino mode is forbidden)
preferentially decay into the lighter (heavier) chargino ($\tf_L$ only)
and the lighter neutralinos $\tz_{1,2}$ (the heavier neutralinos
$\tz_{3,4}$) if $|\mu|$ ($m_{\tg}$) is very large. This is because
$\tw_1, \tz_{1,2}$ ($\tw_2, \tz_{3,4}$) are the sparticles with the
largest gaugino components.

\end{itemize}

{\it Top Squark Decays:} We have seen that $t$-squarks are different in
that ({\it i})~the mass eigenstates are parameter-dependent mixtures of
$\tt_L$ and $\tt_R$, ({\it ii})~$\tt_1$, the lighter of the two states
may indeed be much lighter than all other sparticles (except, of course,
for phenomenological reasons, the LSP) even when other squarks and
gluinos are relatively heavy, and ({\it iii}) top squarks couple to
charginos and neutralinos also via their Yukawa components. As a result
the decay patterns of $\tt_1$ can differ considerably from those of
other squarks. Yukawa interactions may also be important for $b$-squarks
and $\tau$-sleptons if $\tan\beta$ is very large. 

The decay $\tt_1 \to t\tg$ will dominate as usual if it is kinematically
allowed. Otherwise, decays to charginos and neutralinos, if allowed,
form the main decay modes. Since $m_t$ is rather large, it is quite
possible that the decay $\tt_1 \to t\tz_1$ is kinematically forbidden,
and $\tt_1 \to b\tw_1$ is the only tree-level two body decay mode that
is accessible, in which case it will obviously dominate.  If the stop is
lighter than $m_{\tw_1}+m_b$, and has a mass smaller than about 125~GeV
(which, we will see, is in the range of interest for experiments at the
Tevatron), the dominant decay mode of $\tt_1$ comes from the
flavour-changing $\tt_1-\tc_L$ loop level mixing induced by weak
interactions~\cite{HIKASA} and the decay $\tt_1 \to c\tz_1$ dominates
its allowed tree level decays into (at least) four-body final states.
If $m_{\tt_1} \sim 175-225$~GeV, the three-body decays $\tt_1 \to b
W\tz_1$ may be accessible, with the two body decays $\tt_1 \to b\tw_1$
and $\tt_1 \to t\tz_1$ still closed. The rate for this three body decay
then has to be compared with the two body loop decay to assess the decay
pattern of $\tt_1$. Unfortunately, this branching fraction
\cite{WOEHRMAN} is sensitive to the model parameters, and no general
statement is possible.

\subsection{Gluino Decays}

Since gluinos have only strong interactions, they can only decay via
\begin{equation}
\tg \to \bar{q}\tq_{L,R},  q\bar{\tq}_{L,R}, 
\label{eq:gluinodk}
\end{equation}
where the squark may be real or virtual depending on squark and gluino
masses. If $m_{\tg}>m_{\tq}$, $\tq_L$ and $\tq_R$ are produced in equal
numbers in gluino decays (except for phase space corrections from the
non-degeneracy of squark masses). In this case, since $\tq_R$ only
decays to neutralinos, neutralino decays of the gluino dominate. If, as
is more likely, $m_{\tg}<m_{\tq}$, the squark in Eq.~(\ref{eq:gluinodk})
is virtual and decays via Eq.~(\ref{eq:sfermiondk}), so that gluinos
decay via three body modes,
\begin{equation}
\tg \to q\bar{q} \tz_i, q\bar{q'}\tw_i.
\end{equation}
In contrast to the $m_{\tg}>m_{\tq}$ case, gluinos now predominantly
decay~\cite{CAS,BBKT} into charginos because of the large $SU(2)_L$ gauge
coupling, and also into the neutralino with the largest $SU(2)_L$ gaugino
component. For small values of $|\mu|$ ($\ll |M_2|$), these may well be the
heavier chargino and the heaviest neutralino~\cite{BBKT}; if instead
$\mu$ is relatively large, as is generally the case in the mSUGRA
framework, the $\tw_1$ and $\tz_2$ decays of gluinos frequently dominate.

We should also point out that our discussion above neglects
differences between various squark masses. As we have seen in the last
section, however, third generation squarks $\tt_1$ and $\tb_1 \sim
\tb_L$ may in fact be substantially lighter than the other squarks. It
could even be~\cite{NOJ} that $\tg \to \bar{b}\tb_1$ and/or $\tg \to
\bar{t}\tt_1$ are the only allowed two-body decays of the gluino in
which case gluino production will lead to final states with very large
$b$-multiplicity, and possibly also hard, isolated leptons from the
decays of the top quark or the $t$/$b$ squark.
Even if these decays are kinematically
forbidden, branching fractions for decays to third generation fermions
may nonetheless be large
because of enhanced $\tt_1$ and $\tb_1$ propagators (recall
that the decay rates roughly depend on $\frac{1}{m_{\tq}^4}$) with
qualitatively the same effect.

Finally, we note that there are regions of parameter space where
the radiative decay,
\begin{equation}
\tg \to g \tz_i,
\end{equation}
can be important~\cite{BTWRAD}. This decay, which occurs via third
generation squark and quark loops, is typically enhanced relative to the
tree-level decays if the neutralino contains a large $\tilde{h}_u$
component (which has large Yukawa couplings to the top family).

\subsection{Chargino and Neutralino Decays}

Within the MSSM framework,
charginos and neutralinos can either decay into lighter charginos and
neutralinos and gauge or Higgs bosons, or into $\tf \bar{f}$ pairs if
these decays are kinematically allowed. We will leave it as an exercise
for the reader to make a list of all the allowed modes and refer the
reader to the literature~\cite{HHG,BBKMT} for various formulae and
numerical values of the branching fractions.  If these two-body decay
modes are all forbidden, the charginos and neutralinos decay via three
body modes,
\begin{equation}
\tw_i \to  f\bar{f'}\tz_j, \tw_2 \to f\bar{f}\tw_1 \nonumber
\end{equation}
\begin{equation}
\tz_i \to f\bar{f} \tz_j \  or \  f\bar{f'}\tw_1,
\end{equation}
mediated by virtual gauge bosons or sfermions (amplitudes for Higgs
boson mediated decays, being proportional to fermion masses are generally
negligible except when the corresponding Yukawa couplings are enhanced).
Typically, only the lighter chargino and the neutralino
$\tz_2$ decay via three body modes, since the decays $\tz_{3,4} \to
\tz_1Z$ or $\tz_1 h$ and $\tw_2 \to W\tz_1$ are often
kinematically accessible.  Of course if the $\tz_2$ or $\tw_1$ are heavy
enough they will also decay via two body decays: these decays of $\tz_2$
are referred to as ``spoiler modes'' since, as we will see, they
literally spoil~\footnote{The decay to Higgs does not yield leptons,
whereas the decay to $Z$ has additional backgrounds from SM $Z$
sources.}  the clean leptonic signal via which $\tz_2$ may be searched
for.

For sfermion masses exceeding about $M_W$, $W$-mediated decays generally
dominate the three body decays of $\tw_1$, so that the leptonic
branching for its decays fraction is essentially the same as that of the
$W$; {\it i.e. } 11\% per lepton family. An exception occurs when $\mu$
is extremely large so that the LSP is mainly a $U(1)_Y$ gaugino and $\tw_1$
dominantly an $SU(2)_L$ gaugino. In this case, the $W\tw_1\tz_1$ coupling is
considerably suppressed: then, the amplitudes for $\tw_1$ decays
mediated by virtual sfermions may no longer be negligible, even if
sfermions are relatively heavy, and the leptonic branching fractions may
deviate substantially from their canonical value of 11\%.

One may analogously expect that $\tz_2$ decays are dominated by
(virtual) $Z^0$ exchange if sfermion masses substantially exceed
$M_Z$. This is, however, not true since the $Z^0$ couples only to the
Higgsino components of the neutralinos. As a result, if either of the
neutralinos in the decay $\tz_2 \to \tz_1 f\bar{f}$ has small Higgsino
components the $Z^0$ contribution may be strongly suppressed, and the
contributions from amplitudes involving relatively heavy sfermions may be
comparable. This phenomenon is common in the mSUGRA model where $|\mu|$
is generally much larger than the electroweak gaugino masses, and
$\tz_1$ and $\tz_2$ are, respectively, mainly the hypercharge and $SU(2)_L$
gauginos. If, in addition, $m_{\tq} \sim m_{\tg}$, we see from
Eq.~(\ref{eq:squark}) that sleptons are much lighter than squarks, so
that the leptonic decays $\tz_2 \to \ell \bar{\ell} \tz_1$, which lead
to clean signals at hadron colliders, may be considerably
enhanced.~\cite{BT} There are, however, other regions of parameter
space, where sleptons are relatively light, but the amplitudes from
virtual slepton exchanges interfere destructively with the $Z^0$
mediated amplitudes, and lead to a strong suppression of this
decay.~\cite{BCKT,MRENNA} Of course, the branching fraction for the
three-body decay is tiny if two-body ``spoiler modes'' $\tz_2 \to
Z\tz_1$ or $\tz_2 \to h\tz_1$ are kinematically allowed.  For
basically the same reasons the decay $\tz_2 \to \tw_1 f\bar{f'}$ which
is mediated by virtual $W$ exchange, even though it is kinematically
disfavoured, can sometimes be competitive~\cite{BDDT} with the LSP decay
mode of $\tz_2$.~\footnote{Formulae for three body decays of charginos
and neutralinos have been listed by Bartl {\it et. al.}~\cite{BARTL} Their
conventions do not match with ours, so care must be exercised in
transcribing them into a common notation. These formulae do not include
effects of Yukawa interactions which have recently been computed.~\cite{BCDPT}}

We should also mention that, if the parameter $\tan\beta$ is large,
bottom and tau Yukawa interactions can considerably modify \cite{BCDPT}
the decay patterns of charginos and neutralinos. This happens partly
because $\tb_1$ and $\ttau_1$ masses are reduced with respect to those
of other squarks and sleptons, but also because coupling to Higgsino
components of $\tw_1$ and $\tz_1$ is not negligible. For $\tan\beta \agt
25-30$, the branching fraction for the decay $\tw_1 \to \tau\nu \tz_1$
can substantially exceed that of $\tw_1 \to e\nu\tz_1$ or $\tw_1 \to
\mu\nu\tz_1$ decays. Likewise, $\tz_2\to b\bar{b}\tz_1$ may be the
dominant decay mode of $\tz_2$ while the decay $\tz_2\to
\tau\bar{\tau}\tz_1$ can occur much more rapidly than its decay to $e$
or $\mu$. It could even be that the stau becomes so light that the
decays $\tw_1\to \ttau_1\nu$ and $\tz_2\to \ttau_1\tau$ become
accessible, and being the only two-body modes dominate the decay of
charginos and neutralinos.

Finally, we note that there are regions of parameter space where the
rate for the two body radiative decay
\begin{equation}
\tz_2 \to \tz_1 \gamma
\end{equation}
which is mediated by $f\tf$ and gauge boson-gaugino loops may be
comparable~\cite{KOM,HW} to that for the three body decays. These decays are
important in two different cases: ({\it i}) if one of the neutralinos is
photino-like and the other Higgsino-like, both $Z^0$ and sfermion
mediated amplitudes are small since the photino (Higgsino) does not
couple to the $Z$-boson (sfermion), and ({\it ii}) both neutralinos are
Higgsino-like and very close in mass (this occurs for small values of
$|\mu|$); the strong suppression of the three-body phase space then
favours the two-body decay. We mention here that neither of these cases
is particularly likely, especially within the mSUGRA framework.

\subsection{Higgs Boson Decays} 

Unlike in the SM, there is no clear dividing line between the
phenomenology of sparticles and that of Higgs bosons, since as we have
just seen, Higgs bosons can also be produced via cascade decays of heavy
sparticles. Higgs boson decay patterns exhibit~\cite{HHG} a complex
dependence on model parameters.  Unfortunately, we will not have time to
discuss these here, and we can only refer the reader to the literature.
We will, therefore, confine ourselves to mentioning a few points that
will be important for later discussion.

In SUGRA models, all but the lightest Higgs scalar tend to be (but are
not always) rather heavy and so are not significantly produced either in
sparticle decay cascade decays or directly at colliders. An important
exception occurs for very large values of $\tan\beta$ for which
$A$, and hence, also $H$ and $H^{\pm}$ may be within the kinematic
reach of future colliders or even LEP2.

Within the
more general MSSM framework, the scale of their masses is fixed by
$m_A$, which is an independent parameter.  If $m_A$ is large ($ \agt
200$~GeV), 
$h$
(which has a mass smaller than $\sim 130$~GeV) behaves like the SM Higgs
boson, while $H$, $A$ and $H^{\pm}$ are approximately decoupled from
vector boson pairs. The phenomenology is then relatively simple: the
decay $h \to b\bar{b}$ which occurs via $b$-quark Yukawa
interactions dominates, unless charginos and/or neutralinos are also
light; then, decays of $h$ into neutralino or chargino pairs,
which occur via the much larger gauge coupling, is dominant unless
$\tan\beta$ is very large. The
invisible decay $h \to \tz_1\tz_1$, is clearly the one most
likely to be accessible, and has obvious implications for Higgs
phenomenology. These supersymmetric decay modes are even more likely for
the heavier Higgs bosons, particularly if their decay to $t\bar{t}$
pairs is kinematically forbidden; this is especially true for $h$
which cannot decay to vector boson pairs, but also for $H$ since its
coupling to $VV$ pairs ($V=W,Z$) is suppressed when it is heavy.
The decays $A \to h Z$ and $H \to
hh$ can be important, while $H \to AA$
is usually inaccessible.  Finally, charged Higgs bosons $H^+$ mainly
decay via the $t\bar{b}$ mode unless this channel is kinematically
forbidden. Then, they mainly decay via $H^+ \to Wh$, or if this
is also kinematically forbidden, via $H^+ \to c\bar{s}$ (
$\tan\beta \alt 1.2$)
or $H^+ \to
\bar{\tau}\nu$ ($\tan\beta> 1.2$).

\section{Sparticle Production at Colliders}\label{sec:prod}

Since $R$-parity is assumed to be conserved, sparticles can only be pair
produced by collisions of ordinary particles. At $e^+e^-$ colliders,
sparticles (such as charged sleptons and sneutrinos, squarks and
charginos) with significant couplings to either the photon or the
$Z$-boson can be produced via $s$-channel $\gamma$ and $Z$ processes,
with cross sections comparable with $\sigma(e^+e^- \to \mu^+\mu^-)$,
except for kinematic and statistical factors. Selectron and electron
sneutrino production also occurs via $t$-channel neutralino and
chargino exchange, while sneutrino exchange in the $t$-channel will
contribute to chargino pair production.  Neutralino production, which
proceeds via $Z$ exchange in the $s$-channel and selectron exchange in
the $t$ and $u$ channels, may be strongly suppressed if the neutralinos
are gaugino-like and selectrons are relatively heavy. Cross section
formulae as well as magnitudes of the various cross sections may be
found {\it e.g.\/} in Baer {\it et. al.}~\cite{BBKMT}

\begin{figure}
\centerline{\psfig{file=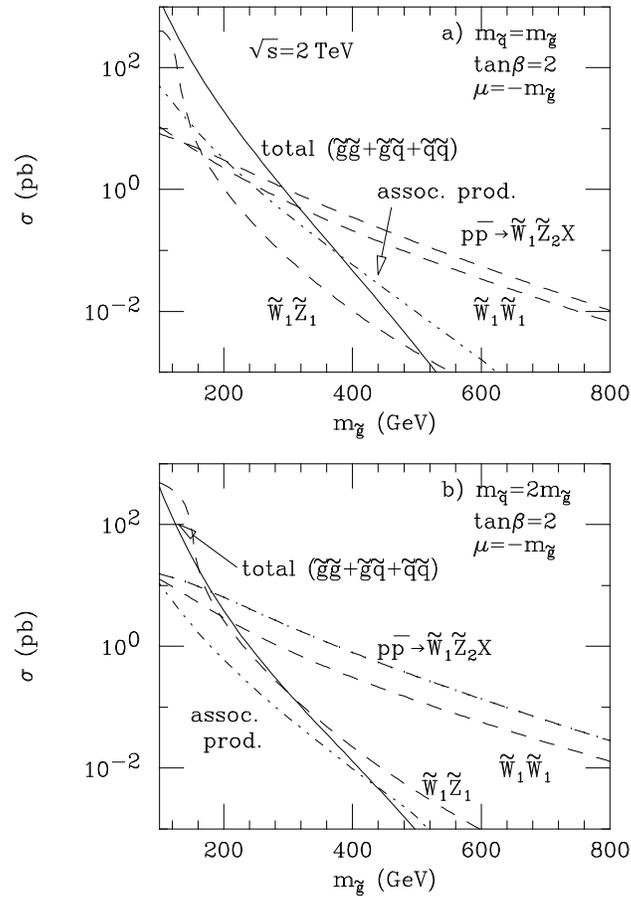,height=13cm,angle=90}}
\caption[]{ Total cross sections for various sparticle production
processes by $p\bar{p}$ collisions at $\protect\sqrt{s}=2$~TeV.}
\label{figtevcs}
\end{figure}
At hadron colliders, the situation is somewhat different. Since
sparticle production is a high $Q^2$ process, the underlying elementary
SUSY process is the inelastic collision of quarks and gluons inside the
proton.~\cite{STAND} In other words, it is the hard scattering
partonic cross section that is computable within the SUSY
framework. This cross section is then convoluted with parton
distribution functions to obtain the inclusive cross section for SUSY
particle production. Thus, unlike at electron-positron colliders, only a
fraction of the total centre of mass energy is used for sparticle
production. The balance of the energy is wasted in the underlying low
$p_T$ event which only contaminates the high $p_T$ signal of interest.

Squarks and gluinos, the only strongly interacting sparticles, have the
largest production cross sections unless their production is
kinematically suppressed.  These cross sections~\cite{SQGLPROD} are
completely determined in terms of their masses by QCD and do not depend
on the details of the supersymmetric model. QCD corrections to these
have also been computed.~\cite{ZERW} Squarks and gluinos can be also be
produced~\cite{ASSPROD} in association with charginos or neutralinos via
diagrams involving one strong and one electroweak vertex. Finally,
$\tw_i$ and $\tz_j$ can be produced by $q\bar{q}$ annihilation via
processes with $W$ or $Z$ exchange in the $s$-channel, or squark
exchange in the $t$ (and, for neutralino pairs only, also the $u$)
channel.

\begin{figure}
\centerline{\psfig{file=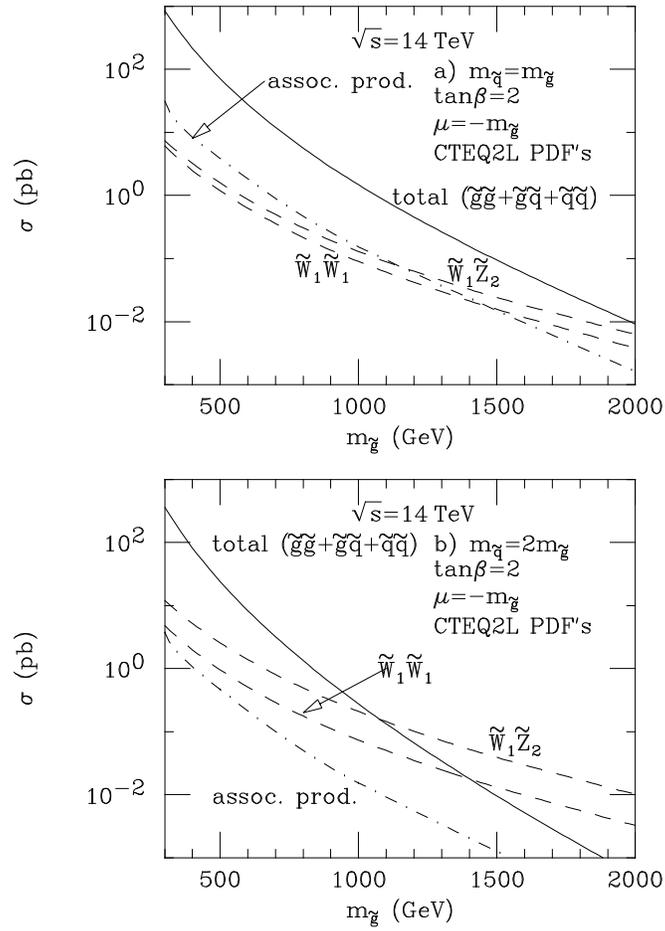,height=13cm,angle=90}}
\caption[]{Total cross sections for various sparticle production
processes by $pp$ collisions at $\protect\sqrt{s}=14$~TeV.}
\label{figlhccs}
\end{figure}

The cross sections for various processes at a 2~TeV $p\bar{p}$ collider
(corresponding to the energy of the Main Injector (MI) upgrade of the
Tevatron) are illustrated in Fig.~\ref{figtevcs}, while those for a
14~TeV $pp$ collider (the LHC) are shown in
Fig.~\ref{figlhccs}.  We have illustrated our results for ({\it
a})~$m_{\tq}=m_{\tg}$, and ({\it b})~$m_{\tq}=2m_{\tg}$ and fixed other
parameters at the representative values shown. These figures help us
decide what to search for. While squarks and gluinos are the obvious
thing to focus the initial search on, we see from Fig.~\ref{figtevcs}
that at even the MI (and certainly at any higher luminosity upgrade that
might be envisioned in the future), the maximal reach is likely to be
obtained via the electroweak production of charginos and neutralinos,
provided of course that their decays lead to detectable
signals.~\footnote{This conclusion crucially depends on the validity of
the gaugino mass unification condition Eq.~(\ref{eq:gaugino}) which
implies that gluinos are much heavier than $\tw_1$ and $\tz_2$.}  In
contrast, we see from Fig.~\ref{figlhccs} that gluino (and, possibly,
squark) production processes offer the best opportunity for SUSY
searches at the LHC for gluino masses up to 1~TeV (recall that this is
roughly the bound from fine-tuning considerations~\cite{FINE}) even if
squarks are very heavy.

\section{Simulation of Supersymmetry Events}\label{sec:sim}

Once produced, sparticles rapidly decay into other sparticles until the
decay cascade terminates in a stable LSP. It is the end products of
these decays that are detectable in experiments and which, it is hoped,
can provide experimental signatures for supersymmetry. The evaluation of
these signatures obviously entails a computation of the branching
fractions for the decays of all the sparticles, and further, keeping
track of numerous cascade decay chains for every pair of parent
sparticles. Many groups have constructed computer programs to calculate
these decay processes. For any set of MSSM parameters~\footnote{The
current version of ISAJET (v.~7.29) allows the reader to input independent soft
SUSY breaking masses for each of the sfermion multiplets as well as
independent masses for the three gauginos. Unless the user explicitly
specifies, ISAJET assumes that sfermions of the first two generations with the
same gauge quantum numbers have the same soft SUSY breaking masses; it
also incorporates gaugino mass unification as the default. 
In addition the user has to specify $\mu$,
$\tan\beta$, $m_A$ along with the three third generation $A$-terms. 
In other words, all thirty MSSM parameters but the six $A$-terms for the first
two generations, which are usually irrelevant for phenomenology can be
independent inputs. This 
allows for simulation of a large variety of theoretical scenarios.}
(or alternatively, for a SUGRA parameter set (\ref{eq:sugra})), a public
access program known as ISASUSY (ISASUGRA) which can be extracted from
the Monte Carlo program ISAJET~\cite{ISAJET} lists all sparticle and
Higgs boson masses as well as their decay modes along with the
corresponding partial widths and branching fractions.

Event generator programs provide the link between the theoretical
framework of SUSY which provides, say, cross sections for final states
with quarks and leptons, and the long-lived particles such as $\pi$,
$K$, $\gamma$, $e$, $\mu$ {\it etc.} that are ultimately detected in
real experiments. Many authors have combined sparticle production and
decay programs to create parton level event generators which may be
suitable for many purposes. More sophisticated generators include other
effects such as parton showers, heavy flavour decays, hadronization of
gluons and quarks, a model of the underlying event, {\it etc.} These
improvements have significant impact upon detailed simulations of, for
instance, the jets plus isolated multi-lepton signal from squark and
gluino production at the LHC. General purpose SUSY event generators available
today are: ISAJET~\cite{ISAJET},
SPYTHIA~\cite{Mrennagen} and SUSYGEN~\cite{Kats}
($e^+e^-$ collisions only). A detailed discussion of these programs and
their virtues and shortcomings is beyond the scope of these Lectures. 
We will instead refer the interested
reader to the literature~\cite{GEN} and to the documentation that accompanies these
codes. 

\section{Observational Constraints on Supersymmetry}\label{sec:CONSTRAINTS}

The non-observation of any supersymmetric signal at either
LEP~\cite{LEPSUSY} or at the Tevatron~\cite{ETMBND,DILEPBND,DZEROSTOP}
provides the most direct lower limits on sparticle masses. Indirect
limits may come from virtual effects of SUSY particles on rare processes
({\it e.g.\/} flavour changing neutral currents or proton decay) or from
cosmological considerations such as an over-abundance of LSP's,
resulting in a universe that would be younger than the age of
stars. While these indirect limits can be important, they are generally
sensitive to the details of the model: non-observation of loop effects
could be a result of accidental cancellation with some other new physics
loops (so care must be exercised in extracting limits on sparticle
masses); proton decay~\cite{PDK} is sensitive to assumptions
about GUT scale physics while the cosmological constraints~\cite{COSM}
can be simply evaded --- the price is the 
loss of a promising dark matter candidate ---
by allowing a tiny violation of $R$-parity
conservation which would have no impact on collider searches. We do not
mean to belittle these constraints which lead to important bounds in any
{\it given} framework (for instance, minimal SUGRA $SU(5)$), but should
also recognise that these bounds are likely to be more model-dependent
than direct constraints from collider experiments. It is, however, only
for reasons of time that we will confine ourselves to direct limits from
colliders.

\setcounter{footnote}{0} 

The cleanest limits on sparticle masses come from experiments at
LEP. The agreement~\cite{LEPC} of $\Gamma_Z$ with its expectation in the
SM gives~\cite{WIDTH} essentially model-independent lower limits of
30-45~GeV on the masses of charginos, squarks, sneutrinos and charged
sleptons whose couplings to $Z^0$ are fixed by gauge symmetry. These
limits~\footnote{The same considerations also exclude spontaneous
$R$-violation via a vev of a doublet sneutrino because the associated
Goldstone boson sector would then have gauge couplings to $Z^0$ and make
too large a contribution to $\Gamma_Z$.}  do not depend on how
sparticles decay. Likewise, the measurement of the invisible width of
the $Z^0$ which gives the well-known bound on the number of light
neutrino species, yields a lower limit on $m_{\tnu}$ only 2-5~GeV below
$\frac{M_Z}{2}$ if the sneutrino decays invisibly via $\tnu \to
\nu\tz_1$, even if only one of the sneutrinos is light enough to be
accessible in $Z^0$ decays.~\footnote{Experiments searching for
neutrino-less double beta decay are designed to detect the recoil of the
nucleus. If stable sneutrinos are the LSP and their density is large
enough to form all of the galactic dark matter their flux would be high
enough to be detectable via elastic scattering from nuclei in these
experiments.  As a result, sneutrinos with masses between 12-20~GeV and
about 1~TeV are excluded.~\cite{NONU} The Kamiokande
experiment~\cite{KAM}, from a non-observation of high energy solar
neutrinos produced by the annihilation of gravitationally accumulated
sneutrinos in the sun exclude 3~GeV$\leq m_{\tnu} \leq $25~GeV. These
limits, when combined with the LEP bounds clearly disfavour the
sneutrino as the stable LSP.}  In contrast, the bounds on neutralino
masses are very sensitive to the model parameters because for large
$|\mu|$, as we have already pointed out, the neutralino may be
dominantly a gaugino with strongly suppressed couplings to the $Z^0$.

LEP experimentalists also perform direct searches for sparticles whose
decays frequently lead to extremely characteristic final
states.~\cite{CHEN} For instance, slepton (squark) pair production
followed by the direct decay of the sfermion to the LSP leads to a pair
of hard, acollinear leptons (jets) together with $\pslt$. Chargino
production can lead to events with acollinear jet pairs, a lepton + jet
+ $\pslt$ and also acollinear leptons + $\pslt$. Such event topologies
are very distinctive and do not occur in the SM if the centre of mass
energy is below the $WW$ threshold. Thus the observation of
just a handful of such events would suffice to signal new physics.
During the past year the LEP energy has been increased in steps from
$M_Z$ to 130-140~GeV to 161~GeV and beyond. 
Currently, the highest energy of LEP operation is 172~GeV. For $\sqrt{s}
> 2M_W$, $WW$ events contaminate the SUSY signal. Strategies for
separating the signal from SM background are discussed in the next
Section. 

From a
non-observation~\cite{YIBIN} of any SUSY events in the
data sample of 11~$pb^{-1}$ that has been accumulated at
172~GeV, lower limits $m_{\tw_1} > 84-86$~GeV, $m_{\te_R} \agt
70$~GeV (the bound on the smuon mass is a little weaker)
have already been deduced. A $t$-squark below 63-75~GeV,
depending on the stop mixing angle is also excluded, assuming $\tt_1 \to
c\tz_1$. Finally, assuming that the GUT unification
condition for $M_1$ and $M_2$ is valid, the L3 and ALEPH 
collaborations have deduced a 95\% lower
limit, $m_{\tz_1}> 24.6$~GeV, if $m_{\tnu} \geq 200$~GeV.
With a larger data sample (as will be expected in the next run)
LEP2 experiments should be able to probe charged sparticles up to 80-90\%
of the kinematic limit.

The search for squarks and gluinos is best carried out at hadron
colliders by searching for $\eslt$ events from $\tq\tq$, $\tg\tq$ and
$\tg\tg$ production. The final states from the cascade decays of gluinos
and squarks leads to events consisting of several jets plus possibly
leptons and $\eslt$.  For an integrated luminosity of about
10-20~$pb^{-1}$ on which the analyses of the Run IA of the CDF and D0
experiments are based, the classic $\eslt$ channel offers the best hope
for detection of supersymmetry. The non-observation of $\eslt$ events
above SM background expectations (after cuts to increase the signal
relative to background) allows Tevatron experimentalists \cite{ETMBND}
to infer a lower limit of 173~GeV on $m_{\tg}$. This bound improves to
229~GeV if squarks and gluinos are assumed to have the same mass.  Since
then, the CDF and D0 experiments have collectively accumulated about
$\sim$~200~$pb^{-1}$ of integrated luminosity, and should begin to be sensitive
to various multilepton signatures which we will discuss when we address
prospects for SUSY searches in the future. We should mention though that
already with the data sample of Run I, non-observation of any events in
the $dileptons + jets +\eslt$ channel allows these collaborations to
infer bounds~\cite{DILEPBND} very close to bounds from the $\eslt$
searches. The D0 Collaboration has also excluded~\cite{DZEROSTOP}
$\tt_1$ with a mass between 60 and 90~GeV, assuming $\tt_1 \to c\tz_1$
and that $m_{\tt_1}-m_{\tz_1}$ is sizeable.

Before closing this section, we briefly remark about potential
constraints from ``low energy'' experiments, keeping in mind that these
may be sensitive to model assumptions. The measurements of the inclusive
$b \to s\gamma$ decay by the CLEO experiment~\cite{CLEO} and its
agreement with SM expectations~\cite{BSGSM} constrain~\cite{BSGSUSY} the
{\it sum} of SUSY contributions to this process.~\footnote{The related
process $b \to s\ell\bar{\ell}$ has also been discussed.~\cite{GOTO}}
Supersymmetry also allows for new sources of CP violation~\cite{CPV} in
gaugino masses or $A$-parameters. These phases, which must be smaller
than $\sim 10^{-3}$ in order that the electric dipole moment of the
neutron not exceed its experimental bound, are set to zero in the MSSM.

Finally, we note that because SUSY, unlike technicolour, is a decoupling
theory, the agreement of the LEP data with SM expectations
is not hard to accommodate. We just have to make the sparticles heavier
than 100-200~GeV. This would, of course, make it difficult to accommodate
``anomalies'' in the LEP data unless some sparticles are rather light. 
The anomalies of yesteryear, however,
seem to be fading away.

\section{Searching for Supersymmetry at Future Colliders and
Supercolliders}\label{sec:Future}

\subsection{$e^+e^-$ Colliders}

We saw in the last Section that the LEP2 collider has been successfully
operated above the $WW$ threshold.  During the next run due to begin in
July 1997, experiments should accumulate $\sim 100$~$pb^{-1}$ of
integrated luminosity at $\sqrt{s}=184$~GeV, and so, should be able to
probe charginos and sleptons up to about 85-90~GeV.  The signals for
sparticles are much the same as discussed in the last Section. The
significant difference is that while SM backgrounds can be easily
removed below the $WW$ threshold, the separation of the SUSY signal from
$W$-pair production requires more effort. This should not be very
surprising since the $W$ is a heavy particle and its decays can lead to
both acollinear dilepton + $\eslt$ as well as $jets+\ell +\eslt$ and
$jets + \eslt$ event topologies. Another possible complication to be
kept in mind as we search for heavier sparticles is that cascade decay
channels may begin to open up. This should not pose too much of a
problem, however, since the energy has been increased in stages. For
example, one would expect to see chargino production before the
production of sleptons which are heavy enough to decay to charginos sets
in.

Signals for sparticle production at LEP2 have been studied in great
detail~\cite{CHEN,DION,GRIVAZ,LEPREP} assuming that sparticles decay directly
to the LSP. Below the $WW$ threshold, they are readily detectable in
exactly the same way as at LEP. Above that, the production of $W^+W^-$
pairs, which has a very large cross section $\sim$18~$pb$ (compared to
0.2~$pb$ for smuons and $\sim 10$~$pb$ for charginos with mass about
$M_W$) is a formidable background. The situation is not as bad as it may
appear on first sight. For $WW$ events to fake sleptons, both $W$'s have
to decay to the particular flavour of leptons, which reduces background
by two orders of magnitude. Further rejection of background may be
obtained by noting that while slepton events are isotropic, the leptons
from $W$ decay exhibit strong backward-forward asymmetry. Thus by
selecting from the sample of acollinear $\mu^+\mu^-$ events those events
where the fast muon in the hemisphere in the $e^-$ beam direction has
the opposite sign to that expected from a muon from $W$ decay, it is
possible to reduce the background by a factor of five, with just 50\%
loss of signal.

The strategy for charginos is more complicated~\cite{GRIVAZ} and will
not be detailed here. We will only mention that here the clean
environment of electron-positron colliders plays a crucial role. The
idea is to make use of the kinematic differences between the two-body
decay of the $W$ into a massless neutrino, and the three body decay of
the chargino into the massive LSP.  Using the cuts detailed in
Ref.~\cite{GRIVAZ}, it should be possible to detect charginos up to
within a few GeV from the kinematic limit in the mixed lepton plus jet
channel. 

Neutralino signals, as we should by now anticipate, are
sensitive to model parameters. A recent analysis~\cite{BAEREE,LEPREP} within
the framework of the SUGRA models describes strategies to optimize these
signals, and also separate them from other SUSY processes. This analysis
also discusses signals from cascade decays of sparticles.

Higher energy electron-positron colliders will almost certainly be
linear colliders, since synchroton radiation loss in a circular machines
precludes the possibility of increasing the machine energy significantly
beyond that of LEP2. Several laboratories are evaluating the prospects
for construction of a 300-500~GeV collider, whose energy may later be
increased to 1~TeV, or more: these include the Next Linear Collider
(NLC) program in the USA, the Japanese Linear Collider (JLC) program in
Japan, the TESLA and CLIC programs in Europe, and VLEPP in the former
Soviet Union.  The search for the lightest charged sparticle, be it the
chargino or the slepton (or perhaps the $\tt_1$) should proceed along
the same lines~\cite{JLC,MUR,CONF,FUJII} as at LEP2 and discovery should be
possible essentially all the way to the kinematic limit. Of course,
because production cross sections rapidly decrease with energy, a
luminosity of 10-30~$fb^{-1}/yr$ will be necessary.  

While these studies have conclusively demonstrated that the lightest of
the ``visible'' sparticles will be easily discovered at the Linear
Collider, cascade
decays have to be incorporated for a study of the heavier sparticles.
Recent studies \cite{BMT,SNOWNLC} within the mSUGRA framework have
examined the prospects for discovering various sparticles at a 500~GeV
Linear Collider.  It is shown that with an integrated luminosity of
$\sim 20$~$fb^{-1}$, it should be possible to discover charged sleptons
(and also sneutrinos if they decay visibly), charginos, $\tt_1$ and
squarks~\cite{FINNELL} essentially all the way up to the kinematic limit
even if these do not directly decay to the LSP. A machine with a centre
of mass energy of about 700-1000~GeV should be able to search for
charginos up to 350-500~GeV, and so, assuming the gaugino mass
unification condition, will cover the parameter space of weak scale
supersymmetry.

It is also worth mentioning that one can exploit~\cite{JLC,MUR} the
availability of polarized beams to greatly reduce SM backgrounds: for
example, the cross section for $WW$ production which is frequently the
major background is tiny for a right-handed electron beam. While the
availability of polarized beams and the clean environment of $e^+e^-$
collisions clearly facilitates the extraction of the signal, we will see
below that these capabilities play a really crucial role for the
determination of sparticle properties which, in turn, serves to
discriminate between models. The availability of polarization does not,
however, appear to be crucial for SUSY discovery.~\cite{BMT}

We should stress that $e^+e^-$ colliders are ideal
facilities to search for Higgs bosons.~\cite{SOPC} At LEP2, one can
typically search for Higgs bosons with a mass up to about
$\sqrt{s}-100$~GeV; an $e^+e^-$ collider operating at 300~GeV would be
virtually guaranteed to find one of the Higgs bosons if the MSSM
framework, with its weakly coupled Higgs sector, is correct, although it
may not be possible to distinguish this from the Higgs boson of the
SM. Indeed if no Higgs boson is discovered at a 500~GeV Linear Collider,
many accepted ideas about unification of interactions may have to be
re-evaluated. 
In contrast, we will see that if the LHC is operated only at its lower
value of the design luminosity (10~$fb^{-1}/yr$), the discovery of a Higgs
boson cannot be guaranteed even with relatively optimistic (but not
unrealistic) assumptions about detector capabilities. It appears
that several years of LHC operation with a luminosity of
100~$fb^{-1}/yr$ are necessary to reasonably 
ensure the discovery of at least one
of the MSSM Higgs bosons.

\subsection{Future Searches at Hadron Colliders}

{\it Tevatron Upgrades}

The CDF and D0 experiments have together already collected an integrated
luminosity of about 200~$pb^{-1}$, to be compared with
10-20~$pb^{-1}$ for the data set on which the $\eslt$ analyses described
in the last Section were based. It is thus reasonable to explore whether
an analysis of this data can lead to other signatures for supersymmetry.
Of course, the size of the data sample will increase by yet another
order of magnitude after about a year of MI operation, and by
significantly more if the TeV33 upgrade, with its design luminosity of
$\sim 10-25~fb^{-1}/yr$, comes to pass.

{\it Gluinos and Squarks:} While the increase in the data sample will
obviously result in an increased reach via the $\eslt$
channel~\cite{KAMON,SNOWTEV}, we have already seen that the cascade decays of
gluinos and squarks lead to novel signals ($n$ jets plus $m$ leptons
plus $\eslt$) via which one might be able to search for SUSY. Since the
gluino is a Majorana particle, it decays with equal likelihood to
positive or negative charginos: the leptonic decays of the chargino can
then lead to events with two isolated same-sign (SS) charged
leptons~\cite{BKP,BGH,BTW} together with jets plus $\eslt$.
 
\begin{figure}
\centerline{\psfig{file=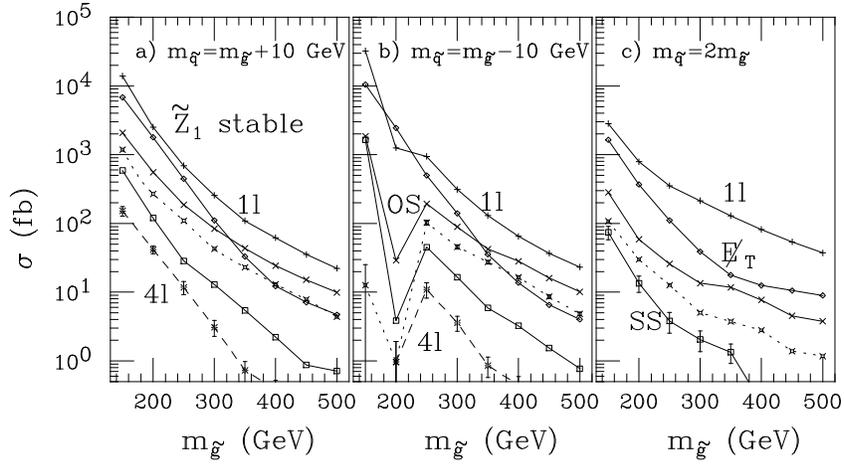,height=7cm,angle=90}}
\caption[]{Cross sections (in $fb$) at the Tevatron
($\protect\sqrt{s}=1.8$~TeV) for various event topologies after cuts
described in Ref.~\protect\cite{RPV} from which this figure is taken. We
take $\mu=-m_{\tg}$, $\tan\beta=2$, $A_t=A_b=-m_{\tq}$ and
$m_A=500$~GeV. The $\eslt$ cross sections are labelled with
diamonds, the 1-$\ell$ cross sections with crosses, the $\ell^+\ell^-$
cross sections with x's and the SS ones with squares.  The dotted curves
are for the $3\ell$ cross sections while the dashed curves show the
cross sections for $4\ell$ events.  For clarity, error bars are shown
only on the lowest lying curve; on the other curves, these error bars
are significantly smaller. The $m_{\tg}=150$~GeV case in
$b$ is excluded even by the LEP constraints on the $Z$ width, since
in this case the sneutrino mass is just 26~GeV.}
\label{figRPV}
\end{figure}

If instead of one of the charginos we have a leptonically decaying
neutralino, trilepton event topologies result. While other topologies
will also be present, the SS and $3\ell$ events are especially
interesting because (after suitable cuts~\cite{RPV,BKTRAP}) the SM {\it
physics} backgrounds~\footnote{In addition, there are always
detector-dependent instrumental backgrounds from misidentification of
jets or isolated pions as leptons, mismeasurement of the sign of the
lepton charge {\it etc.} that a real experimentalist has to contend
with.} are estimated to be 2.7~$fb$ and 0.7~$fb$, respectively, for
$m_t=175$~GeV.  The corresponding signal cross sections, together with
the cross sections in other channels, are illustrated in
Fig.~\ref{figRPV} which has been obtained using ISAJET 7.13. These
include signals from all sparticle sources, not just gluinos and
squarks. We see that while the cross sections in the relatively
background-free $3\ell$ and
SS event topologies are indeed tiny, Tevatron experiments should just
about be reaching the sensitivity to probe SUSY via these
channels. Although the experimental groups have yet to analyse their
data for these rare signals, it is reassuring that they have performed
the analysis of the multijet plus dilepton plus~$\eslt$ channel and
obtain bounds \cite{DILEPBND} similar to those from the $\eslt$ analysis.
This may be viewed as evidence that these multi-lepton analyses are feasible.

Before closing, we remind the reader that for large values of
$\tan\beta$ charginos and neutralinos preferentially decay into the
third generation particles. In fact, we have seen that they may even
exclusively decay to the stau family. In this case, multilepton signals
may be greatly reduced, although there are potentially new
signatures~\cite{BCDPT} involving $b$-jets and $\tau$-leptons via which
to search for SUSY. Prospects for these searches are currently under
investigation.

{\it Charginos and Neutralinos:} The electroweak production of charginos
and neutralinos, we have seen, offers yet another channel for probing
supersymmetry, the most promising of which is the hadron-free trilepton
signal from the reaction $p \bar{p} \to \tw_1\tz_2 X$, where both the
chargino and the neutralino decays leptonically.  In fact, we saw in
Fig.~\ref{figtevcs} that for very large integrated luminosities, this
channel potentially offers the maximal reach for supersymmetry (since
the opposite sign (OS) 
dilepton signal from $\tw_1 \overline{\tw_1}$ production suffers
from large SM backgrounds from $WW$ production).  It was first
emphasized by Arnowitt and Nath~\cite{AN} that, with an integrated
luminosity of $\sim 100~pb^{-1}$, this signal would be observable at the
Tevatron even if resonance production of $\tw_1\tz_2$ is suppressed. A
subsequent analysis~\cite{BT} showed that the signal may even be further
enhanced in some regions of parameter space due to enhancements in the
$\tz_2$ leptonic branching fractions, as discussed in
Sec.~\ref{sec:decays}. Detailed Monte Carlo
studies~\cite{BKTWINO,KAMON,BCKT,MRENNA,SNOWTEV} including effects of
experimental cuts were performed to confirm that Tevatron experiments
should indeed be able to probe charginos via this channel. Indeed
analyses~\cite{CDFWINO} by the CDF and D0 collaborations, from a
non-observation of a signal in this channel, have obtained upper limits
on the trilepton cross section. The resulting limit on the chargino mass
is below that obtained from LEP. This analysis may be viewed as leading
to the best (direct) lower limit on $m_{\tz_2}$ over some ranges of model
parameters. It is possible that in
the future Tevatron experiments will probe parameter regions that may
not be accessible at LEP2.  

We stress that the trilepton signal, which depends on the
neutralino branching fractions, is sensitive to the model parameters,
and it is not possible to simply state the reach in terms of the mass of
the chargino. For favourable values of parameters, experiments at the MI
should be able to probe charginos heavier than 100~GeV, corresponding to
$m_{\tg} \agt 300-350$~GeV (at TeV33, up to $\sim$500-600~GeV where the
two-body spoiler decays of $\tz_2$ become accessible, and for light
sleptons, even up to 600-700~GeV); on the other hand, there are other
regions of parameter space where the leptonic branching fraction of the
neutralino is strongly suppressed~\cite{BCKT,MRENNA}, and there is no
signal for charginos as light as 45-50~GeV even at the TeV33 upgrade of
the Tevatron. The signal may also be suppressed \cite{BCDPT} because
charginos and neutralinos decay exclusively into stau (and $\tnu_{\tau}$).
Thus, while this channel can probe significant regions of
the parameter space of either the MSSM or SUGRA, the absence of any
signal in this channel will not allow one to infer a lower limit on
$m_{\tw_1}$.

{\it Top Squarks:} We have seen that $\tt_1$, the lighter of the two top
squarks, may be rather light so that it may be pair-produced at the
Tevatron even if other squarks as well as the gluino are too heavy. If
its decay to chargino is allowed, the leptonic decay of one or both
stops leads to events with one or two isolated leptons together with
jets plus $\eslt$ --- exactly the same event topologies as for the top
quark search. Thus $t\bar{t}$ production is the major
background~\cite{BDGGT} to the $t$-squark search. Because
$\sigma(t\bar{t}) \sim 10 \sigma(\tt_1\bar{\tt_1})$ for a top and stop
of the same mass, stop signals are detectable at the Tevatron only if
the stop is considerably lighter than the top. In an early analysis
\cite{BST}, it was shown that with an integrated luminosity of around
100~$pb^{-1}$, the stop signal should be detectable at the Tevatron if
$m_{\tt_1} \alt 100$~GeV, provided $b$-jets can be adequately
tagged.~\footnote{M.~Mangano (private communication) has also performed
this analysis with a more detailed simulation and more realistic
assumptions about $b$-tagging capabilities.}  Since then, LEP
experiments have obtained strong bounds on the chargino mass, leaving
only a small range of parameters where this analysis is applicable. The
D0 collaboration \cite{D0STOP} have searched their data sample of $\sim
75$~$pb^{-1}$ for $\tt_1\bar{\tt_1} \to b\tw_1 b\tw_1 \to bb e^+e^-$
events and found that an order of magnitude larger data sample is needed
to attain the sensitivity that is needed for this search.~\footnote{The
sensitivity would be improved when channels involving $e$ and $\mu$ are
analysed and combined. Enhanced $b$-tagging capability that should be
available during the next run will also improve the sensitivity.}

If the chargino is heavy, the stop will instead decay via $\tt_1 \to c
\tz_1$ and stop pair production will be signalled by dijet plus $\eslt$
events, and hence looks like the squark signal, but without any cascade
decays.~\cite{BST} An 
analysis by the D0 collaboration~\cite{DZEROSTOP} excludes
60~GeV~$<m_{\tt_1}<$90~GeV if the LSP mass is smaller than 25-50~GeV.

Mrenna {\it et. al.},~\cite{MRENNA}
using cuts optimized to detect heavier stops, find that, with an
integrated luminosity of 2~$fb^{-1}$, it should be possible to explore
stops as heavy as 160~GeV if they decay via the chargino mode. They
claim a reach of 200~GeV with a data sample of 25~$fb^{-1}$ that may
be available at TeV33. If chargino is too heavy for the decay
$\tt_1 \to b\tw_1$ to be accessible but the stop mass is in the
180-250~GeV range, the three body decay $\tt_1 \to bW\tz_1$
may be kinematically accessible. Since its branching ratio is sensitive
to model parameters,~\cite{WOEHRMAN} one has to examine how this decay
compares to the loop decay $\tt_1 \to c\tz_1$ in order to assess the
viability of this signal.

{\it Sleptons:} The best hope for slepton detection appears~\cite{SLEP}
to be via the clean OS dilepton plus $\eslt$ channel. But even here,
there is a large irreducible background from $WW$ production as well as
possible contamination of the signal from other SUSY sources such as
chargino pair production. It was concluded that at the MI it would be
very difficult to see sleptons from off-shell $Z$ production, {\it i.e.}
if $m_{\tell} \agt 50$~GeV. Sleptons in this mass range are obviously
already excluded by LEP experiments. At TeV33, experiments would probe sleptons
with masses up to about 100~GeV, given an integrated luminosity of
25~$fb^{-1}$.

{\it SUSY Searches at the LHC}

While it is certainly possible that SUSY may be discovered at a
luminosity upgrade of the Tevatron, we have seen that there are
parameter ranges where a SUSY signal may evade detection even if
sparticles are not very heavy. In order to cover
the complete parameter-space of weak scale SUSY either a linear collider
operating at a centre of mass energy $\sim 0.7- 1$~TeV or the LHC is
necessary.

We see from Fig.~\ref{figlhccs} that, at the LHC, squarks and gluinos
dominate sparticle production up to gluino masses beyond 1~TeV. It is thus
reasonable to focus most attention on these although, of course, signals
should be looked for in all possible channels. For reasons of brevity,
and because the ideas involved in LHC searches are qualitatively similar
to those described above, we will content ourselves with just presenting
an overview of the LHC reach, and refer the interested reader to the
vast amount of literature that already exists for details.

As before, the cascade decays of gluinos and squarks result in $n$-jet
plus $m$-leptons plus $\eslt$ events~\cite{BTW2,ATLAS,CMS} where $m=0$
corresponds to the classic $\eslt$ signal. Of the multilepton channels,
the SS and $m \geq 3$ channels suffer the least from SM backgrounds. It
is worth keeping in mind that at the LHC, many different sparticle
chains contribute to a particular event topology, and further, that the
dominant production mechanism for any particular channel depends on the
model parameters. For instance, gluino pair production (with each of the
gluinos decaying to a chargino of the same sign) is generally regarded
as the major source of SS dilepton events; notice, however, that the
reaction $pp \to \tb_L\bar{\tb}_L X \to t{\tw_1}^- \bar{t}{\tw_1}^+ X$
may also be a copious source of such events, since now the leptons can
come either from top or chargino decays (recall that we had noted that
$\tb_L$ may be relatively light).  It is, therefore, necessary to
simultaneously generate all possible sparticle processes in order to
realistically simulate a signal in any particular event topology. This
is possible using ISAJET. However, this raises another issue which is
especially important at the LHC. If we see a signal in any particular
channel, can we uncover its origin? We will return to this later, but
for now, focus ourselves on the SUSY reach of the LHC.

The ATLAS collaboration~\cite{ATLAS} at the LHC has done a detailed
analysis of the signal in the $\eslt$ as well as in the SS dilepton
channels. They found that gluinos as light as 300~GeV should be easily
detectable in the $\eslt$ channel.~\footnote{The capability of LHC experiments
to detect a signal from relatively light gluinos is important to ensure
that there is no window of masses where SUSY signals may escape
detection at both the Tevatron MI and the LHC.}  
Then requiring rather stiff cuts,
$\eslt > 900$~GeV, $p_T(jet_1,jet_2,jet_3)>200$~GeV,
$p_T(jet_4)>100$~GeV along with a cut $S_T>0.2$ on the transverse
sphericity, they find that with an integrated luminosity of
10~$fb^{-1}$, it should be possible to search for gluinos
with a mass up to 1.3~TeV (2~TeV) for $m_{\tq}=2m_{\tg}$
($m_{\tq}=m_{\tg}$). This reach is altered by about $\pm300$~GeV if the
integrated luminosity is changed by an order of magnitude. Very similar
results for the reach in the $\eslt$ channel have also been
obtained~\cite{BCPT} within the context of the SUGRA framework, although
the event selection criteria used are quite different. In the same-sign
dilepton channel, the ATLAS collaboration~\cite{ATLAS} concludes that
the reach of the LHC will be 900-1400~GeV (for $m_{\tq}=2m_{\tg}$) or
1100-1800~GeV (for $m_{\tq}=m_{\tg}$), where the lower (higher) number
corresponds to a luminosity of 1~$fb^{-1}$ (100~$fb^{-1}$). 

Prospects for SUSY detection in the SS and other multilepton channels
(with or without real $Z$ bosons) have also been discussed.~\cite{BTW2}
An analysis of the multilepton signals within the context of SUGRA
models has also been performed.~\cite{BCPT2} 
The greatest SUSY reach is obtained in
the $1\ell$ channel. The reason is that a large fraction of events from
a heavy gluino or squark contain at least one lepton from cascade
decays. By exploiting the presence of hard jets and large $\eslt$ in
SUSY events, these authors have been able to devise cuts to reduce SM
backgrounds from $W \to \ell\nu$ and $t\bar{t}$ production to a
manageable level. Assuming an integrated luminosity of 10~$fb^{-1}$,
a gluino mass reach of $\sim$2.3 (1.6)~TeV is
claimed for $m_{\tg} \simeq m_{\tq} \ (1.5m_{\tq})$.  More importantly,
it has been shown that for $\mhf \leq 400$-500~GeV ($m_{\tg}\leq
1$-1.3~TeV), there should be an observable signal in several (OS, SS,
$3\ell$) channels if $\tan\beta \alt 10$. Otherwise, any signal in the
$\eslt$ or $1\ell$ channel cannot be attributed to gluino and squark
production in the mSUGRA framework. For very large values of
$\tan\beta$, sparticle decay patterns may be considerably modified, and
the situation needs to be reassesed. We do not, however, expect SUSY
would evade detection at the LHC.

Within the SUGRA framework, the LHC should, in the clean trilepton
channel, be able~\cite{LHCWINO,PHD} to probe $\tw_1\tz_2$ production all
the way up to where spoiler modes of $\tz_2$ become accessible if $\mu <
0$ and $\tan\beta$ is not too large. Then it is possible to find a set
of cuts that cleanly separate the $\tw_1\tz_2$ event sample from both SM
backgrounds as well as other sources of SUSY events. This will prove to be
important later. For positive values of $\mu$, signals are readily
observable for rather small and large values of $m_0$; in the
intermediate range 400~GeV~$ \alt m_0 \alt 1000$~GeV, this signal is
suppressed because of the suppression of the leptonic $\tz_2$ branching
fraction emphasized earlier.

As at Tevatron upgrades, the OS dilepton channel offers~\cite{AMET,SLEP}
the best opportunity for slepton searches.  At the LHC, it should be
possible to detect sleptons up to about 250-300~GeV, although excellent
jet vetoing capability will be needed to detect the signal for the
highest masses.

The SUSY reach of possible future hadron colliders is summarized in
Table~1.
\begin{table}[t]
\small
\caption[] {Estimates of the discovery 
reach of various options of future hadron colliders. The signals have 
mainly been computed for negative values of $\mu$. We expect that the reach
in especially the $all \to 3\ell$ channel will be sensitive to the sign of
$\mu$.}

\vskip 0.4cm
~\footnotesize
\begin{tabular}{c|cccc}
&Tevatron II&Main Injector&TeV33&LHC\\
Signal&0.1~fb$^{-1}$&1~fb$^{-1}$&10~fb$^{-1}$&10~fb$^{-1}$\\
&1.8~TeV&2~TeV&2~TeV&14~TeV\\
\hline
$\eslt (\tq \gg \tg)$ & $\tg(210)/\tg(185)$ & 
$\tg(270)/\tg(200)$ & $\tg(340)/\tg(200)$  & $\tg(1300)$ \\ 

$l^\pm l^\pm (\tq \gg \tg)$ & $\tg(160)$ & $\tg(210)$ & 
$\tg(270)$  & \\ 

$all \rightarrow 3l$ $(\tq \gg \tg)$ & $\tg(180)$ & 
$\tg(260)$ & $\tg(430)$ & \\ 

$\eslt (\tq \sim \tg)$ & $\tg(300)/\tg(245)$ & 
$\tg(350)/\tg(265)$ & $\tg(400)/\tg(265)$ & 
$\tg(2000)$ \\ 

$l^\pm l^\pm (\tq \sim \tg)$ & $\tg(180-230)$ & $\tg(320-325)$ & 
$\tg(385-405)$  & $\tg(1000)$ \\ 

$all \rightarrow 3l$ $(\tq \sim \tg)$ & $\tg(240-290)$ & 
$\tg(425-440)$ & $\tg(550^)$  & 
$\stackrel{>}{\sim}\tg(1000)$ \\ 


$\tilde{t}_1 \rightarrow c \tz_1$ & $\tilde{t}_1(80$--$100)$ &
$\tilde{t}_1 (120)$ & $\tt_1(150)$   &\\

$\tilde{t}_1 \rightarrow b \tw_1$ & $\tilde{t}_1(80-100)$ &
$\tilde{t}_1 (120)$ & $\tt_1(180)$   &\\ 

$\Theta(\tilde{t}_1 \tilde{t}_1^*)\rightarrow \gamma\gamma$ &
--- & --- & ---  & $\tilde{t}_1 (250)$\\

$\tl \tl^*$ & $\tl(50)$ & $\tl(50)$ & $\tl(100)$ & 
$\tl(250$--$300)$

\end{tabular}
\end{table}

Several comments  are worth noting:
\begin{itemize}

\item In some places, two sets of numbers are given for the reach. These
correspond to results from different analyses more fully described in
the review~\cite{DPF} from where this Table is taken. Basically, the
more conservative number also requires the signal to be larger than 25\%
of the background, in addition to exceeding the 5$\sigma$ level. Also,
the two analyses do not use the same cuts.

\item The multilepton rates in the Table are shown for negative values
of $\mu$ and $\tan\beta=2$.  For other parameters, especially for $\mu >
0$, the trilepton rates may be strongly suppressed due to a suppression
of the $\tz_2$ branching fraction discussed above. Notice also that at
TeV33, the reach in the leptonic channels exceeds that in the $\eslt$
channel.  At the TeV33 upgrade, hadronically quiet trilepton events may
be observable all the way up to the spoiler modes for favourable ranges
of model parameters. It is, however, important to remember that
supersymmetry may escape detection at these facilities even if
sparticles are relatively light.

\item At the LHC, gluinos and squarks are detectable to well beyond
1~TeV in the $1\ell$ and $\eslt$ channels, and up to 2~TeV if their masses are
roughly equal. Thus the LHC should be able to probe the complete
parameter space of weak scale SUSY, at least within the assumed
framework. Moreover, there should be observable signals in the
multilepton channels if a signal in the $\eslt$ channel is to be attributed
to supersymmetry.

\item Tevatron upgrades will not probe sleptons significantly beyond the
reach of LEP2, whereas the LHC reach may be comparable to that of the
initial phase of linear colliders.

\item Tevatron upgrades should be able to detect $\tt_1$ with a mass up
to 120~GeV at the MI~\cite{BST}, and up to 150-180~GeV at
TeV33.~\cite{MRENNA,SNOWTEV} It has also been pointed out, assumming
that $\tt_1 \to c\tz_1$ is its dominant decay, that it should be
possible~\cite{DN} to search for $\tt_1$ at the LHC via the two photon
decay of the scalar $\tt_1\bar{\tt_1}$ bound state, in much the same way
that Higgs bosons searches (to be discussed next) are carried out.

\end{itemize}

{\it Higgs Bosons:} At the LHC, MSSM Higgs
bosons are dominantly produced~\cite{HHG} by $gg$ fusion (via loops of
quarks and squarks), and for some parameter ranges, also via $gg \to b\bar{b}H$
reactions. Vector boson fusion, which in the SM dominates these other
mechanisms for large Higgs masses ($m\agt 900$~GeV) is generally
unimportant, since the couplings of heavy Higgs bosons to $VV$ pairs is
suppressed. Unfortunately, we do not have much time to discuss various
strategies that have been suggested~\cite{HIGGS} for the detection of
the Higgs sector of SUSY. Over much of the parameter space, all the
Higgs bosons except the lightest neutral scalar, $h$, are too
heavy to be of interest, although for some ranges of MSSM parameters,
signals from $H \to \gamma\gamma, \tau\bar{\tau}, \mu\bar{\mu}, 4\ell$
and $A\to \gamma\gamma, \tau\bar{\tau}, \mu\bar{\mu}$ may be
observable. The two photon decay mode is the most promising channel for
$h$ detection at the LHC. The regions of parameter space where
there is some signal for an MSSM Higgs boson either at the LHC or at
LEP2 have been nicely summarized in the technical report of the CMS
Collaboration,~\cite{CMS} assuming that sparticles are too heavy to be
produced via Higgs boson decays.~\footnote{This is not necessarily a
good assumption. The branching fractions for the SUSY decays can be
quite substantial for large regions of parameter space, and can reduce
the signals via which the Higgs bosons are usually searched for and
increase the parameter space hole referred to below.~\cite{BISSET1}
Sometimes, however, they lead to novel signals for Higgs boson searches
which can then refill~\cite{BISSET2} some of the hole region.  Of
course, for almost all cases where SUSY Higgs decays are important, it
should also be possible to detect the sparticles at the LHC. It is only
Higgs boson detection that may be more difficult.}  The most striking
feature of their analysis is that despite optimistic detector
assumptions, if the LHC is operated only at its low design luminosity
option,
there are regions of parameter space where {\it there may
be no signal for any of the Higgs bosons either at the LHC or at LEP2.}
Part of this hole may be excluded by analyses of rare
decays such as $b\to s\gamma$ mentioned earlier. 

It has been
suggested~\cite{GUN} that Higgs boson signals may also be detectable in
this hole region via $t\bar{t} h$ production where a lepton from
$t$ decay may be used to tag the event so that $h$ can then be
detected via its dominant $b\bar{b}$ decay. This would require efficient
$b$ tagging with the high luminosity option for the LHC. Whether this is
technically possible is not clear at this time. A different
analysis~\cite{FROID} indicates that the with an integrated luminosity of
300~$fb^{-1}$ the LHC would be ``guaranteed'' to find at least
one of the MSSM Higgs bosons over essentially the entire range of
parameters not covered by LEP2 via signals in
the $h \to \gamma\gamma$ and $A,H \to \tau\tau$ channels if the data
from the ATLAS and CMS experiments are combined. It is instructive to note
that there are substantial
regions of parameter space where it would be possible to detect more
than one Higgs boson of the MSSM. 

It is clear that Higgs boson searches
at the LHC pose a formidable experimental challenge. In contrast, these
would be relatively easy at a linear
collider, the first of many examples of the complementary nature of
these facilities.

\subsection{The SUSY Reach of Various Facilities: A Recapitulation}

\begin{figure}
\centerline{\psfig{file=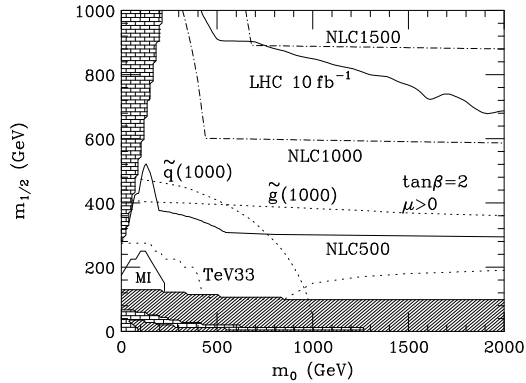,height=5cm,angle=90}}
\caption[bth]{The SUSY reach for various facilities as given by the mSUGRA
model. We take $\tan\beta=2$, $A_0=0$ and $\mu > 0$.}
\label{fig:SUGRA}
\end{figure}

Because the mSUGRA model is completely determined by just four
parameters, it provides a compact framework for comparing the prospects
for SUSY detection via disparate SUSY processes and at different
experimental facilities. Within this framework, the scale of sparticle
masses is mainly determined by $m_0$ and $\mhf$ so that
the $m_0-\mhf$ plane, for fixed values of $A_0$, $\tan\beta$ and
$\sgn\mu$ provides a convenient panorama for comparing the capabilities
of various future facilities, as shown in Fig.~\ref{fig:SUGRA}. Here, we
have chosen $\tan\beta=2$, $A_0=0$ and $\mu> 0$. Except perhaps for very
large values of $\tan\beta$, the qualitative features illustrated 
are only weakly sensitive to this choice.

The bricked (hatched) region is excluded by theoretical
(experimental) constraints.~\cite{BEAR} The region below the lines
labeled MI and TeV33 is where experiments at the Tevatron should be
able to discover SUSY, assuming an integrated luminosity of 2 and
25 fb$^{-1}$.  The discovery region is a
composite of several possible discovery channels, although the $\eslt$
and clean $3\ell$ channels dominate the reach.  To help orient
the reader, we have also shown contours for gluino and squark masses
of 1~TeV.

The upper solid line of Fig.~\ref{fig:SUGRA} shows the boundary of the
corresponding region at the LHC \cite{BCPT2} which essentially coincides
with the discovery region in the $1\ell\ +$ jets $+\eslt$ channel.
Similarly, the solid line labeled NLC500 denotes the reach of the NLC
operating at $\sqrt{s}=0.5$ TeV, as obtained using ISAJET \cite{BMT}.
It consists of three parts: the horizontal portion at $\mhf \sim$ 300
GeV essentially follows the $m_{\tw_1}=250$~GeV contour, while the
rising portion below $m_0=200$~GeV follows the $m_{\te_R}=250$~GeV
contour. The reach drops when $m_{\te_R} \simeq m_{\tz_1}$ because the
daughter electron becomes too soft. An observable signal from $e^+e^-
\to \tz_1\tz_2$ makes up the intermediate portion of the contour.  The
dashed-dotted contours mark the boundaries of the region where $\tw_1$
and/or $\te_R$ are kinematically accessible at NLC1000 or 1500.
Although new backgrounds from two-to-three- and four-particle production
processes have not been evaluated, we believe that this region closely
approximates the reach of the NLC operating at these higher energies.

The following observations are worthy of emphasis:

\begin{itemize}

\item We again see that though TeV33 can probe an interesting region of
parameter space, there is a significant range of parameters, consistent
with qualitative upper bounds on sparticle masses from fine tuning
arguments, where there is no observable signal.  
The discovery reach of TeV33 is not
overwhelmingly larger than that of the Main Injector (MI).

\item The large reach of the LHC is evident from this figure.
We also see that the LHC provides a significant safety margin over the
upper limits expected from fine-tuning arguments. 

\item For the purposes of assessing the SUSY reach (and for this purpose
alone), we see that an $e^+e^-$ collider operating between $1-1.5$~TeV
has a reach similar to that of the LHC.

\end{itemize}

\section{Beyond SUSY Discovery: More Ambitious Measurements}\label{sec:amb}

We have seen that if the minimal SUSY framework that we have adopted is
a reasonable approximation to nature, experiments at supercolliders
should certainly be able to detect signals for physics beyond the SM. If
we are lucky, such signals might even show up at LEP2 or at Tevatron
upgrades.  We will then have to figure out the origin of these signals.
If the new physics is supersymmetry, it is likely (certain at the LHC)
that there will simultaneously be signals in several channels. While the
observation of just one or two of these signals would convince the
believers, others would probably demand stronger evidence.  It is not,
however, reasonable to expect that we will immediately detect all (or
even several of) the super-partners. Thus, it is important to think
about just what information can be obtained in various experiments,
information that will help us to elucidate the nature of the underlying
physics. Towards this end, we would like to be able to:

\begin{itemize}

\item measure any new particle's masses and spins, and
\item measure its couplings to SM particles; these would serve to pin
down its internal quantum numbers.
\end{itemize}
More ambitiously, we may ask:
\begin{itemize}
\item Assuming that the minimal framework we have been using is correct,
is it possible to measure the model parameters? Is it possible to
actually provide tests for, say, the mSUGRA framework, and thus
also test the assumptions about the physics at the GUT or Planck scale
that are part and parcel of this picture?

\item At hadron colliders, especially, where several new sparticle
production mechanisms may be simultaneously present,~\footnote{Since the
energy of any linear collider is likely to be increased in several steps
to the TeV scale, one may hope that this will be less of a problem
there. The lighter sparticles will be discovered first. Knowledge about
their properties thus obtained should facilitate the untangling of the
more complex decays of heavier sparticles.} is it possible to untangle
these from one another?

\item As mentioned in Sec.~1, like any other (spontaneouly
broken) symmetry, supersymmetry, though softly broken, implies
relationships between the various couplings in the theory.  Is it
possible to directly test supersymmetry by experimentally verifying
these coupling constant relationships?
\end{itemize}

\subsection{Mass Measurements}

{\it $e^+e^-$ colliders:} The clean environment of $e^+e^-$ colliders as
well as the very precise energy of the beam allows for measurements of
sparticle masses. We will briefly illustrate the underlying ideas with a
simple example. It is easy to show that the total cross section for
smuon production has the characteristic $P$-wave threshold dependence,
\begin{displaymath}
\sigma(\tmu\bar{\tmu}) \propto (1-\frac{4{m_{\tmu}}^2}{s})^\frac{3}{2},
\end{displaymath}
and further, that the energy distribution of the daughter muon from the
decay of the smuon, assuming only direct decays to the LSP, is flat and
bounded by~\cite{ST}
\begin{equation}
\frac{m_{\tmu}^2-m_{\tz_1}^2}{2(E+p)} \leq E_{\mu} \leq 
\frac{m_{\tmu}^2-m_{\tz_1}^2}{2(E-p)},
\label{eq:ENDPT}
\end{equation}
with $E$($p$) being the energy (momentum) of the smuon.  We thus see
that the energy dependence of the smuon cross section gives a measure of
the smuon mass, while a measurement of the end points of the muon energy
spectrum yields information about $m_{\tmu}$ as well as $m_{\tz_1}$.  

Of course, theoretically these relations are valid for energy and
momentum measurements made with ideal detectors without any holes and
with perfect energy and momentum resolutions. In real detectors there
would be smearing effects as well as statistical fluctutations. It has
been shown,~\cite{CHEN} taking these effects into account, that with an
integrated luminosity of 100~$pb^{-1}$, experiments at LEP2 should be
able to determine the smuon mass within 2-3~GeV. At Linear Colliders
such as the JLC, an integrated luminosity of 20~$fb^{-1}$ will be
sufficient~\cite{MUR,JLC,FUJII} to determine the smuon and LSP masses to
within 1-2~GeV. The availability of polarized beams is extremely useful
to obtain pure signal samples. 

If sleptons are heavy but charginos
light, a study of the reaction $e^+e^- \to \tw_1\overline{\tw_1} \to
jj\tz_1+ \ell\nu\tz_1$ should allow the determination of $m_{\tw_1}$ and
$m_{\tz_1}$ with a precision of $\alt 3$~GeV, both at LEP2 (where an
integrated luminosity of about 1~$fb^{-1}$ would be
necessary~\cite{STRASSLER}) and at
the JLC. A good jet mass resolution is crucial. Finally, it has also
been shown~\cite{FINNELL} that with the availability of beam
polarization at linear colliders it should be possible to determine
squark masses with a precision of $\sim 5$~GeV even if these decay via
MSSM cascades, assuming that their decays to gluino are not allowed: 
in particular, it should be possible to determine the
splittings between $L$- and $R$-type squarks with good precision.

Precision mass measurements are also possible~\cite{BMT,SNOWNLC} at
Linear Colliders even if sparticles cascade decay. Since the end-points
of the lepton energy spectrum in Eq.~(\ref{eq:ENDPT}) do not depend on the
stability of the daughters, the end points of the electron energy
spectrum from the process $e^+e^- \to \tnu_e\bar{\tnu_e} \to
e^-\tw_1^+e^+\tw_1^- \to e^+e^-\mu^{\pm}jj+\eslt$ provides an
opportunity for simultaneous measurement of $m_{\tnu_e}$ and
$m_{\tw_1}$. With a left-handed electron beam (conservatively assumed to
be 80\% polarized) these masses are shown to be measureable to better
than 1.5\%. A measurement of the $b$-jet energy in is a $e^+e^- \to
\tt_1\bar{\tt_1} \to b\tw_1 b\tw_1$ events has been shown to yield the
stop mass to $6\%$ (better if $m_{\tw_1}$ is independently
determined). For a discussion of other mass measurements including from
three body decay of the chargino, we refer the reader to the original
literature.~\cite{BMT,SNOWNLC}.

{\it Hadron Colliders:} Can one say anything about sparticle masses from
SUSY signals at hadron colliders? Despite the rather messy event
environment,
it is intuitively clear that this should be possible if one can isolate
a single source of SUSY events from both SM backgrounds as well as from
other SUSY sources: a study of the kinematics would then yield a measure
of sparticle masses within errors determined by the detector resolution.
We have already seen that it is indeed possible to isolate a relatively
clean sample of $p\bar{p} \to \tw_1\tz_2 \to \ell\bar{\ell}\ell' +
\eslt$ events at the LHC.  The end-point of the $m_{{\ell}\bar{\ell}}$
distribution, it has been shown,~\cite{LHCWINO,PHD} yields an accurate
measure of $m_{\tz_2}-m_{\tz_1}$. 

At the 1996 Snowmass Workshop, an mSUGRA model parameter set
$(m_0,\mhf, A_0,\tan\beta, \sgn\mu) = (200, 100, 0 , 2, -1)$ (all
mass parameters are in GeV), which leads to a rather light sparticle
spectrum,
was chosen to facilitate a comparison of
the capabilities of TeV33, NLC and the LHC for studying SUSY. We will
refer to this as the Snowmass Point.
The LHC subgroup \cite{SNOWLHC} showed
that for this point at least, it is possible to isolate a very pure
sample of gluino pair events. They identified an important 
decay chain $\tg \to b
\tb_1 \to bb\tz_2 \to bb\ell^+\ell^-\tz_1$ which has a combined branching
fraction of 25\%. Thus gluino pair production leads to very distinctive
events with multiple $b$-jets and several leptons.
Even allowing for $b$-tagging efficiency and other experimental cuts, they
estimated that there should be $\sim$272K such 
events per LHC year! This enormous data sample enabled
them to infer that a determination of the end point of the dilepton mass
distribution will only be limited by the error in the absolute
calibration of the electromagnetic calorimeter, which they estimated to be
50~MeV. These authors also went on to show how $m_{\tg}-m_{\tb_1}$ can be
determined to 10\% for this value of mSUGRA parameters. 

A measurement of the gluino mass would be especially important at hadron
colliders since gluinos cannot be pair produced by tree-level processes
at $e^+e^-$ colliders. The best technique for this has been
proposed by Paige \cite{PAIGE} also at the 1996 Snowmass Workshop. He
showed that the distribution in the variable,
\begin{displaymath}
M_{eff}=|p_{T1}|+|p_{T2}|+|p_{T3}|+|p_{T4}|+\eslt
\end{displaymath}
defined as the scalar sum of the transverse energies of the four hardest
jets plus $\eslt$, yields a measure of the gluino/squark mass scale
defined as $M_{SUSY} = \min(m_{\tg},m_{\tu_R})$ to a precision of about
10\%. The choice of $m_{\tu_R}$ to represent the squark mass scale is
arbitrary. Since this method appears to require only a moderately clean
SUSY sample, it should be possible to determine $M_{SUSY}$ for gluinos
and squarks as heavy as 1.5~TeV after about three years of low
luminosity LHC operation.

Hinchliffe {\it et. al.}~\cite{HINCH} studied several different
mSUGRA cases to assess
the prospects for other SUSY measurements at the LHC. They showed, for
instance, that it
might be possible to measure the mass of $h$ produced in gluino
and squark cascades via its $b\bar{b}$ decay. They estimate a precision
of $\pm 1$~GeV on this measurement. 
For details about this, and techniques for other measurements, we
refer the reader to this important study.

\subsection{Determination of Spin at $e^+e^-$ colliders}

If sparticle production occurs via the exchange of a vector boson in the
$s$-channel, it is easy to check that the sparticle angular distribution
is given by, $\sin^2\theta$ for spin zero particles, and
$E^2(1+\cos^2\theta)+m^2\sin^2\theta$ for equal mass spin $\frac{1}{2}$
sparticles. Thus if sparticles are produced with sufficient boost, the
angular distribution of their daughters which will be relatively
strongly correlated to that of the parent, should be sufficient to
distinguish between the two cases.  Chen {\it et. al}~\cite{CHEN} have
shown that, with an integrated luminosity of 500~$pb^{-1}$, it should be
possible to determine the smuon spin at LEP2. A similar
analysis~\cite{JLC} has been performed for a 500~GeV linear collider.

\subsection{Tests of the mSUGRA Framework and Determination of Model
Parameters} 

We begin by recalling that within the mSUGRA GUT framework,
the four parameters, $m_0$, $\mhf$, $\tan\beta$ and
$A_0$, together with the sign of $\mu$ completely determine all
the sparticle masses and couplings. Since the number of observables can
be much larger than the number of parameters, there must exist relations
between observables which can be subjected to experimental tests. In
practice, such tests are complicated by the fact that there are
experimental errors, and further, it may not be possible to cleanly
separate between what, in principle, should be distinct observables;
{\it e.g.}  cross sections for $\eslt$ events from $\tg\tg$, $\tg\tq$
and $\tq\tq$ sources at the LHC.

Because of the clean experimental environment, the simplicity of the
initial state and the availability of polarized beams, many tests can
be most cleanly done at $e^+e^-$ colliders, where we have already seen
that it is possible to determine sparticle masses with a precision of
1-2\%. The determination of the selectron and smuon masses will allow us
to test their equality $m_{\te_L}=m_{\tmu_L}$, $m_{\te_R}=m_{\tmu_R}$ at
the percent level~\cite{MUR,JLC,FUJII} --- the same may be done with staus,
though with a somewhat smaller precision. This is a test of the assumed
universality of slepton masses. A comparable precision is obtained even
if the sleptons do not directly decay to the LSP.~\cite{BMT}

A different test may be possible if both $\tell_R$ and $\tw_1$ are
kinematically accessible and a right-handed electron beam is
available. It is then possible to measure $m_{\tz_1}$, $m_{\tw_1}$,
$\sigma_R(\tell_R\bar{\tell_R})$ and $\sigma_R(\tw_1\overline{\tw_1})$
(note that the chargino cross section for right-handed electron beams
has no contribution from sneutrino exchange!). These four observables
can then be fitted to the four MSSM parameters $\mu$, $\tan\beta$ and
the electroweak gaugino masses $M_1$ and $M_2$.  In
practice~\cite{MUR,JLC}, while $\mu$ may be rather poorly determined if
the chargino is dominantly a gaugino, $\frac{M_1}{M_2}$ is rather
precisely obtained so that it should be possible to test the gaugino
mass unification condition at the few percent level, given 50~$fb^{-1}$
of integrated luminosity.~\footnote{Feng and Strassler~\cite{STRASSLER}
have shown that, with 1~$fb^{-1}$ of data, a test of this relation at
the 20\% level may also be possible at LEP2.} It may further be possible
to determine $m_{\tnu_e}$ by measuring chargino production with a
left-handed electron beam. This is of interest because the difference
between the squared masses of the sneutrino and $\tell_L$ is a direct
test of the $SU(2)_L$ gauge symmetry for sleptons.  For further details and
other interesting tests, we refer the reader to the original
literature.~\cite{MUR,JLC,FUJII}

One would naively assume that analogous tests are much more difficult at
the LHC. This is because measurement of individual
sparticle masses (as opposed to mass differences), which as we have just
seen allows us to directly test various 
mSUGRA assumptions at Linear Colliders, appears to be difficult.
One straightforward approach is to search for correlations
amongst various signals that might be observed (along with bounds on
signals that are not seen) at various colliders. To make these
correlations apparent, it is
convenient to display various signals as predicted by the model
in the $m_0-\mhf$ plane for fixed sets
of values of $\tan\beta$ and $A_0$. One would then attempt to zero in on
the parameters by looking for regions of the plane for which the model
predictions agree with the rates for all signals that are seen in
experiments at the Tevatron, LEP2 and the LHC. One would also check
that predictions for the rates for signals that are {\it not} seen are
indeed below the sensitivity of these experiments. Such a plot would
resemble Fig.~\ref{fig:SUGRA} except with many more contours.

A systematic study of how LHC data could serve to test the mSUGRA
framework was begun by Baer {\it et. al.}~\cite{BCPT2} 
A measurement of $m_{\tg}$ or
$m_{\tz_2}-m_{\tz_1}$ as discussed above would roughly fix $\mhf$, while
a measurement of $m_{\tq}$ (or the gluino squark mass difference) would
yield an estimate of $m_0$. A determination of the relative cross sections
for different event topologies can also yield information about the
underlying parameters. For instance, the ratio of the cross sections for
multilepton plus jets plus $\eslt$ events and for multijet plus $\eslt$
events without leptons is significantly larger if the leptonic branching
fractions of charginos and neutralinos is enhanced. Indeed such an
observation would suggest that sleptons are relatively light (probably
even light enough to be produced via cascade decays of gluinos and
squarks) which, in turn, may lead us to infer that $m_0$ is not very
large.  It was shown~\cite{BCPT2} that while $m_0$ and $\mhf$ may be
roughly determined from LHC data, a determination of $\tan\beta$ or
$A_0$ is difficult.  The LHC subgroup study~\cite{SNOWLHC} of the
Snowmass Point showed that it should be possible to determine $m_0$,
$\mhf$ and $\tan\beta$ with roughly the same precision as at Linear
Colliders.~\footnote{For this study, it was reasonable to assume that
the lighter Higgs boson mass (which is 68~GeV) would be measured at
LEP2. This constraint helps to pin down the value of $\tan\beta$ for
this set of input parameters.} In
fact, their determination of $\mhf$ from the neutralino mass difference
was more precise than the one obtained by the NLC Group. In contrast,
the direct measurement of slepton masses yielded a better determination
of $m_0$ at the NLC. 

The reader may wonder whether the precision
obtained by the LHC study is somehow special to the 
particular choice of parameters.
To some extent, this is indeed the case. The sparticle spectrum for this
point is so light that the SUSY event rate is enormous, so that it is
possible to make extremely stringent cuts to isolate clean signal
samples. Prospects for these measurements for other parameter choices
were examined by Hinchliffe {\it et. al.}~\cite{HINCH} It was found that
$\mhf$ could typically be measured to a few percent --- even for the
extreme case of gluinos and squarks as heavy as 1~TeV, a precision of
10\% was obtained. The precision with which other parameters can be
extracted depends on the scenario. It was found that, for the most part,
it is possible to obtain allowed ranges of $\tan\beta$ and
$m_0$. Sometimes two solutions were obtained, in which case more detailed
measurements would be necessary to discriminate between them. These
studies underscore the capabilities of LHC experiments and contain the
first steps towards an effective measurement strategy at the LHC.

At the LHC, it should be possible to {\it falsify} the mSUGRA framework
(or, for that matter, any other framework specified by a small number of
parameters) 
by a standard $\chi^2$ analysis. If it is not possible to find a
consistent set of parameters that accommodates all the data (and we are
convinced that our experimental friends have not made a mistake!) the
assumed framework will need to be modified or discarded. 
If this turns out to be the case, the direct measurements of several 
sparticle masses (and as we shall
see, some mixing angles) at Linear Colliders will directly
point to the correct theoretical picture. The broader issue of how to
proceed from LHC data to determine the underlying theory appears less
obvious. 

\subsection {Identifying Sparticle Production Mechanisms at the LHC}

At $e^+e^-$ colliders where the centre of mass energy is incrementally
increased, it may be reasonable to suppose that it is unlikely (except,
perhaps, for the sfermion degeneracy expected in the mSUGRA model) that
several particle thresholds will be crossed at the same time. It would
thus be possible to focus on just one new signal at a time, understand
it and then proceed to the next stage.  The situation at the LHC will
be quite different. Several sparticle production processes
will simultaneously occur as soon as the machine turns on, so that
even if it is possible to distinguish new physics from the SM, the issue
of untangling the various sparticle production mechanisms will remain.
For example, even if we attribute a signal in the $\eslt + jets$ channel
to sparticle production, how would we tell whether the underlying
mechanism is the production of just gluinos or a combination of gluinos and
squarks?~\footnote{Here, we tacitly assume that squarks will not be much
lighter than gluinos.}

Some progress has already been made in this direction. We have already
seen that the $\tw_1\tz_2$ source of trileptons can clearly be isolated
from other SUSY processes. The opposite sign dilepton signal from
slepton production is probably distinguishable from the corresponding
signal from chargino pair production since the dileptons from slepton
production always have the same flavour.~\footnote{The extent to which
this channel is contaminated by other SUSY sources has not been
explicitly checked.}
To tell whether squarks are being produced in addition to gluinos, at
least two distinct strategies have been suggested. The first makes use
of the fact that there are more up quarks in the proton than down
quarks.  We thus expect many more $\tu_L\tu_L$ and $\tg\tu_L$ events as
compared to $\td_L\td_L$ and $\td_L\tg$ events at the LHC. As a result,
any substantial production of squarks in addition to gluinos will be
signalled~\cite{BTW2} by a charge asymmetry in the same-sign dilepton
sample: cascade decays of gluinos and squarks from $\tg\tq$ and $\tq\tq$
events lead to a larger cross section for positively charged same sign
dileptons than for negatively charged ones. This has since been
confirmed by detailed studies by the ATLAS collaboration~\cite{ATLAS}
where the SS dilepton charge asymmetry is studied as a function of
$\frac{m_{\tg}}{m_{\tq}}$, and shown to monotonically disappear as this
ratio becomes small. Another method~\cite{BCPT} for distinguishing gluino from
squark {\it and} gluino production relies on
the jet multiplicity in the $\eslt$ sample. 
The idea is to note that $\tq_R$, which are produced as
abundantly as $\tq_L$, frequently decay directly to the LSP via $\tq_R
\to q\tz_1$ and so lead to only one jet (aside from QCD radiation). In
contrast, gluinos decay via $\tg \to q\bar{q}\tw_i$ or $\tg \to
q\bar{q}\tz_i$, so that that gluino decays contain two, and frequently
more, jets from their cascade decays.  Thus the expected jet
multiplicity is lower if squark production forms a substantial fraction
of the $\eslt$ sample.  Of course, since $\langle n_{jet} \rangle$ (from
gluino production) depends on its mass, some idea of $m_{\tg}$ is
necessary for this strategy to prove useful. A detailed
simulation~\cite{BCPT} shows that the mean value of the $n_{jet}$
distribution increases by about $\frac{1}{2}$ unit, when the squark mass
is increased from $m_{\tq}=m_{\tg}$ by about 60-80\%. 

Cascade decays of gluinos and squarks can result in the production
of the Higgs bosons of supersymmetry. It is, therefore, interesting to
ask whether these can be detected in the data sample which has already
been enriched in SUSY events. Neutral Higgs bosons might be
detectable~\cite{BTW2} via an enhancement of the multiplicity of central
$b$-jets in the $\eslt$ or same sign dilepton SUSY samples. Some care
must be exercised in drawing conclusions from this because such
enhancements may also result because third generation squarks happen to
be lighter than the other squarks.\footnote{These may be directly
produced with large cross sections or may lead to enhancement of gluino
decays to third generation fermions as discussed in
Sec.~\ref{sec:decays}.}  It has also been shown~\cite{BCPT,HINCH} that it may
also be possible to reconstruct a mass bump in the $m_{b\bar{b}}$
distribution if there is a significant branching fraction for the decay
$\tz_2 \to h\tz_1$ and $h$ is produced in events with no
other $b$-jets, since otherwise we would have a large combinatorial
background. The charged Higgs boson, if it is light enough, may be
identifiable via the detection of $\tau$ lepton enhancements in SUSY
events~\cite{BTW2} or even in $t \to bH^+$ decays;~\cite{FROID} 
it should, however, be kept in mind that such light
charged Higgs bosons also contribute to the $b \to s\gamma$ decays.

We also saw that in the recent case studies for the LHC
\cite{SNOWLHC,HINCH} it was frequently possible to isolate specific SUSY
production and decay chains by judicious choice of cuts. An example of
this is gluino pair production which yielded a measurement of the mass
difference, 
$m_{\tg}-m_{\tb_1}$, as discussed above.  It appears that while the
complexity and variety of potential SUSY signals precludes us from
writing down a general algorithm that can be used to identify the
production mechanisms that may be present in the LHC data sample, by
studying the features of a {\it given} data set we will be able to infer a
considerable amount. These sort of studies have only just begun, and
considerable work remains to be done. In this respect, at least, it
appears that the analysis of data from Linear Colliders will be simpler.

We stress that the complex cascade decay chains of gluinos and squarks
may be easier to disentangle if we already have some knowledge about the
masses and couplings of the lighter charginos and neutralinos that are
produced in these decays. While it is indeed possible that $\tw_1$ may
be discovered at LEP2 and that its mass is determined there, it is
likely that we may have to wait for experiments at the Linear Collider
to be able to pin down the couplings, and, perhaps, even for discovery
of $\tw_1$ and $\tz_2$.~\footnote{This could especially be the case if
the spoiler decay modes of the neutralino are accessible.}
In this case, a reanalysis of the LHC data in
light of new information that may be gained from these experiments may
prove to be very worthwhile: it may thus be necessary to archive this
data in a form suitable for subsequent reanalysis. Once again, we see
the complementary capabilities of $e^+e^-$ and hadron colliders.

\subsection{Direct Tests of Supersymmetry}

We have already seen that supersymmetry, like any other symmetry,
implies relationships~\footnote{These relations are corrected by
radiative corrections which are generally expected to be smaller than a
few percent.~\cite{HIK}} between various dimensionless couplings in the
theory even if it is softly broken. For example, the
fermion-sfermion-gaugino (or, since the Higgs multiplet also forms a
chiral superfield, the Higgs-Higgsino-gaugino) coupling is completely
determined by the corresponding gauge coupling. A verification of the
relation would be a direct test of the underlying supersymmetry. We
emphasize that such a test would be essentially model independent as it
relies only on the underlying global supersymmetry, and not on any
details such as assumptions about physics at the high scale or even the
sparticle content. In practice, such tests are complicated by the fact 
that the gauginos (or the
Higgs bosons $h_u$ and $h_d$ and the corresponding Higgsinos, or for
that 
matter, the sfermions $\tf_L$ and $\tf_R$) are not
mass eigenstates, so that the mixing pattern has to be disentangled
before this test can be applied. This will require an accurate
measurement of several observables which can then be used to disentangle
the mixing and also simultaneously to measure the relevant coupling.

Feng {\it et. al.}~\cite{FMPT} have argued that such a test can be performed
via a determination of chargino properties. As we have seen, the charged
gaugino and the corresponding Higgsino can mix only if electroweak
symmetry is broken. This is the reason why the off-diagonal terms in the
chargino matrix of Eq.~(\ref{eq:chargino})
are equal to $-\sqrt{2}M_W\cos\beta$ and
$-\sqrt{2}M_W\sin\beta$, respectively. Assuming that the chargino is a
mixture of just one Dirac gaugino and one Dirac Higgsino, the most
general mass matrix would contain four parameters: the two diagonal
elements and the two off-diagonal ones. These latter can always be
parametrized by $-\sqrt{2}M_W^{\chi}\cos\beta^{\chi}$ and
$-\sqrt{2}M_W^{\chi}\sin\beta^{\chi}$.  It is the SUSY constraint on the
Higgs-Higgsino-gaugino coupling that forces $M_W^{\chi}=M_W$.  
A determination of four independent quantities that depend only on these
parameters could then be used to see if the SUSY constraint
$M_W^{\chi}=M_W$ is valid. How best to do this determination depends on
what the underlying parameters are. Here we will merely say that
these SUSY tests can be done at about the 30\% level at a 500~GeV Linear
Collider, and refer the reader to the original paper ~\cite{FMPT} for
further information.~\footnote{A somewhat different test of SUSY which
yields a similar precision when the chargino is mainly gaugino-like has
also been discussed in this study. J.~Feng and N.~Polonsky have observed
that this latter test becomes an order of magnitude more precise if the
sneutrino mass is independently known to a few GeV.}

A more precise test has recently been suggested by Nojiri {\it
et. al.}~\cite{MIHOKO} These authors suggest that an accurate
measurement of the cross section and angular distribution of electrons
produced via $e^+e^- \to \te_R^+\te_R^-$ could lead to a 2\% measurement
of the electron-selectron-hypercharge gaugino coupling if an integrated
luminosity of 100~$fb^{-1}$ is accumulated at Linear colliders. In any SUSY
model, this coupling, at tree level, should be equal to the $U(1)_Y$
gauge coupling $g'$ up to a Clebsch-Gordan coefficient. 
It is instructive to note that a 2\% measurement of
this coupling begins to be sensitive to radiative corrections which, in
turn, would be sensitive to sparticles that may be beyond the kinematic
reach of the machine. 

We are not aware of any proposal for an analogous test at the LHC.

\section{Beyond Minimal Models}\label{sec:nonmin}

Up to now, we have mainly confined our analysis to the MSSM framework. Even
then, as we saw in Sec.~\ref{sec:framework}, the unmanageably large number
of free parameters required us to make additional assumptions in order
to obtain tractable phenomenology. It is clearly impractical to
seriously discuss the phenomenology of the many possible extensions of the MSSM
framework that have been considered. 
Here, we will first list some of the ways in which this
framework may be modified, leaving it to the reader to figure out the
implications for phenomenology. Thinking about this will also help to
view our previous discussion in proper perspective. We will then select
two of these modifications (for reasons explained below) and discuss their
phenomenological implications in more detail.

The MSSM framework may be extended or modified in several ways, roughly
arranged in order of increasing ``non-minimality''.

\begin{enumerate}
\item We may give up the exact universality of the gaugino masses at the
GUT scale. Threshold corrections due to unknown GUT, and perhaps even
gravitational, interactions would certainly yield model-dependent
corrections~\cite{HALLGAUG} which preclude exact unification.  It is
also conceivable that there is, in fact, no grand unification at all,
but the apparent unification of couplings inferred from LEP experiments is a
result~\cite{IBAN} of string type unification; in this case, the gaugino
masses unify at the string scale which is generally somewhat larger than
$M_{GUT}$.  We have already
noted that even in SUGRA type models, we do not really know the exact
scale at which scalar masses unify. 

We also remark that the additional
assumption~\cite{WSFN} of minimal kinetic energy terms is crucial for
obtaining universal scalar masses.  Any deviation from the assumed
university of scalar masses will, at the very least, modify the
conditions of radiative symmetry breaking. The pattern of scalar masses
may also be modified by $D$-terms if the gauge group contains additional
factors. Naively, one would think that if these additional symmetries
are broken at a sufficiently large scale, this would have no impact upon
low energy physics. This is not always the case.~\cite{DTERM} These
terms can significantly alter the pattern of sparticle masses, and
hence, impact upon production and decays of sparticles. Detailed
measurements of scalar masses and branching ratios in future experiments
can potentially~\cite{SNOWTH} lead to the 
discovery of new
physics, at energy scales to which we may not have direct access at
colliders during our lifetimes.

\item $R$-parity may be explicitly broken by superpotential interactions
$g_1$ and $g_2$ in Eq.~(\ref{eq:supL}) and Eq.~(\ref{eq:supB}).

\item SUSY breaking may possibly occur at relatively low energies and
not at $\sim 10^{10}$~GeV as in SUGRA type models where 
gravity is the messenger of
SUSY breaking.
Models~\cite{NELS} where SUSY breaking occurs at relatively low energy ($\sim
10-100$~TeV) and is communicated by ordinary gauge interactions
have recently received a lot of attention. In these Gauge Mediated Low
Energy Supersymmetry Breaking (GMLESB) models the mass patterns, and hence
the phenomenology,
are considerably different from mSUGRA.

\item There could be additional chiral superfields even in the low
energy theory: new generations
(with heavy neutrinos), additional Higgs multiplets, or a right-handed
sneutrino superfield. We certainly do not need new generations or new
Higgs doublets, as they may spoil the apparent unification of couplings.
Higgs fields in higher representations cause additional problems if they
develop a VEV. Higgs singlets cannot be logically
excluded, and are interesting because they allow for new quartic Higgs
boson couplings, though one would have to understand what keeps them
from acquiring GUT or Planck scale masses. A singlet right-handed
sneutrino (note that this is {\it not} the superpartner of the usual
neutrinos) is an interesting possibility since it occurs in $SO(10)$ GUT
models, and also, because it allows for spontaneous breaking of
$R$-parity conservation.~\cite{VALLE}

\item Finally, we could consider models with extended low energy gauge
groups --- either left-right symmetric models~\cite{MOHA} or models with
additional $Z$ bosons.~\cite{HR}

\end{enumerate}

For want of time, we will confine ourselves to items (2) and
(3) above. This is not because these extensions are necessarily
more likely to be correct than the other. We consider $R$-parity
violation because there are no sacred symmetry principles that forbid
these interactions. We incorporated $R$-parity
conservation only because we were motivated to do so for phenomenological
reasons. The conservation of $R$-parity gave us valuable freebies (such as a
candidate for cold dark matter) but this does not mean that it is
necessarily right. We will soon see that it is possible to build
perfectly acceptable models where $R$-parity is not conserved.

On another note, we will see that models where SUSY breaking occurs at
the $PeV$ scale and is communicated to SM particles and their
superpartners by gauge interactions are, like mSUGRA, very economic in
the sense that the low energy theory can be simply parametrized.
These models are very ambitious in that their goal is not
only to include a mechanism for transmission of SUSY breaking, but also
to obtain SUSY breaking dynamically. While this goal is yet to be
realized in a compelling manner, they represent a viable
alternative to the conventional picture. The resulting collider
signatures, however,
differ in important ways from mSUGRA expectations. From our point
of view, this alone is reason enough to pay special attention to this
framework.

\subsection{$R$-parity Violation}
In some sense, including $R$-parity violating interactions results in
the minimal extension of the MSSM because it does not require the
introduction of any new particles. Notice, however, that a general
analysis of this requires the introduction of 48 new parameters in the
superpotential of Eq.~(\ref{eq:supL}) and (\ref{eq:supB}): 3$\mu'$'s,
9~$\lambda$'s, 27~$\lambda'$'s and 9~$\lambda''$'s. There are relations
amongst these couplings in theories with larger symmetries; {\it e.g.}
GUTs.  Many (but not all) products of the baryon- and lepton-number
violating couplings are strongly constrained~\cite{PROBIR} by the
non-observation of proton decay. In fact, it is usually assumed that
only one of $B$ or $L$ violation is possible, since (assuming sparticles
are heavier than the proton) the only spin $\frac{1}{2}$ particles into
which the proton can decay are leptons; {\it i.e.} the proton will be
stable if either $B$ or $L$ is conserved.  In phenomenological analyses,
it is customary (and in light of the large number of new parameters,
convenient) to assume that one of the couplings dominates. Even so,
several of the couplings are strongly constrained.

In a very nice analysis, Barger {\it et. al.}~\cite{BARGRVIOL} have
studied the implications from various experiments --- $\beta$-decay
universality, lepton universality, $\nu_{\mu} e$ scattering, $e^+e^-$
forward-backward asymmetries and $\nu_{\mu}$ deep-inelastic scattering
--- for these new interactions. They find strong constraints on the
lepton-number violating couplings, assuming~\footnote{Constraints from
non-observation of $\mu \to e\gamma$ or $\mu \to 3e$ decays and $\mu N
\to e N$ processes are much stronger if, say, both $e$ and $\mu$ number
violating interactions are large.} that only one of the couplings is
non-zero: for instance, they find that of the $\lambda$-type couplings,
only $\lambda_{131}$ and $\lambda_{133}$ can exceed 0.2 (compare this
with the electromagnetic coupling $e=0.3$) for a SUSY scale of 200~GeV,
though several $\lambda'$ and many more of the $\lambda''$ interactions
can exceed this value. Dimopoulos and Hall~\cite{DH} have, from the
upper limit on the mass of $\nu_e$, obtained a strong bound ($<
10^{-3}$) on $\lambda_{133}$.  The same argument yields a stringent
bound~\cite{GOD} on $\lambda'_{133}$ and a less restrictive, but
significant, bound on $\lambda'_{122}$.  First generation baryon number
violating interactions are strongly constrained from the non-observation
of $n-\bar{n}$ oscillations or $NN \to K\bar{K} X$.~\cite{ZWIRNER}
Constraints~\cite{PROBIR} from rare $B$ processes such as $B^+ \to
K^+\bar{K}^0$ as well as neutral $K$ and $D$ meson mixing limit
$\lambda''$ couplings involving the third family. These couplings are
also constrained by the precision measurements~\cite{BHATT} of $Z^0$
properties at LEP.

The reason to worry about all this is that if $R$-parity is not
conserved, both sparticle production cross sections as well as decay
patterns may be altered.  For instance, if $\lambda'$ interactions are
dominant (with $i=1$), squarks can be singly produced as resonances
~\footnote{The H1 and ZEUS experiments at the HERA collider have
reported an excess of events in high energy $e^+p \to e^+ +X$ scattering
at very high values of $Q^2$. Many theorists have suggested that this
may be an indicator of some novel physics, a popular
interpretation being an $s$-channel resonance in positron-quark
scattering. This could be a spin zero particle with lepton and quark
quantum numbers, the lepto-quark, or a scalar quark with $\lambda'$ type
$R$-violating interactions. It is not clear, however that the results of
the two experiments are mutually any more compatible~\cite{DRHERA} than
the reported deviation from the SM. Unfortunately, it will take a couple
of years for the situation to be clarified.} in $ep$ collisions at
HERA,~\cite{DREINER} or in the case of $\lambda''$ interactions, at
hadron colliders.~\cite{HDE} The production rates will, of course, be
sensitive to the unknown $R$-parity violating couplings. Likewise, if
$R$-parity violating couplings are large compared to gauge couplings,
these $R$-violating interactions will completely alter sparticle decay
patterns.

Even if all the $\lambda$'s are too small (relative to gauge couplings)
to significantly affect the production and decays of sparticles (other
than the LSP), these interactions radically alter the phenomenology
because the LSP decays visibly, so that the classic $\eslt$ signature of
SUSY is no longer viable. Even so, sparticle detection should not be a
problem in the clean environment of $e^+e^-$ colliders. 
In fact, LEP should be able to probe
regions of parameters not explorable in the MSSM since signals from LSP
pair production can now be detected.~\cite{ALEPH}

The viability of SUSY detection at hadron colliders would clearly be
sensitive to 
details of the model.  Two extreme cases where the LSP decays either
purely leptonically into $e$'s or $\mu$'s and neutrinos via
$\lambda$-type couplings, or when it always decays into jets via
$\lambda''$ couplings have been examined for their impact on
Tevatron~\cite{RPV,DP} and LHC~\cite{DGR,BEARRPV} searches for supersymmetry.
The signals, in the former case, are spectacular since the decays of
each LSP yields two leptons in addition to any other leptons from direct
decays of $\tw_1$ or $\tz_2$ produced in the gluino or squark cascade
decays.  With an integrated luminosity of 100~$pb^{-1}$ that has already
been accumulated, experiments at the Tevatron should be able to probe
gluinos as heavy as 500-600~GeV.  In the other case where the LSP decays
purely hadronically, gluino and squark detection is much more difficult
than in the MSSM. The reason is that the $\eslt$ signal is greatly
reduced since neutrinos are now the only sources of $\eslt$. In fact, if
squarks are heavy, there may well be no reach in this channel even at
the Main Injector. Further, the multilepton signals from cascade decays
are also degraded because the jets from LSP decays frequently spoil the
lepton isolation. Indeed if squarks are heavy, none of the SUSY signals
would be observable in this run of the Tevatron; even the Main Injector
will then not probe gluino masses beyond $\sim$200~GeV (350~GeV, if
$m_{\tq}=m_{\tg}$). 

At the LHC, attention had mainly been 
focussed~\cite{DGR} on the same-sign dilepton signal from gluino pair
production.  In the case where the LSP decays purely leptonically, the
gluino mass reach exceeds 1~TeV. A natural question to ask is what
happens if the LSP only decays hadronically into three jets. In
particular, is it possible that SUSY can escape detection in LHC
experiments even though sparticle masses are below 1~TeV? In a recent
study~\cite{BEARRPV} using ISAJET, it was shown that even in this
unfavourable case, experiments at the LHC would be able to detect SUSY
signals in the $1\ell$, $\ell^+\ell^-$, $\ell^{\pm}\ell^{\pm}$ and
$3\ell$ plus multijet channels if gluinos or squarks are lighter than 1~TeV.
This study assumed that sparticle masses and mixing patterns are exactly
as in the mSUGRA model, the only difference being that the LSP decayed
into three quarks via the $\lambda''_{221}$ coupling. It thus seems
unlikely that weak scale supersymmetry will escape detection at the LHC.

\subsection{Gauge-Mediated Low Energy Supersymmetry Breaking}

The set-up in this class of models is similar in several respects to the
SUGRA type models that we discussed in
Sec.~\ref{sec:sugra}. Supersymmetry is again assumed to be dynamically
broken in a hidden sector of the theory. This sector is coupled to a
``messenger sector'' (which then feels the effects of SUSY breaking) by
a new set of gauge interactions.  Some particles in the messenger sector
are assumed also to have SM gauge interactions. These then induce soft
SUSY breaking masses for the sfermions, gauginos and the Higgs
bosons. The effective SUSY breaking scale for the observable sector is
thus suppressed by $M_{mess}$ rather than $M_{Planck}$ as for
gravity-mediated SUSY breaking. We thus expect this scale to be $\sim
\frac{\alpha}{4\pi} \times \mu_s^2/M_{mess}$, where $\mu_s$ is the
induced SUSY breaking scale in the messenger sector, and $\alpha$ is the
fine structure constant for the relevant SM gauge interaction. The
effective SUSY breaking scale in the observable sector can be
100-1000~GeV even if $\mu_s$ and $M_{mess}$ are as small as tens to
hundreds of TeV.

If the effective SUSY breaking scale as well as the particle content of
the low energy theory is the same in GMLESB and mSUGRA models, 
what difference does all this make?
The main difference is that the mass of the gravitino, 
$m_{\tG} \sim  \frac{\mu_s^2}{M_{Planck}}$
is comparable to the weak scale in SUGRA type models,~\footnote{We
ignore the possibility that it can be fine-tuned to be much
smaller.~\cite{ELLIS}} but is tiny if $M_{mess}$ is small. For instance,
for $M_{mess} \sim 100$~TeV, $m_{\tG} \sim 1$~eV. But if gravitinos
interact only with gravitational strength, why should we care? The
point~\cite{FAYET} is that gravitinos, like $W$ bosons, get their mass
via the (super)-Higgs mechanism. As a result,
the coupling of the longitudinal component of
the gravitino of energy $E$ is enhanced by a factor $\frac{E}{m_{\tG}}$ 
in exactly the same way that the coupling of the longitudinal $W$ boson
is enhanced by $\frac{E}{M_W}$. In other words, the effective
``dimensionless'' coupling of longitudinal gravitinos is 
$\sim\frac{E}{M_{Planck}} \times \frac{E}{m_{\tG}}$, where the first
factor corresponds to the usual gravitational coupling and the second
factor to the additional enhancement. For $E \sim 100$~GeV
(corresponding to the weak scale) and $m_{\tG} = 1$~eV, it is easy to
check that this coupling is about $10^{-6}$. The width of a particle of
mass 100~GeV that decays into its superpartner and a longitudinal gravitino
via this coupling is $\sim 10^{-10}$~GeV, corresponding to a lifetime of
$\sim 10^{-13}$~seconds! {\it Thus interactions of very light 
longitudinal gravitinos
are obviously relevant for particle physics, and often even for collider
phenomenology.} 

We do not have the time to delve into details of this class of
models.~\footnote{Recently, there have been several attempts to merge
the hidden and messenger sectors together so that we do not require
three artificially separated sectors. We refer the interested reader to
the literature.~\cite{MURAYAMA}} We will, instead, focus our attention on the simplest version of
this model which we will use to illustrate the differences in the
phenomenology. We refer the reader to Dimopoulos {\it
et. al.}~\cite{THOMAS} for details of this framework as well as variants
of the minimal model. This paper also spells out the underlying
assumptions.

The messenger sector of the minimal GMLESB model
consists of one set of ``quark'' and ``lepton''
superfields in the \underline{5}+\underline{5*} representation of
$SU(5)$ (the inclusion of a complete representation ensures that the
successful prediction of $\sin^2\theta_W$ is not disturbed) coupled to a
singlet via a superpotential $f = \lambda_1\hat{S}\hat{q}\hat{\bar{q}} +
\lambda_2\hat{S}\hat{\ell}\hat{\bar{\ell}}$. The scalar and auxiliary
components of $\hat{S}$ acquire VEVs via their interactions with the
hidden sector, the latter signalling the
breakdown of SUSY. SM gauge interactions induce masses for the gauginos
of the observable sector via one loop quantum corrections. If $\langle F
\rangle \ll \langle S \rangle^2$, these are
given by,~\cite{NELS}
\baeq
\begin{equation}
m_{\tilde{\lambda}_i}=\frac{\alpha_i}{4\pi}\Lambda,
\end{equation}
where $\Lambda = \frac{\langle F \rangle}{\langle S \rangle}$.
The chiral scalars feel the effects of SUSY breaking only via these
gaugino masses, so that SUSY breaking squared scalar masses, which are induced
as two-loop effects, are given by,
\begin{equation}
m_{scalar}^2=2\Lambda^2 \left [ C_3(\frac{\alpha_3}{4\pi})^2 +
C_2(\frac{\alpha_2}{4\pi})^2
+\frac{3}{5}(\frac{Y}{2})^2(\frac{\alpha_1}{4\pi})^2\right ],
\end{equation}
\eaeq
with $\alpha_1$
given in terms of the usual hypercharge coupling $g'$ by
$\alpha_1=\frac{5}{3}\frac{g'^2}{4\pi}$, $C_3=\frac{4}{3}$ for
colour triplets and zero for colour singlets while $C_2=\frac{3}{4}$
for weak doublets and zero for weak singlets. SUSY breaking $A$- and
$B$-parameters are induced only at two-loop order and are small. 

We see that the gaugino masses obey the GUT relation (\ref{eq:gaugino})
although the underlying physics is quite different.  It is also
straightforward to see that squarks are heavier than gluinos in this
minimal framework.  The sfermion mass patterns are quite different from
those in mSUGRA. The squarks are the heaviest, followed by $\tell_L$
followed by $\tell_R$: numerically, $m_{\tq}^2:m_{\tell_L}^2:m_{\tell_R}^2
\simeq 11.6:2.5:1.$  We should regard these masses as being defined at
the scale $M_{mess}$ and evolve these to the weak scale as before. Of
course, if $M_{mess} \sim 100$~TeV, the effect of the evolution on these
masses is  not
as important as in mSUGRA. Radiative breaking nonetheless occurs as
before because the Higgs boson mass parameters at the messeneger scale
are much smaller than the corrseponding $t$ and $b$-squark masses. Since
the $A$-parameter is only induced at higher loops, we can take $A$ to be
zero at $Q=M_{mess}$. Also as before, we eliminate~\footnote{This may
not yield a small value of $B$ at the messenger scale. If we incorporate
the constraint $B(M_{mess}) \simeq 0$, then we will find that
$\tan\beta$ is large.~\cite{BABU} We do not include this constraint for
reasons discussed elsewhere.~\cite{GMLESB}} $B$ in favour of $\tan\beta$
so that the model is completely specified by the parameter set,
\begin{displaymath}
(\Lambda, \tan\beta, M_{mess}, \sgn\mu)
\end{displaymath}
We see that $\Lambda$ sets the scale for sparticle masses, and is thus
the most important of these parameters. As in the mSUGRA framework, we
expect that the phenomenology is not very sensitive to $\tan\beta$ or
$\sgn\mu$. The messenger mass $M_{mess}$ only serves to determine the scale
at which the mass relations are to be used as boundary conditions, so
that any dependence on it is, presumably, logarithmic. 

It is instructive to note that the gauge-mediation ansatz automatically
guarantees that squarks (and also sleptons) with the same gauge quantum
numbers will have the same mass. In particular, the first two
generations of $\tq_L$ (also, separately, $\tq_R$) are degenerate
{\it without invoking the need for additional symmetries such as the
global $U(N)$ that was needed in mSUGRA}. These models are, therefore,
advertised as naturally being free of flavour changing neutral current
problems. That this is being somewhat oversold can be seen by noting that
the messenger ``leptons'' and one of the MSSM Higgs doublets has the
same quantum numbers. There is, therefore, no reason why the messenger
``sleptons'' cannot couple to the quarks in the same way as the
usual Higgs scalar. Then, we will have more than one scalar
with Yukawa coupling to the same quarks, which, as is well
known,~\cite{GLASHOW} leads 
to FCNC problems. Even in these models a discrete symmetry seems
necessary to prevent these couplings.~\footnote{Recall, however, our
discussion about continuous global symmetries in
Sec.~\ref{sec:sugra}.} A definite disadvantage of this
framework {\it vis \`a vis} mSUGRA is that we lose $\tz_1$ as a candidate for
cold dark matter.

Phenomenologically, the major difference comes from the fact that
$\tz_1$ which is frequently lighter than all sparticles other than the
gravitino, is now unstable, and can decay via $\tz_1 \to \gamma\tG$, and
possibly also via $\tz_1 \to Z\tG$ or $\tz_1 \to H_i\tG$ ($H_i=h,H,A$). 
The branching
fraction for other sparticles to directly decay to the gravitino are
small since, as we saw, the effective dimensionless gravitino coupling
was much smaller than any of the gauge couplings. Thus heavy sparticles
cascade decay to lighter sparticles exactly as in the MSSM (with masses
and mixing angles 
fixed to be as given by the GMLESB model) until the next to lightest
superparticle (NLSP) is reached. The NLSP is, however, unstable and,
depending on the messenger scale,~\cite{GMLESB,BAGGER} may decay inside the
detector.~\footnote{If the NLSP lives long enough so that it decays outside
the detector, collider phenomenology will be essentially the same 
as in the MSSM.}
If this is $\tz_1$, it will decay via
$\tz_1 \to \gamma\tG$; the gravitino escapes undetected, so that SUSY
event topologies are characterized by multijet plus multilepton plus
$\eslt$ {\it together with two photons} (not both of which will be
necessarily detected in the experimental apparatus) in the final
state. It is worth mentioning that the lifetime of $\tz_1$ can be rather
long so that the photons need not emerge from the primary vertex in the
event. In fact, the gap between the primary and secondary vertices can
yield a measure of the messenger scale. At the LHC where event rate is
generally not a problem, it may be experimentally easier to measure this gap
via the subdominant decay $\tz_1 
\to \tG e^+e^-$ of the neutralino.

The CDF experiment has seen one event which caused a considerable amount
of excitement amongst the champions~\cite{CHAMPS} of these
models. Specifically, they saw an event with an isolated $e^+e^-$ pair
together with a pair of hard, isolated photons and $\eslt$. This event
was interpreted as the selectron pair production with the subsequent
decay $\te \to e\tz_1 \to e \gamma \tG$ of each selectron. To see if
this explanation is viable, Baer {\it et. al.}~\cite{GMLESB} computed the cross
sections in the various event topologies that would be expected at the
Tevatron as a function of $\Lambda$ which, we remind the reader, sets
the scale of sparticle masses. As in the mSUGRA framework, they
found that multijet plus multileptons (plus photon) events occur at a
significantly larger rate than clean multilepton events without jet
activity. If $\Lambda$ is adjusted so that one $e^+e^-\gamma\gamma$
event is expected in the current run of the Tevatron, they showed that
several tens of $\gamma\gamma$ plus multijet plus multilepton and
isolated $\gamma$ plus multijet plus multilepton events should also have
been present after experimental cuts. SM backgrounds to these
characteristic event topologies are small, and it is difficult to
imagine how these events could have escaped detection. This conclusion,
it is argued, is valid even if the messenger sector is more complicated.
The GMLESB explanation of the CDF event, therefore, appears to be
unlikely. Indeed a very recent analysis by the D0
Collaboration~\cite{GMD0} finds that the $\eslt$ spectrum for the $pp \to
\gamma\gamma + X$ channel at the Tevatron appears to be in complete
agreement with SM expectation. In particular, there is no excess at the high
$\eslt$ end as would be expected in the GMLESB framework.

Before closing, we should also mention that the NLSP need not
necessarily be $\tz_1$. Since it is unstable, the cosmological
constraints that we have discussed do not apply so that it may even be
charged. In fact, for large values of $\tan\beta$, $\ttau_1$ is often
the NLSP and decays via $\ttau_1 \to \tau\tG$. In this case, all SUSY
events would contain multiple, isolated $\tau$ leptons in the final
state instead of photons and collider signals would be correspondingly
altered.~\cite{DUTTA}

\section{Concluding Remarks}\label{concl}

We have seen that experiments at the LHC should be able to explore
essentially the whole parameter space of weak scale supersymmetry if we
require that sparticles provide the degrees of freedom that stabilize
the electroweak symmetry breaking sector.  While experiments at Tevatron
upgrades or LEP2 will explore substantial regions of this parameter
space, and maybe even discover sparticles, a non-observation of any
signal should not be regarded as disheartening: the expected mass scale
is several hundred GeV up to a TeV, and so may well not be accessible
except at supercolliders. Electron-positron linear colliders, with a
centre of mass energy of 500-1000~GeV should also be able to discover
sparticles (almost certainly so if the frequently assumed unification
condition for gaugino masses is correct). Linear colliders are the ideal
facility for the discovery and subsequent detailed study of Higgs
bosons.~\footnote{Although we have not discussed muon colliders in these
Lectures, it is worth mentioning that because of the larger value of
$m_{\mu}$, MSSM (and SM) Higgs boson can be produced as $s$-channel
resonances at these machines. It has been shown~\cite{VERNON} that at a
500~GeV muon collider, $h$
should be distinguishable from the SM Higgs boson over a wide range of
parameters, and further, that it should be possible to discover $H$ and
$A$ if their mass is smaller than $\sqrt{s}$. The integrated luminosity,
beam resolution, as well as machine and detector features that are
needed for these measurements have been delineated in this study to
which we refer the interested reader.}

The mSUGRA model that we have described in Sec.~\ref{sec:sugra}
provides a consistent and calculable framework for SUSY phenomenology.  
It is consistent with all accelerator, astrophysical and cosmological
data, with grand unification, and can incorporate
(though not explain) the observed pattern of electroweak symmetry
breaking. Furthermore, because SUSY is a decoupling theory in
that virtual effects of sparticles become suppressed if their masses are
much larger than $M_Z$, the observed agreement of the SM with LEP
constraints is simply incorporated.  These models also provide a natural
candidate for cold dark matter. 

We have seen, however, that the mSUGRA framework is based on
extrapolation of the symmetries of physics at very high scales. It is
important to keep in mind that one or more of its underlying
assumptions may prove to be incorrect. This is especially important when
considering the design of future high energy physics facilities. While
it is reasonable to use the model as a guide, it is important to examine
just how sensitively the various signals depend on these assumptions.
The important thing, however, is that these assumptions will be testable
in future experiments. For instance, even a partial determination of the
pattern of sparticle masses, about which we will get information from
experiments at the LHC and at Linear Colliders, will serve to guide us
to the physics of SUSY breaking. We will be able to learn whether the
ideas underlying mSUGRA, GMLESB, or other
alternatives~\cite{PESKIN,SNOWTH} that we have not been able to discuss
are correct. All experimental measurements --- not just sparticle masses
--- will be useful for this purpose. We may learn about gaugino-Higgsino
mixing via knowledge of their decay patterns, while a study of third
generation sfermions may provide information~\cite{MIH} about their
intra-generational mixing
(which again may serve to discriminate between models). 

Experiments at supercolliders are essential both for
a complete exploration of the entire parameter space, and for the
elucidation of any new phenomena that might be discovered. Experiments
at the LHC and TeV Linear Colliders will unambiguously
discover or exclude weak scale supersymmetry. 
Together, these facilities will allow a comprehensive
study of sparticle properties (if SUSY is discovered)
which, in turn, will yield information about
physics at higher energy scales. Even if SUSY turns out not to be The
Answer, we will almost certainly learn something new, and probably
unexpected, in these experiments.

Before closing, we should remind ourselves that SUSY theories, in spite
of all the attention that they have received, are not a panacea.
Supersymmetry really
addresses a single (but very important) issue: how is electroweak
symmetry broken?  It does not shed {\it any} light on the other
shortcomings of the SM. For example, SUSY has nothing to say about the
pattern of fermion masses and mixings, the replication of
generations, the choice of the gauge group or of the particle multiplets.
While there are new sources of $CP$ violation in SUSY
theories, it is fair to say that SUSY models do not really explain the
origin of this. Finally, even in SUSY theories, the cosmological
constant needs to be severely fine-tuned to be consistent with observation.
Supersymmetric theories also cause new problems not
present in the SM. We should ask:
\begin{itemize}
\item Why are baryon and lepton number conserved at low
energy when it is possible to write dimension four
$SU(3)_C \times SU(2)_L \times U(1)_Y$ invariant interactions that allow for
their non-conservation? Perhaps this tells us something about
symmetries at the high scale.

\item Why is the supersymmetric parameter $\mu \sim
M_{Weak}$? 

\item What is the origin of SUSY breaking and why are SUSY breaking
parameters fifteen orders of magnitude smaller than the Planck scale?

\item Why are $CP$ violation and FCNC from new SUSY sources so small? 

\end{itemize}

We do not know the answers to these and probably several other
questions, although many interesting suggestions exist in the literature.
Perhaps clues to some of these questions lie in the unknown
mechanism of SUSY breaking. We really need guidance from experiment in
order to know which directions are fruitful for theory to pursue. 
We should, of course,
always keep open the possibility that it is not supersymmetry, but some
totally different mechanism that is responsible for stabilizing the
electroweak scale. Only experiments can tell whether weak scale
supersymmetry is realized in nature. What is clear, however, is that the
exploration of the TeV scale will provide essential clues for
unravelling the nature of electroweak symmetry breaking interactions.
We must look to see what we find.
\begin{center}
{\bf ACKNOWLEDGEMENTS} 
\end{center}
I thank Sergio Novaes and Jo\~ao Barata for inviting me to
lecture at this
School, and for the magnificent arrangements in Campos do Jord\~ao.  I am
also grateful to Oscar Eboli for facilitating my trip to Brazil, as
well as
for arranging my visit to the University of S\~ao Paulo. I thank them
all, and also many new friends, for their gracious hospitality.
Memories of Tucupi and Rubaiyat are still fresh. I thank H.~Baer,
V.~Barger, M.~Drees and T.~ter Veldhuis for
their valuable comments on the manuscript. It is also a pleasure to
thank numerous collaborators and colleagues for the many discussions on
the subject of supersymmetry. Without their input, these Lectures would
not have been possible. This research is supported, in part, by the
U.S. Department of Energy grant DE-FG-03-94ER40833.

\end{document}